\documentclass[10pt,a4paper]{article}

\usepackage[latin1]{inputenc}
\usepackage[english,conditional,light]{draftcopy}
\usepackage[english]{babel}
\usepackage[none]{hyphenat}
\usepackage[pdftex]{graphicx}
\usepackage{amssymb}
\usepackage{amsmath}
\usepackage{xspace}
\usepackage{ifthen}
\usepackage{color}
\usepackage{layout}
\usepackage{caption}
\usepackage{here}
\usepackage{picins}
\usepackage{url}
\usepackage[pdftex,colorlinks=false,pdfborder=0 0 0, breaklinks=true,bookmarks=true,bookmarksopen=true,
bookmarksnumbered=true,pdfstartview={XYZ}]{hyperref}
\usepackage{commands}
\usepackage{longtable}
\usepackage{floatflt}
\usepackage{wrapfig}
\usepackage{nicefrac}

\usepackage[body={450pt,700pt}]{geometry}

\begin{document}

\title{The observation of Extensive Air Showers from Space}

\author{
M. Pallavicini, 
R. Pesce, 
A. Petrolini and 
A. Thea.
\\
Dipartimento di Fisica dell'Universit\`a di Genova and INFN,
\\
via Dodecaneso 33, 
I-16146, 
Genova, 
Italy.
}

\maketitle

\begin{abstract}

We summarise some basic issues relevant to the optimisation and design
of space-based experiments 
for the observation of the Extensive Air Showers 
produced by Ultra-High Energy Cosmic Particles interacting with the atmosphere.
A number of basic relations is derived and discussed with a twofold
goal: defining requirements for the experimental apparatus and
estimating the exptected performance.

\end{abstract}

%------------------------------------------------------------------------------
\newboolean{full}
\setboolean{full}{false}
\ifthenelse{\boolean{full}}		{\typeout{full=yes}}		{\typeout{full=no}}
%------------------------------------------------------------------------------

\newcommand {\ensm}			{\ensuremath}
\newcommand {\DD}[1]                    {\ensm{\mathinner{\Delta#1}}}
\newcommand {\abs}[1]			{\ensm{\left\lvert{#1}\right\rvert}}
\newcommand {\ve}[1]			{\ensm{\boldsymbol{#1}}}
\newcommand {\defuni}[1]		{\ifmmode \mathrm{#1} \else ${#1}$ \fi}
\newcommand {\um}[1]			{\defuni{\; #1}}
\renewcommand {\sci}[2]                   {\mbox{\ensm{ #1 \! \cdot \! 10^{#2} }}}
\newcommand {\scix}[1]                  {\mbox{\ensm{ 10^{#1} }}}
\newcommand {\gcm }			{\um{g/cm^2}}
\renewcommand {\degr}			{\ensuremath{^{\circ}}\xspace}
\renewcommand {\Cel}			{\um{\!\!\!\mbox{ }^{\circ}\mathrm{C}}}

\newcommand {\pton}[1]			{\ensm{\left(  #1 \right)}}
\newcommand {\pqua}[1]			{\ensm{\left[  #1 \right]}}
\newcommand {\pgra}[1]			{\ensm{\left\{ #1 \right\}}}

\newcommand {\minispazio}		{\;\;\;}
\newcommand {\spazio}			{\;\;\;\;\;\;}
\newcommand {\punto}			{\minispazio .}
\newcommand {\virgola}			{\minispazio ,}

\newcommand {\veps}			{\varepsilon}

\newcommand {\beq}			{\begin{equation}}
\newcommand {\eeq}			{\end{equation}}

\renewcommand{\deriv}[2]                  {\cfrac{\diffl{#1}}{\diffl{#2}}}
\newcommand{\derivsq}[2]                {\cfrac{\difflsq{#1}}{\diffl{#2}^2}}
\newcommand{\derivsqmix}[3]             {\cfrac{\difflsq{#1}}{\diffl{#2}\diffl{#3}}}

\renewcommand{\Fnumb}		{\ensuremath{f \! / \! \#}\xspace}
\renewcommand{\Cherenkov}	{Cherenkov\xspace}
\renewcommand{\FoV}		{FoV\xspace}
\renewcommand{\Xmax}		{\ensuremath{X_{\mathrm{M}}}\xspace}
\renewcommand{\Xris}		{\ensuremath{X_{\mathrm{R}}}\xspace}
\renewcommand{\Nmax}		{\ensuremath{N_{\mathrm{M}}}\xspace}
\renewcommand{\Gmax}		{\ensuremath{\gamma_{\mathrm{M}}}\xspace}
\renewcommand{\UHECR}		{{\cal UHECR}\xspace}
\newcommand{\UHECP}		{{\cal UHECP}\xspace}
\renewcommand{\EAS}		{{\cal EAS}\xspace}
\renewcommand{\EUSO}		{{\cal EUSO}\xspace}

\newcommand {\com}[1]   
	    {\ifthenelse{\boolean{full}}
            {{	
		  {{ \bf\it #1 }}
	    }}    
	    { }} 

\renewcommand{\FIXME}[1]{\com{\textbf{\textcolor{blue}{\newline\emph{TO CHECK !!!}}}{#1}\xspace}}

\newcommand{\xyz}[1]{\com{\textbf{\textcolor{blue}{\newline\emph{ADD/CHANGE :::}}}{#1}\xspace}}

\newcommand{\FoVAzimuthAngle}{\ensm{\phi}}

\renewcommand{\FoVShowerZenith}		{\ensuremath{\psi_{z}}\xspace}
\renewcommand{\FoVShowerAzimuth}	{\ensuremath{\psi_{a}}\xspace}
\renewcommand{\Nmax}			{\ensuremath{N_{0}}\xspace}

\tableofcontents

%--------------------------------------------------------------------------------
\section{Introduction}
\label{sec:Intro}
%--------------------------------------------------------------------------------

Ultra-High Energy Cosmic Particles (\UHECP) with energies in excess of
$\approx \scix{19} \um{eV}$ hit the Earth with a very low flux of about
one particle $\cdot \um{km^{-2}\cdot sr^{-1}\cdot millennium^{-1}}$, for
particles with energy $ E\gtrsim\scix{20} \um{eV} $~\cite{Abraham:2008ru}.

The observation of \UHECP and the interpretation of the related
phenomenology is one of the most challenging topics of modern
High-Energy Astro-Particle Physics.  Direct detection is impossible at
these energies, due to the exceedingly low flux, but \UHECP can be
detected by observing the Extensive Air Showers (\EAS) produced by the
interaction of the primary particle with the Earth atmosphere.
See~\cite{Westerhoff,Watson,Petrera:2008rf}, and references therein, for recent
reviews on these topics.

The ground-based Pierre Auger Observatory (PAO)~\cite{Watson:2008zzb} is
currently taking data: its south site in Argentina and its forthcoming
north site in the US will provide in the next few years, a clear
understanding of many important topics~\cite{Abraham:2007si,
Abraham:2008ru}.  However it is likely that the next generation of
experiments for the study of \UHECP, after PAO, will be space-based, in
order to increase the event statistics by exploiting the larger
instantaneous geometrical aperture which can be obtained by a space
observatory with respect to ground based experiments.

The aim of this paper is to discuss and summarise a few key issues
relevant to the design and optimisation of space-based experiments for
the observation of \UHECP.  Analytical and semi-analytical relations
will be presented and discussed, in order to roughly set the scenario
for space-based observation of \EAS.

A first discussion of these arguments can be found in~\cite{arisaka},
which was a starting point for many of the results derived here.
  
The results we present are a basic input to the design and optimisation
of the experimental apparatus.  Obviously these results cannot replace a
full Monte-Carlo simulation, for detailed studies. However they are
exceedingly useful to improve the basic understanding, for a fast
outlook and for a rough cross-check of the detailed Monte-Carlo
simulation results.  The basics of the optimisation of the experiment
design can be easily understood using the results we will present,
leaving the hard work of a full Monte-Carlo simulation to a second
phase.  In fact the very many important parameters affecting the
performance do not allow to perform a full Monte-Carlo simulation of all
the possible cases: the results presented here provide therefore valuable starting
points for detailed Monte-Carlo simulations.

A full Monte-Carlo simulation was developed in the framework of the
\EUSO Collaboration: ESAF~\cite{esaf1,esaf2}, allowing to extract
detailed predictions.
In fact most of the work presented in this paper is the result of the
development of the \EUSO experiment~\cite{EUSO}, a path-finder mission
of the European Space Agency (ESA), which successfully completed its
phase A study in 2004 but was frozen afterwards due to programmatic and
financial reasons.  A large part of the scientific community is now
looking forward to both the path-finder Japan-lead mission
JEM-EUSO~\cite{jemeuso} and to the most challenging mission
(super-EUSO~\cite{supereuso}) in the framework of the ESA Cosmic Vision
program 2015-2025~\cite{cosmicvision}.

The results presented in this paper were extensively used in the concept
study of the \EUSO~\cite{EUSORedBook} and super-\EUSO~\cite{supereuso}
experiments.

In fact the design of a space-based apparatus for the observation of
\UHECP is a very challenging task with very little design margins, as it
will be clear from the results we present. It is therefore important
both to optimise the design from the beginning and to set safe design
margins since the beginning of the concept study itself.

%--------------------------------------------------------------------------------
\section{The observational approach}
\label{sec:Approach}
%--------------------------------------------------------------------------------

John Linsley~\cite{linsley}, in 1982, first suggested 
that the Earth atmosphere at night, 
viewed from space, constitutes 
a huge calorimeter for remotely observing  \UHECP (SOCRAS).
Yoshiyuki Takahashi,
together with John Linsley and Livio Scarsi, resurrected the original idea in 1995, when the
concept seemed to be close to be technically feasible (Space AirWatch). 
Since then a number of proposals and studies were carried on, including the 
OWL~\cite{owl} (Orbiting Wide-angle Light-collectors) project, 
the TUS/KLYPVE~\cite{tus} project and \EUSO~\cite{EUSO}. 
In more recent times the JEM-EUSO~\cite{jemeuso} Collaboration aims to propose again the \EUSO
concept on the International Space Station 
while the super-EUSO proposal~\cite{supereuso}, for a most challenging next-generation
experiment,
has been recommended by ESA for technological developments.

A space-based experiment can detect the near-UV air scintillation light isotropically
produced during the \EAS development in the atmosphere by the interaction of
the \EAS secondary particles with the air molecules. The measurement of the
scintillation light allows to record the \EAS development. 
Additional information can be gathered 
by observing the \Cherenkov light diffusely reflected at the Earth surface 
(by land, sea or clouds). The Earth atmosphere plays the role of a giant
calorimeter, although it is passive, continuously changing and outside the
human control. 

This approach is complementary to ground-based observations.  In fact \EAS
develop close to the Earth surface. Therefore, thanks to the large average
distance from the \EAS of an Earth orbiting apparatus, a large \FoV apparatus
can watch a huge target of atmosphere, so that the space-based observation is
best suited for observing very low fluxes and/or particles weakly interacting
with the atmosphere.  Nevertheless, due to the larger distance from the \EAS
with respect to ground-based experiments, the signal is much fainter and
therefore the \EAS reconstruction is worse than at ground-based
experiments.

The required apparatus is an Earth-watching 
large aperture, large \FoV, fast and highly pixelised 
digital camera for detecting near-UV single Photo-ns 
superimposed on a huge background,
capable of three to five years of operation in space, at least.

Any \EAS is seen by the apparatus as a point moving on its Focal Surface (FS) with a
direction and an angular velocity depending on the \EAS primary direction with respect to the
line-of-sight. 
These characteristics allow one to distinguish the \EAS from
the various types of background (see section~\ref{subsec:NoiseBackground}), 
because, typically, those 
have a different space-time development.

\begin{figure}[htbp]
\centering
	\includegraphics[width=\textwidth]{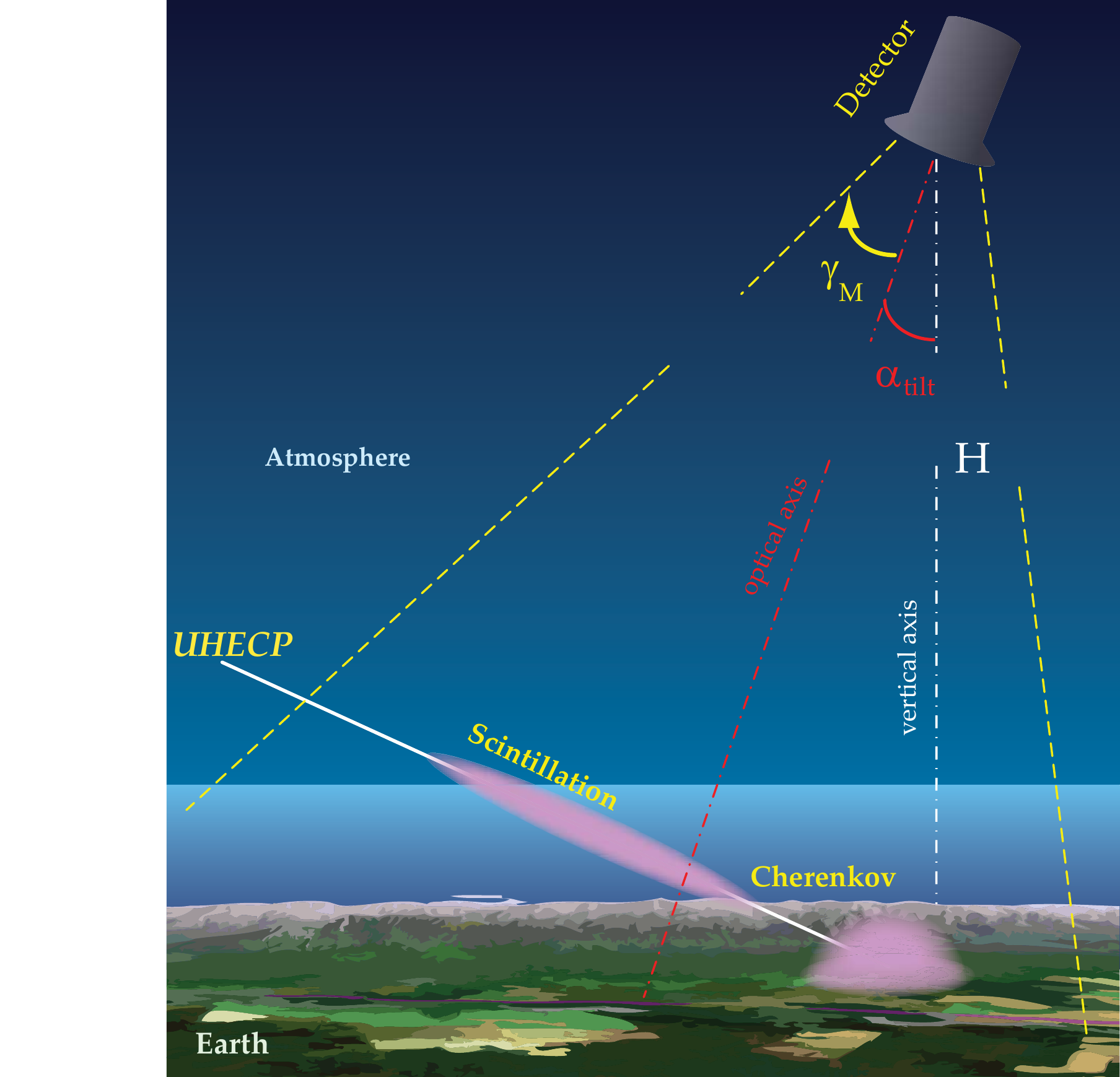}
	\caption{AirWatch observational approach: $H$ is the
	orbital height, $\Gmax$, the \FoV half-angle and
	\TiltAngle is the tilt angle between the optical axis and the
	local nadir.}
	\label{fig:approach}
\end{figure}

The general scheme of the observational approach is shown in figure~\ref{fig:approach}.

%--------------------------------------------------------------------------------
\section{Scientific requirements}
\label{sec:SciReqs}
%--------------------------------------------------------------------------------

The following typical scientific requirements for \UHECP observation
from space are assumed
for indication purposes only (see~\cite{supereuso}). Depending on the scientific
objectives different requirements might be used by adapting the
relations to the case considered.

\begin{itemize}

\item 
All sky coverage.

\item 
Spatial granularity and resolution of the \EAS image projected at the Earth:
$\DD{\ell} \sim 0.5 \um{km} $ or better, in order to ensure a good enough
reconstruction of the \UHECP.

\item
Sampling time of the \EAS signal of order of $\DD{t} \sim \um{\mu s}$ or
better, in order to ensure a good enough
reconstruction of the \UHECP.

\item 
Angular resolution on the reconstructed primary \UHECP direction: 
$\DD{\psi} \lesssim (1\degr \div 3\degr)$
for a large enough subsample of events, in order to allow source
identification and taking into account the deviation induced by magnetic
fields on charged particles.

\item 
Energy resolution: $\DD{E}/E \lesssim 0.3$.

\item 
Resolution on the depth of the \EAS maximum measurement 
$\Delta{\Xmax} \lesssim 50 \um{g/cm^2}$ (accounting for the intrinsic variability of the \EAS development).

\item 
Energy threshold: $ E_\mathrm{TH} \approx \sci{1}{19} \um{eV} $,
with a flat efficiency plateau at $E \gtrsim  E_\mathrm{TH} $ 
to keep systematic effects well under control.
This alos ensures 
a fair overlap with the energy spectrum observed by ground-based experiments.

\item
Capability to measure \EAS with energies up to 
$ E_\mathrm{MAX} \approx \scix{21} \um{eV} $.

\item
Instantaneous geometrical aperture, $ \GeoAperture $, one order of magnitude larger than
currently existing and/or planned ground-based experiments:
$ \GeoAperture \gtrsim \scix{6} \um{km^2 \cdot sr}$. 

\end{itemize}

%--------------------------------------------------------------------------------
\subsection{The experimental apparatus}
%--------------------------------------------------------------------------------

The design of an apparatus for the \UHECP detection from space is a very
challenging task, because of the low expected rate of events, the faint
signal from any \EAS and the tight technical constraints (especially on
mass, power, volume and telemetry) imposed on a space 
experiment as well as the harsh space environment.

The engineering is very complex and the design has a very strong
impact on the scientific performances. A careful optimisation is therefore
mandatory.

%--------------------------------------------------------------------------------
\subsection{Architecture of the instrument}
%--------------------------------------------------------------------------------

The required experimental apparatus is made of the following parts.

\begin{itemize}
\item
The main digital camera, which is assumed to have cylindrical symmetry
around the optical axis, operating in the near-UV, a large aperture,
large FoV, fast and pixelized single Photo-n detector, consisting of:
\begin{itemize}
\item
the main reflective deployable optics; it consists of:
\begin{itemize}
\item
the main mirror: a large, lightweight, segmented, nearly spherical, deployable
mirror;
\item
the corrector plate on the entrance pupil which needs to be deployable as well;
\item
the optical filters;
\item
active control mechanism for both the mirror and the corrector plate;
\item
supporting structure.
\end{itemize}
\item
The Photo--Detector (PD) on the focal surface (FS) of the optics; it consists of:
\begin{itemize}
\item
a bidimensional array of sensors;
\item
the light-collection system on the sensor;
\item
the f/e electronics chip and ancillary electronics;
\item
the housing;
\item
the back-end, trigger and on-board data-handling electronics.
\end{itemize}
\end{itemize}
\item
The Atmospheric Monitoring System (AM), capabile to monitor the relevant
atmospheric properties by a suitable instrumentation.
%%%\begin{itemize}
%%%\item
%%%a dedicated LIDAR;
%%%\item
%%%an infrared camera;.
%%%\end{itemize}
\item
The Monitoring, Alignment and Self-Calibration system (MAC).
\item
A Radio Pulse Detection system (RPD).  
\item
The Central Control Unit (CCU), providing the intelligence to all the systems.
\item
The system and other ancillary parts including:
\begin{itemize}
\item
the mechanical structure, including the external protection;
\item
the thermal control system;
\item
the power system, including solar panels and batteries;
\item
the shutter;
\item
the optical baffle.
\end{itemize}
\item
The ground-support equipment.
\end{itemize}

%--------------------------------------------------------------------------------
\subsubsection{Mission requirements}
%--------------------------------------------------------------------------------

The following mission requirements are envisaged.

\begin{itemize}

\item
An accurately designed orbit (see~\ref{sec:Orbit}).
The ground-track of the satellite orbit needs to be optimized according
to the following parameters: minimization of natural and man-made light
entering the FoV (Sun, Moon and any other light source); 
maximization of the rate of passages above a few fixed points at the
Earth surface, for both exploiting ground calibration sources and hybrid
observations and cross-calibrations with ground-based experiments. 
A free-flyer can give many degrees of freedom in the choice of an
optimized orbit and the orbit strongly affects both the performance and
the operations. A careful orbit optimization is a complex task requiring
a full phase-A study. However it is not considered to be a critical
issue at this stage as the orbital parameters can be varied to a large
extent. 

\item 
Observation of a mass of atmosphere as large as possible, attaining a good
enough geometrical acceptance which implies a large enough \FoV.

\item 
No \FoV obstruction and no parasitic lights coming from any other spatial
device nor from the Earth.

\item
Dimensions after deploying shall be dictated by the optics, including
the optical baffle and the optical shutter.

\item
Light-tightness:
the interior of the instrument shall be light tight such that the
parasitic lights impinging onto the Photo--Detector will be two orders of
magnitude less than the expected night-glow background rate
in order not to spoil the energy resoluzion.

\item
Thermal control shall be provided to stabilize the large surfaces of the
instrument with large power consumption.

\item
Electrical requirements: the scientific operations will have a low duty cycle. It is envisaged
that during 2/3 of the orbit the Instrument will be in a standby status
of low-power consumption. The power consumption quoted for operations is
thus required during roughly 1/3 of the orbit only.

\item
Attitude: approximate nadir pointing; pointing accuracy is not a
critical factor provided the absolute direction of the instrument axis is
known/measured for off-line use. It is assumed nadir pointing to within
a few degrees with pointing direction known offline to within 
$\Delta \chi \simeq 0.5 \degr $, well below the
expected angular resolution of the instrument.

\item
Telemetry and telecommanding: the expected event rate will be high and
it will depend on the final orbit, affecting the event rates. 
A precise estimation is currently lacking.

\item
Required lifetime: five years minimum, ten years goal.

\end{itemize}

%--------------------------------------------------------------------------------
\subsubsection{Requirements for the digital camera}
%--------------------------------------------------------------------------------

The basic requirements for the apparatus follow.

\begin{itemize}

\item
Single Photo-n detection in the wavelength range 
$\mathrm{WR} \equiv \pgra{ 330 \um{nm} \div 400 \um{nm} }$
to detect the air scintillation signal.
Shorter wavelengths suffer absorption from the ozone layer.

\item 
The faintness of the signal requires 
high photon collection capabilities and photon detection efficiency, as well as
low noise (both intrinsic and externally generated).
Small cross-talk and after-pulse rates are required to keep a good enough energy
resolution.

\item 
A dynamic range spanning some three orders of magnitudes in the \EAS energy,
that is in the signal, in order to cover the range from the required threshold energy up to 
$ E_\mathrm{MAX}$.

\item 
An efficient and selective trigger system, to achieve a good background
rejection on-board, and a powerful on-board data handling system.

\item 
Complete modularity of the apparatus to reduce the risk of single point
failures.

\item 
Compatibility with the constraints relevant to a space mission including mass,
power, volume, telemetry, as well as the many environmental factors.

\end{itemize}

%--------------------------------------------------------------------------------
\section{Definitions and assumptions}
\label{sec:GenAss}
%--------------------------------------------------------------------------------

We will use some simplifying assumptions in order to deduce the basic relations 
and use analytical or semi-analytical
relations whenever possible.

%--------------------------------------------------------------------------------
\subsection{Some geometry definitions}
%--------------------------------------------------------------------------------

The \EAS properties depend, to a first approximation, on the primary particle identity, on its energy
$E$ and zenith angle, $\theta$.
The features of the \EAS image also depend on the location of the \EAS image inside the \FoV of the apparatus, 
as defined by the field-angle $\gamma$ of a specific point of the \EAS image,
and the azimuth angle of the \EAS image projected at the Earth surface. 

\begin{figure}[htb]
\begin{center}
\includegraphics[width=0.7\textwidth]{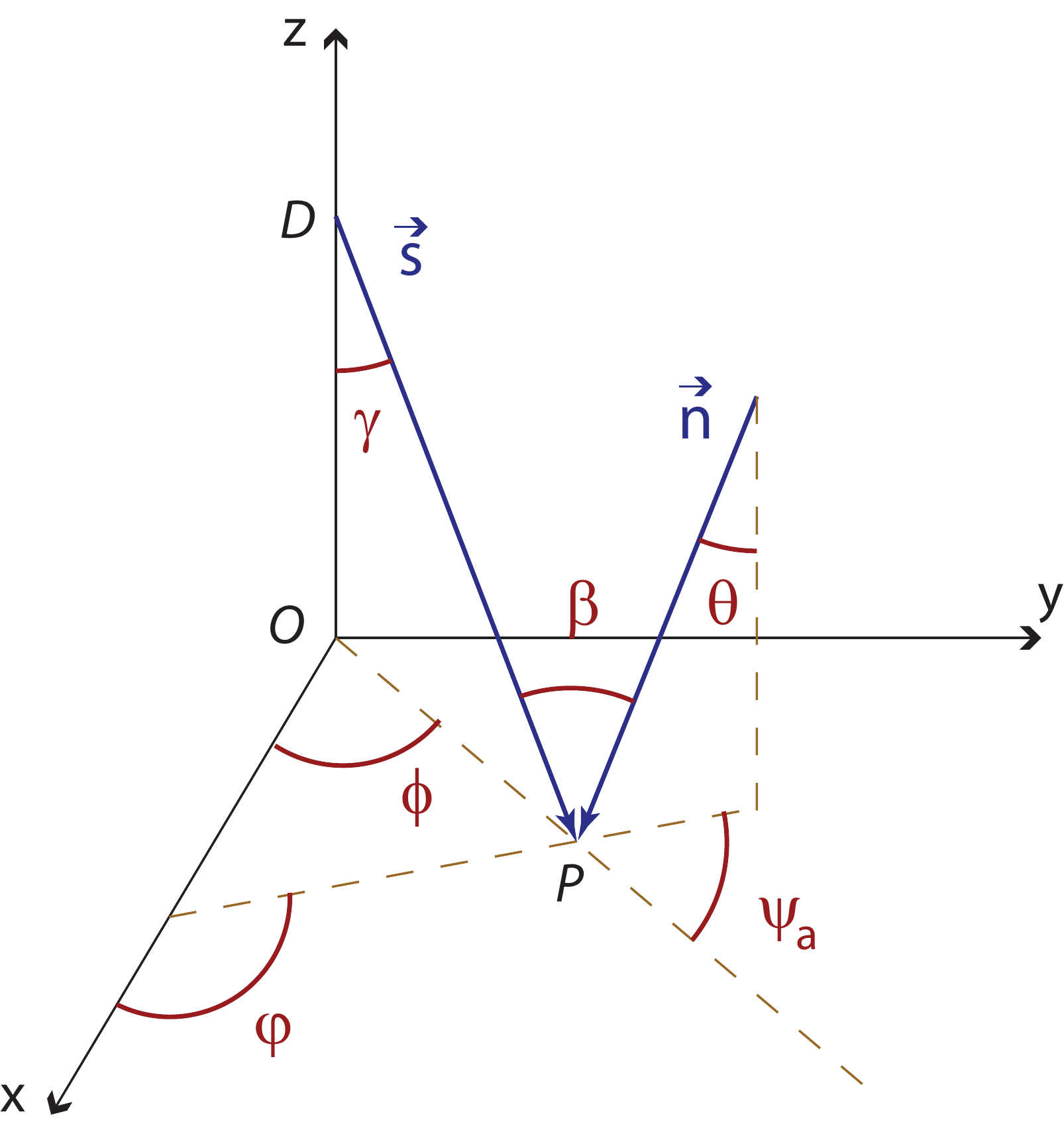}
\end{center}
	\caption{Definition of the angles in the \FoV with respect to a
	global reference frame $Oxyz$. The point $D$ is the detector position while the point $P$ is the impact point on the EAS
	on the Earth surface.}
\label{fig:FoVShowerDir}
\end{figure}

The angle between the latter and the
radial direction in the \FoV is called \FoVShowerAzimuth. Let the \EAS direction azimuth angle, with respect 
to a global reference system $Oxyz$, be $\varphi$ and the azimuth angle of the line of sight
in the \FoV be \FoVAzimuthAngle. Clearly $\FoVShowerAzimuth = \varphi -
\FoVAzimuthAngle$ (figure~\ref{fig:FoVShowerDir}). 
The the unit vectors of the \EAS direction $\hat{n}$ and of the line of sight $\hat{s}$ in the frame $Oxyz$
are respectively
\begin{gather}
	\hat{n}= 
	\{ - \sin \theta  \cos \varphi  ,\;- \sin \theta  \sin \varphi ,\;-\cos \theta\} \virgola 
	\nonumber \\ 
	\hat{s}= 
	\{ \sin \gamma  \cos \FoVAzimuthAngle  ,\; \sin \gamma  \sin \FoVAzimuthAngle ,\;-\cos \gamma\} 
	\punto \nonumber
\end{gather}

For an apparatus with cylindrical symmetry, there is no dependence on
\FoVAzimuthAngle, but only on \FoVShowerAzimuth.

Let $\beta$ be the angle between the \EAS direction and line of sight $\hat{s}$ from
the apparatus to the actual \EAS location:

\begin{equation}
	\cos(\beta) = 
	\cos (\gamma ) \cos (\theta )-\cos (\FoVShowerAzimuth) \sin (\gamma ) \sin(\theta ) 
\punto
\end{equation}

%--------------------------------------------------------------------------------
\subsection{Reference conditions and general assumptions}
\label{sec:Reference}
%--------------------------------------------------------------------------------

In this paper, unless otherwise specified, we shall use, as a reference,
the conditions and parameters summarised in
table~\ref{tab:ReferenceConditions}.

\begin{table}[htb]
\begin{minipage}[h]{0.99\textwidth}
\begin{center}
\begin{tabular}{ccc}\hline
Quantity                         
	& Reference value  
	& Other values  
	\\ 
	\hline
	\hline
Energy                           
	& $E=\un[10^{19}]{eV}$      
	&               
	\\ 
Zenith angle                     
	& $\theta = 50\degr$  
	& $\theta = 30\degr,\:70\degr$                      
	\\
Azimuth angle			 
	& $\FoVShowerAzimuth=90\degr$  
	& $\FoVShowerAzimuth=0\degr,\: 180\degr$ 
	\\
	\hline
Orbital height of the detector        
	& $H=\un[700]{km}$  
	& $H=\un[(400\div 1000 )]{km}$
	\\
\FoV aperture (half-angle)                    
	& $\Gmax=20\degr$  
	& $\Gmax=15\degr,\: 25\degr$
	\\
Field Angle of observation\footnote{The field angle $\gamma$ is chosen approximatively at half of the \FoV.}
	& $\gamma=15\degr$   
	& $\gamma=10\degr,\:20\degr$
	\\
Tilt angle of the detector      
	& $\TiltAngle=0\degr$ 
	& 
	\\
Total photo-detection efficiency  
	& \PDE=0.1 
	& 
	\\
	\hline
Operating Wavelength Range (WR)	
	& $\un[330]{nm}\leq\lambda\leq\un[400]{nm}$
	& 
	\\
Average atmospheric transmission (in WR)				
	& $ K_\mathrm{atm} \gtrsim 0.4$ 
	& 
	\\
Random Background in WR [$\un{ph\cdot m^{-2}s^{-1} sr^{-1}}$] 
	& $B=\sci{5}{11}$ 
	& $B=\sci{(0.3 \div 1.0)}{12}$ 
	\\
\hline     
\end{tabular}
\end{center}
\caption{The parameters and conditions used as reference in this chapter.} 
\label{tab:ReferenceConditions}
\end{minipage}
\end{table}

The basic parameters affecting the \EAS reconstruciton are: 
the energy, nature and direction of the \EAS,
the angle between the \EAS direction and the line-of-sight from the
apparatus to the \EAS, $\beta$, the field angle of the \EAS image, $\gamma$,
and the altitude of the apparatus above the Earth, $H$.

Moreover the following general assumptions will be used in all the paper.

\begin{enumerate}

\item
The \EAS energy is fixed by the scientific requirements to
have a good superposition with the spectrum observed by ground-based experiments.
Therefore a reference energy $E_\mathrm{ref} \approx \un[\sci{1}{19}]{eV}$ is used.

The \EAS direction and \FoVShowerAzimuth strongly affect 
the kinematics of the \EAS image signal and therefore its observability and observed features.
We will consider \EAS with $\theta \approx 45\degr$, 
which grants a good extension of the \EAS with a negligible
shortening of the longitudinal profile due to the impact with ground. 
In order to describe an average behaviour an angle 
$\FoVShowerAzimuth = \pm 90\degr$ will be used
as a reference (see section~\ref{sec:vislength}).

\item
The discussion will be based on typical hadron-induced \EAS, 
with energy \mbox{$E \approx \sci{(0.1 \div 10)}{20} \um{eV}$}
and a simple apparatus, looking downward the Earth, whose configuration and characteristics will be defined
during the discussion.

\item
The \EAS is geometrically modeled as a point moving on a straight line
at the speed of light, as, basically, any \EAS is seen from any
realistic space-detector as a unidimensional object, on all practical
purposes.  A reference \EAS energy $E_\mathrm{ref} \approx \sci{1}{19}
\um{eV}$ is used.

For several reasons, to be discussed later, 
we only consider \EAS with a zenith angle 
$ 30\degr \lesssim \theta \lesssim 70\degr $.
A very inclined \EAS requires in any case a diferent treatment thna
an almost vertcal \EAS.

One should note that when observing from space the case might arise of up-going
\EAS, that is \EAS with zenith angle larger than $90\degr$.

\item
Exponential density profile of the Earth atmosphere (assuming an isothermal atmosphere)
as a function of the height, $h$, above the sea level~\cite{allen}:    
\begin{equation}
\label{atmodensity}
    \rho(h) = 
    \rho_0 \exp\pqua{-\cfrac{h}{h_0}}
    \quad\textrm{with}\quad 
    \rho_0=\un[1.2249]{kg/m^3}
    \quad\textrm{and}\quad 
    h_0=\un[8.4]{km}
    \quad 
    .
\end{equation}

In some calculations we have also used, as a cross-check, 
the Linsley's parametrisation of the US Standard Atmosphere~\cite{USStdAtmo,US-STD-76}.

\item 
The flat Earth approximation is used whenever applicable, that is when
horizontal distances are negligible with respect to the Earth radius.
When this approximation is not good enough a spherical Earth with radius
$R_{\oplus}= 6371 \um{km}$ will be used.

The flat Earth approximation does not substantially affect the results
on the \EAS development for zenith angle $\theta\lesssim 70\degr$.  In
fact, for $\theta = 70\degr$, the difference between the exact and the
approximated value of the
\EAS slant depth at ground is about $\sim 30 \um{g/cm^2}$, as shown in
figure~\ref{fig:depthdifference}, which is smaller than the desired
resolution 
on the depth of the \EAS maximum: $\Delta{\Xmax}$.

Therefore the following approximate relation between the quota above the
Earth surface, $h$, and the distance $\ell$ measured along a straight
line with zenith angle $\theta$ will be assumed:
\begin{equation}
\label{flatearth}
    \deriv{\ell}{h} = \frac{1}{\cos\theta}
\quad .
\end{equation}

The exact relation for a spherical Earth, in terms of the Earth radius $R_{\oplus}$, is:

\begin{equation}\label{sphericalearth}
\deriv{\ell}{h} = \frac{h+R_{\oplus}}{\sqrt{h^2+2R_{\oplus}h+R_{\oplus}^2\cos^2\theta}}
\approx \left( 1- \frac{h}{R_{\oplus}}\tan^2\theta \right)\sec\theta 
\virgola
\end{equation}

clearly showing the condition allowing to neglect the sphericity of the
Earth: $\ell,h \ll R_{\oplus} $.

It is also worth remembering, for future uses, the definition of $X$
(the so called grammage): $\deriv{X}{\ell}=\rho $, plus
the appropriate boundary condition on $X$ at some $\ell$.

\begin{figure}
    \centering
        \includegraphics[width=0.80\textwidth]{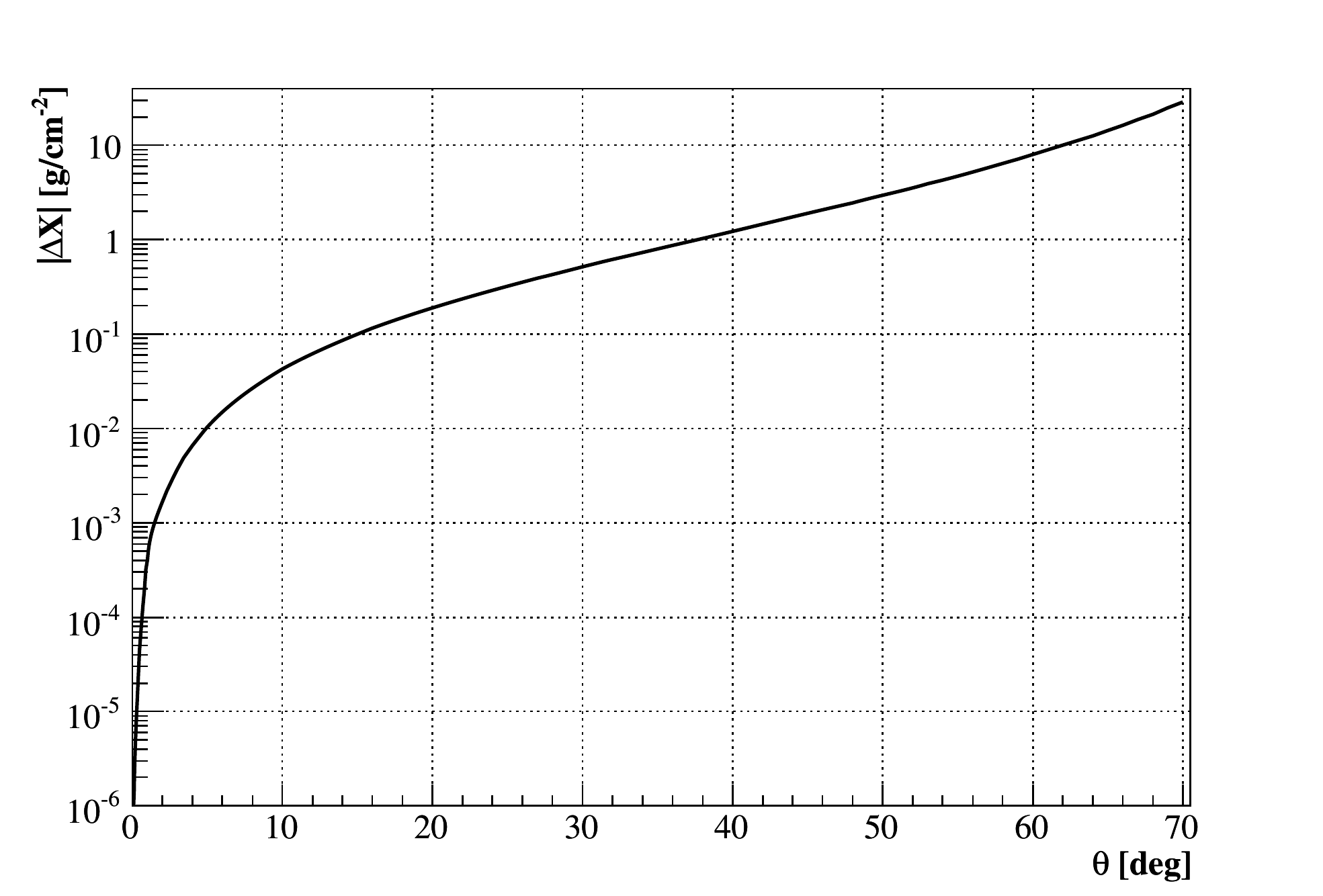}
    \caption{Absolute value of the difference $|\Delta X|$ between the true
    slant depth at ground and the one in flat Earth approximation versus the zenith
    angle.} \label{fig:depthdifference}
\end{figure}

\item 

The Gaisser-Hillas (GH) parametrisation~\cite{gaisser-hillas,Pryke} of hadron-induced \EAS parameterises 
the \EAS longitudinal profile as a gamma function:
\begin{equation}\label{eq:gaisser-hillas}
\begin{split}
    N(X) & = 
	\Nmax 
        \pton{ \frac{ X - X_0 }{ \Xmax - X_0 } }^{ \pton{ \Xmax - X_0 } / \lambda } 
	\exp\pqua{ \pton{ \Xmax - X } / \lambda } 
	\\
	& =
	\Nmax 
        \pton{ \frac{ X - X_0 }{ \Xris } }^{  \Xris  / \lambda } 
	\exp\pqua{ \pton{ \Xris + X_0 - X } / \lambda } 
	\virgola
	\\
	&
	\spazio\text{for $X \geq X_0$}
\end{split}
\end{equation}
where 
$N(X)$, the number of charged particles at the slant depth $X$, 
is expressed in terms of 
\Nmax, the number of charged particles at the \EAS maximum,
\Xmax, the depth of the \EAS maximum,
$X_0$, the starting point of the \EAS development, 
$ \Xris \equiv \Xmax - X_0 $
and the parameter
\mbox{$\lambda \simeq \un[65]{\um{g/cm^2}}$}.
The second expression is used to make explicit the invariance
of the \EAS longitudinal profile with respect to shifts of the first
interaction point $X_0$.

The \textit{pseudo-age} of an \EAS at point $X$, is defined, closely following the age
definition for an electromagnetic shower, as
\beq\label{eq:age}
   s(X) \equiv \cfrac{3 \pton{X-X_0}}{ \pton{X-X_0}+ 2 \pton{\Xmax-X_0}} 
   \virgola
\eeq
in terms of the actual slant depth and the slant depth, $X_0$, of the
first interaction point.

In our calculations (see~\cite{Pryke}) $X_0$ can be fixed at the first
interaction point and the $\lambda$ parameter is rather energy
independent and very similar for both proton and iron induced \EAS in
the energy range we are studying.  Moreover $\Nmax = \alpha E$ with
$\alpha \simeq 0.6/\mathrm{GeV}$~\cite{stanev}.  We can also roughly
assume that for a primary proton \EAS $X_0\approx\gcmsq[35]$ and for a
primary iron \EAS $X_0\approx\gcmsq[10]$ with no significant dependence on the energy.

In practice the parameter $X_0$ poorly correlates with the actual first
interaction point, $X_1$. Fixing $X_0=X_1$ when fitting the EAS profile
with the GH parametrization, produces a worse fit with \Xmax 
reduced by about \gcmsq[10].

Typical values for the \EAS parameters, as taken from~\cite{Pryke}, 
are summarised in table~\ref{tab:EASparameters}.
The values of \Xmax are the average value for different interaction models
(MOCCA-Internal, MOCCA-SIBYLL, CORSIKA-SYBILL and CORSIKA-QGSJET).
Note that these values are affected by $X_0$ 
when using the parametrisation of equation~\eqref{eq:gaisser-hillas}. 
See~\cite{Pryke} for details.

\begin{table}[htb]
    \centering
        \begin{tabular}{ccc|ccc}                                           
	\hline
            Particle & Energy [eV]	& $\Xmax$ [$\un{g/cm^2}$]	& Particle & Energy [eV]	& $\Xmax$ [$\un{g/cm^2}$]	\\ 
	    \hline
            Proton   & $\sci{1.0}{19}$  & $824\pm55$	& Iron     & $\sci{1.0}{19}$  & $724\pm20$		\\ 
            Proton   & $\sci{5.0}{19}$  & $853\pm58$	& Iron     & $\sci{5.0}{19}$  & $754\pm20$		\\ 
            Proton   & $\sci{1.0}{20}$  & $880\pm54$	& Iron     & $\sci{1.0}{20}$  & $784\pm19$		\\ 
            Proton   & $\sci{5.0}{20}$  & $907\pm49$	& Iron     & $\sci{5.0}{20}$  & $814\pm19$		\\ 
	    \hline
        \end{tabular}
    \caption{Typical values of \EAS parameters from~\cite{Pryke}.} 
    \label{tab:EASparameters}
\end{table}

The height $h_\mathrm{M}$ of the \EAS maximum as a function of the \EAS zenith angle $\theta$ 
is shown in figure~\ref{fig:hmax} for three different values of \Xmax using 
Linsley's parametrisation of the atmosphere density profile. The
exponential profile for $\Xmax = 900 \um{g/cm^2}$ is also shown for the
sake of comparison: it is clear that the
choice of the atmosphere density profile does not strongly affect
$h_\mathrm{M}$ up to $ \theta \approx ( 60\degr \div 70\degr ) $. 

\begin{figure}[htb]
    \centering
        \includegraphics[width=0.80\textwidth]{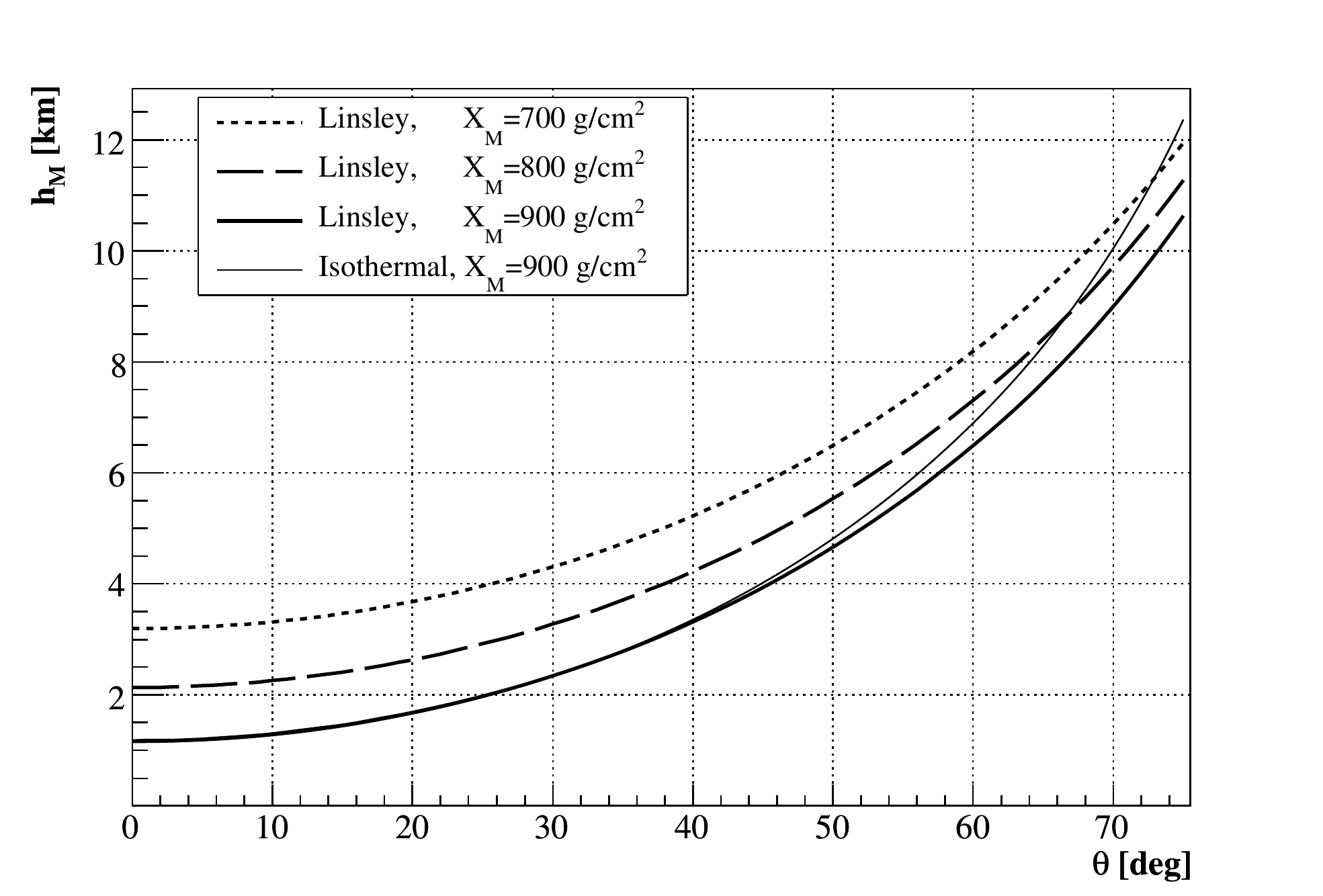}
    \caption{The height of the \EAS maximum versus the \EAS zenith angle.}
    \label{fig:hmax}
\end{figure}

The lateral \EAS profile is not accounted for, as it is hardly accessible when
observing from space.

\item 

A typical hadron-induced \EAS of $E = \sci{(0.1 \div 10)}{20} \um{eV}$
with $\theta\lesssim 70\degr$ is seen on the FS of an orbiting apparatus
as a track not longer than a few degrees, as discussed in detail in
section~\ref{sec:vislength}. Actually this depends on the instantaneous
height of the satellite from ground: a few degrees length must be considered an upper
limit, as it is calculated for $ H \approx 300 \um{km}$, which is a very
small orbital height for any satellite requiring a reasonably long
lifetime~\footnote{This angle, as an upper limit, is easily estimated
computing the \EAS length between the first interaction point and the
ground. The real angle seen by the detector is smaller due to the
background predominance with respect to the signal at the beginning of
the \EAS.}.

We can therefore neglect any possible change of the apparatus properties
as a function of the field-angle for any given \EAS, even if we are
dealing with a large \FoV optics.  Obviously this would not be true for
a very inclined \EAS, which can appear many degrees long and cannot be
easily described in a simple parameterised form.

\item 

The \EAS develops in the atmosphere at an height $ h \lesssim
\un[20]{km} $.  In this part of atmosphere the scintillation yield in
the wavelength range $\mathrm{WR}$ can be considered as nearly
constant~\cite{kakimoto}.  Roughly $ Y \simeq (4.2 \pm 0.2) \um
{photons}\um{particle^{-1}}\um{m^{-1}} $. 
When a better precision is required one can use the yield measurements of~\cite{Nagano:2004am}.

\item
The random background (\RB) is assumed to be uniform, isotropic and constant 
on the space-time scale of the \EAS development.
We assume for the \RB in the
wavelength range $\mathrm{WR}$ a reference value:
\mbox{$ B \approx \sci{ 0.5^{+0.5}_{-0.2} }{12} \um{photons \cdot m^{-2} \cdot s^{-1} \cdot sr^{-1}} $}.

From the current measurements we already know that, under some circumstances, 
it can be up to a factor two larger 
depending on the conditions (including moon phases and cloudiness).

It is assumed that, by continuously measuring the background nearby the \EAS, both in space and in time, 
the underlying \RB can be subtracted away, in real time.
Due to the large \RB this is an essential assumption in order to extract the
faint \EAS signal from the background at energies near $E_\mathrm{TH} $.

The possibility to subtract the background in real-time 
is justified by the rather large expected \RB rate, such that the relative error on the
background estimate is small (see section~\ref{sec:NightGlowEst}).
Therefore an appropriate statistical estimator used to compare the signal to
background is $ S / \sqrt{B} $.

\item 
Atmospheric transmission is naively modeled taking only into account Rayleigh scattering. 
In fact Mie scattering is only important at low altitudes of a few km, 
while the \EAS predominantly develops above a few km
height (see figure~\ref{fig:hmax}).
Mie scattering is therefore ignored at the current level of approximation.
Another factor which might turn out to be important is multiple scattering of
the signal, which is however difficult to study without a full Monte-Carlo
simulation.

A most important issue is the atmospheric transmission as a function of the zenith
angle direction. Table~\ref{ta:AtmoTransm} shows the ratio, $r$, between the
atmospheric transmission from ground to infinity at a zenith angle $\theta$ and
the same quantity in the vertical direction, all quantities being evaluated in
the flat Earth approximation.
The table shows that at $\theta \approx 75\degr$ the atmospheric transmission is reduced
by one order of magnitude with respect to the vertical transmission. This fact is
a crucial one in the evaluation of the effectiveness of a tilting of the
apparatus with respect to the nadir direction, as it will be discussed later.

\begin{table}[htb]
\begin{center}
\begin{tabular}{rc|rc}                                           
\hline\hline   
$\chi$ & $ \mathcal{R} \equiv \frac{\tau_\mathrm{a}[\chi,0]}{\tau_\mathrm{a}[\chi=0,0]} $ &
$\chi$ & $ \mathcal{R} \equiv \frac{\tau_\mathrm{a}[\chi,0]}{\tau_\mathrm{a}[\chi=0,0]} $	\\
\hline\hline
   $  0\degr  $		&	1.000	& $ 45\degr  $		&	0.748\\
   $  5\degr  $		&	0.997	& $ 50\degr  $		&	0.678\\
   $ 10\degr  $		&	0.989	& $ 55\degr  $		&	0.594\\
   $ 15\degr  $		&	0.976	& $ 60\degr  $		&	0.497\\
   $ 20\degr  $		&	0.956	& $ 65\degr  $		&	0.384\\
   $ 25\degr  $		&	0.930	& $ 70\degr  $		&	0.260\\
   $ 30\degr  $		&	0.897	& $ 75\degr  $		&	0.135\\
   $ 35\degr  $		&	0.857	& $ 80\degr  $   & 0.036\\
   $ 40\degr  $		&	0.808	& $ 85\degr  $   & \sci{7}{-4}\\
\hline\hline
\end{tabular}
\end{center}
\caption{Ratio $ \mathcal{R}$ between the
atmospheric transmission from ground to infinity at a zenith angle $\chi$ and
the same quantity at $\chi=0\degr$, all quantities being evaluated in
the flat Earth approximation (for a wavelength $\lambda=\un[337]{nm}$) using equation~\eqref{eq:atmtra}.}
\label{ta:AtmoTransm}
\end{table}

\item
The atmospheric geodesic refraction is normally negligible. Indeed, according
to~\cite{allen}, the correction for the refraction when observing an object
at the Earth surface from an height $H$, at a zenith angle $\chi$ with
respect to the nadir, is $\DD{\chi}\approx 62.37'' \tan\chi$, for
$\chi\lesssim 80\degr$ at a wavelength $\lambda=\un[337]{nm}$. 

So we have $\DD{\chi}\approx 0.008\degr$ at $\chi=25\degr$, that is about the maximum
field-angle in the case of no tilting of the detector optical axis with respect to the nadir.
The effect of the refraction is small with respect to the pixel size
(typically of $\sim 0.1\degr$) and it can be neglected.

However the effect of atmospheric refraction can be huge with a tilted
apparatus,
when observing at large angles with respect to nadir.

\item

Clear sky conditions are assumed. In fact a study taking into account all the
many effects of clouds requires without any doubt a detailed Monte-Carlo simulation.
Therefore we will find best-possible results.

\item 

Basic parameters of the apparatus 
are its altitude above the Earth, $H$, (or, in more general terms, the
orbital parameters) and the tilt angle $\TiltAngle$ between the
optical axis and the local vertical.

The choice of the orbit height of the apparatus requires a trade-off taking into account the
low signal produced by the lower energy \EAS and the requirement to observe a target of atmosphere
as large as possible. Moreover practical and technical constraints may limit the orbit height.

The orbit height of the apparatus, $H$, is assumed to be: 
$ \un[300]{km} \lesssim H \lesssim \un[1000]{km}$ (see section~\ref{sec:Orbit}).
In fact orbits lower than about $ \un[300]{km} $ suffer too much drag 
and require therefore frequent reboosts.
Orbits higher than about $ \un[1000]{km} $ are disfavoured by the fact that the radiation
environment undergoes a substantial change at about $ \un[1000]{km} $ height with the inset of
Van Allen belts, whose high level of trapped radiation may greatly influence the spacecraft.
Moreover orbits higher than about $ \un[1000]{km} $ 
turn out to provide too high a threshold for any reasonably-sized apparatus.

The orbit parameters must be tuned to optimize the expected results:
energy range and instantaneous geometrical aperture can be tuned but
constraints have to be taken into account.

\end{enumerate}

Some of the above assumptions are optimistic with respect to the real
conditions of the experiment (the clear sky assumption, for instance). 
This means that the results we will obtain are necessary, but possibly not sufficient, 
requirements for observations.

\FIXME{ clear sky only or any other assumption is OPTIMISTIC ??????????????}

%--------------------------------------------------------------------------------
\subsection{The air scintillation signal}
%--------------------------------------------------------------------------------

It is worthwhile to recall the expressions giving the air scintillation signal from the \EAS.
Any wavelength dependence is implicit in all the following formulas.

Let $w$ be a linear position coordinate along the \EAS development ($\Delta w = c \DD{t}$),
let $\ve{x}[w] = \ve{x}[w[t]]$ be the current \EAS position and $\ve{y}$
be the current position of the observational apparatus,
let $ Y_e(\ve{x}[w]) = Y_e(\ve{x}[w[t]])$ be the air scintillation yield in the desired wavelength
range at the \EAS position,
let $ N_c(\ve{x}[w]) = N_c(\ve{x}[w[t]])$ be the number of charged particles in the \EAS at $w$ and
let $A$ be the entrance pupil area of the optics
(which is actually the cross-section for a photon hitting the FS when $\gamma=0$).

Let the atmospheric transmission from the generation point to the observation point 
be $ T_{\mathrm{A}}[\ve{x},\ve{y}]$.
The atmospheric absorption due to Rayleigh scattering only can be roughly parameterized as:
\begin{gather}\label{eq:atmtra}
        T_{\mathrm{A}}[\ve{x}[w],\ve{y}]\equiv \tau_\mathrm{a}[\chi,z] \simeq  
	\exp{\left[-\pton{\cfrac{\rho_0 h_0}{\Lambda(\lambda)}}M(\chi)\exp{\pton{-\frac{z}{h_0}}}\right]}
\\
\text{with}\spazio
\cfrac{\rho_0 h_0}{\Lambda} \approx 0.7 
\spazio\text{for}\spazio
\lambda = \un[337]{nm} \virgola
\end{gather}

where $h_\mathrm{0}\approx\un[8.4]{km}$ is the atmosphere scale height at
the sea level (see the~\eqref{atmodensity}),
\mbox{$\rho_0=\un[1.292]{kg}\un{m}^{-3}$} is the atmospheric density at
the sea level, $z$ is the altitude of the photons emission point,
$\Lambda(\lambda)$ is the Rayleigh mean free path at the wavelength
$\lambda$ (see~\cite{Bucholtz:1995}) and $M(\chi)$ is the airmass
function at the zenith angle $\chi$. For $\chi\lesssim 80\degr$,
$M(\chi)\approx\sec\chi$, a better approximation can be found for
example in~\cite{Young:1994}.

Let the overall detection efficiency of the PD, including the f/e
electronics efficiency, be $\veps_{\mathrm{PD}}[\gamma,\mathcal{X}]$; it
is is the probability that a photon reaching the PD will fire the f/e
electronics producing a recorded photon hit. It might depend on many
parameters, $\mathcal{X}$, and it depends on the field-angle $\gamma$,
in particular.  In fact at different field-angles the incidence angles
of the incoming photons may change leading, for instance, to a change in
the intrinsic sensor detection efficiency.

Let the total efficacy\footnote{It is defined as the ratio the number of photons
incident on the entrance pupil per time unit with an angle $\gamma$ and the photon
irradiance [\un{ph}\un{m^{-2}}{\un{s^{-1}}}].}~\cite{bi:LHillman} of the optical apparatus be $ \pton{ A
\veps_{\mathrm{O}}[\gamma] } $ and the total throughput
efficiency\footnote{It is defined as the ratio the number of photons
incident on the entrance pupil per time unit with an angle $\gamma$ and the total photon flux collected by the pupil (that is given by the photon irradiance times the entrance pupil area).}~\cite{bi:LHillman} be $
\veps_{\mathrm{O}}[\gamma]$.  The previous definitions refer to all the
photons reaching any point on the FS and do not include the filter
transmission, $ \eta_{\mathrm{F}}[\gamma]$, which is intentionally kept
as a separate parameter as the filter transmission is one the the key
parameters to play with.

In order to quantify the detectable signal one should build more
appropriate quantities than the total efficacy and total throughput
efficiency.  In fact for any given direction, at an angle $\gamma$ with
respect to the optical axis, only the photons reaching the FS close
enough to the centroid of the distribution of all the photons focused on
the FS are useful for the \EAS reconstruction.  In fact photons too far
from the centroid will contribute to the veiling glare, that is diffused
background on the FS.  This effect might be particularly dangerous
because, thanks to the large \FoV, it is most likely that bright sources
are present somewhere in the \FoV: in case of a large veiling glare
these bright sources might significantly increase the background level
on the whole FS.

A suitable fiducial region (called $\Omega$), around the centroid of the
distribution on the FS of all the photons, must be defined.  The
fiducial region $\Omega$ may be defined in different ways.  The photons
on the FS will then be classified into three classes, as follows: inside
the fiducial region $\Omega$ (transmitted photons); outside the fiducial
region $\Omega$ but detected somewhere on the FS (photons giving rise to
the veiling glare); any other photon (lost or absorbed somewhere).  It
is appropriate to adopt as a fiducial region a region of the order of
the spot-size and/or pixel size.

Let $ \veps_\mathrm{B}[\gamma]$ be the fraction, out of all the photons on the FS, 
falling inside the desired bucket size: this is actually a measure of the encircled energy fraction. 

Define then the \emph{triggering efficacy} as 
$ A\veps_\mathrm{O}[\gamma]\veps_{\mathrm{B}}[\gamma] \equiv A\veps^\prime_\mathrm{O}[\gamma]$ and
the \emph{triggering throughput efficiency} as
$ \veps_\mathrm{O}[\gamma]\veps_\mathrm{B}[\gamma] \equiv \veps^\prime_\mathrm{O}[\gamma]$.

Let $\Delta S[w]\Bigr|_{\Omega}$ (with the subscript $\Omega$) 
remind that the detected signal
depends on the region $\Omega$ on the FS around the the centroid of the spot (bucket)
where photons are considered to be detected photons instead of background.
The photons outside $\Omega$ just contribute to the veiling glare.

The number of signal photons produced by a segment $\Delta w$ of the \EAS, at coordinate $w$, 
and detected by the apparatus is thus:

\begin{equation}\label{eq:Signal}
  \Delta S[w]\Bigr|_{\Omega} = 
        \pton{ \cfrac{ Y_e[\ve{x}[w]]  N_c[\ve{x}[w]]  \Delta w }{4 \pi \pton{\ve{x}[w]-\ve{y}}^2 }  } 
	T_{\mathrm{A}}[\ve{x},\ve{y}]
	\pton{ A \veps_{\mathrm{O}}[\gamma] } 	
	\veps_{\mathrm{B}}[\gamma]
	\eta_{\mathrm{F}}[\gamma]
	\veps_{\mathrm{PD}}[\gamma,\mathcal{X}]
\punto
\end{equation}

$ \Delta S[w]\Bigr|_{\Omega}$ with the subscript $\Omega$ is meant to remind that
those equations are only meaningful after specification
of the region
$\Omega$ on the FS over which photons are considered to be
detected photons, that is photons that are close enough to the centroid of the
image spot.

It is worth noting that when looking at faint signals (a few photons per pixel) 
the errors in formula~\ref{eq:Signal}
will be typically dominated by the Poisson statistical fluctuations, at least
for the faintest \EAS. In fact it will be shown in section~\ref{sec:AngResEst} that for the
faintest \EAS, with a number of detected photons $ N\approx 100 $, one has less
than ten photons per pixel with a relative statistical error of the order of $0.3$,
larger than the expected systematic errors ($\sim 15\%$).\footnote{Obviously when fitting the profile with a given functional dependence the
resulting error will be reduced by the fitting procedure.}

Using the simplifying assumptions listed in section~\ref{sec:GenAss} it is easy
to build a parameterised semi-analytical description of the
relation~\ref{eq:Signal}, which will be used in the rest of the paper (cfr~\cite{Sommers:1995dm,Baltrusaitis:1985mx}):
\begin{equation}
\Delta S[w]\Bigr|_{\Omega} \simeq \frac{A_\mathrm{EP} N_c \veps_\mathrm{T}}{4\pi D^2}\exp\left[-\frac{D}{D_0}\right]Y\DD{w}
\virgola
\end{equation}
where $\veps_\mathrm{T}$ is the total efficiency (optics plus photo-detector and filter), $D$ the distance from the
detector to the fluorescence emission point (it can be assumed as a constant) and $D_0\approx\un[12]{km}$ is the extinction
length of light due to the atmospheric scattering (at $\lambda=\un[337]{nm}$).

%--------------------------------------------------------------------------------
\subsubsection{Single-photon detection factors}
%--------------------------------------------------------------------------------

The atmospheric transmission $ T_{\mathrm{A}}[\ve{x}[w],\ve{y}] $ is the first basic single photon detection
factor.
It is out of control but the experiment design should be optimised to minimise its effects.

The photon collecting power of the optics is:
$A \veps_{\mathrm{O}}[\gamma] \veps_{\mathrm{B}}[\gamma]\eta_{\mathrm{F}}[\gamma]$,
including the effect of PSF and filters. 
It gives the conversion from the incoming number of photons per squared meter,
at an angle $\gamma$ with the optical axis, 
to the number of photons reaching the FS inside the region $\Omega$.
It is the quality factor for the optics as far as the number of collected photons is concerned.
It is the basic and unique sizable parameter to define the optics photon collecting power.

The overall detection efficiency of the PD,
$\veps_{\mathrm{PD}}[\gamma,\mathcal{X}]$, is another basic parameter
which depends on a large number of factors.  These factors are often
very close to one but, when multiplying a large number of them, the
final result may significantly depart from one.  The
Photon-Detection-Efficiency of the sensor is the most significant
parameter which is typically much smaller than one.  For standard PMT,
which have already been tested in the space environment and proved
suitable for applications in space, $\veps_{\mathrm{PD}}=0.12\div
0.15$. Recently several new high quantum efficiency photo-detectors
became available on the market: GaAsP \emph{Hybrid PhotoDiodes}
(HPD)~\cite{Piccioli:2003vq} and Geiger-Avalanche Photo-Diode
(GAPD)~\cite{Buzhan:2003ur,Golovin:2004jt}. These devices, exploiting
the intrinsic quantum efficiency of the
solid state device, aim to reach a $\veps_{\mathrm{PD}}=0.4 \div 0.6$.

The overall capability to detect photons of the instrumental apparatus can be expressed
through the \emph{photo-detection efficacy}, defined as

\begin{equation}\label{eq:PDEfficacyTerms}
\PDEfficacy = A \cdot \veps^\prime_\mathrm{O}[\gamma] \cdot \eta_{\mathrm{F}}[\gamma] \cdot
\veps_{\mathrm{PD}}[\gamma] \punto
\end{equation}

%--------------------------------------------------------------------------------
\subsubsection{Multi-photon detection factors}
%--------------------------------------------------------------------------------

Other more complex factors will affect the \EAS detection,
involving 
multi-photon correlations and correlations between neighbouring pixels, such as:
the read-out electronics and trigger efficiency, $\veps_{\mathrm{TRI}}$,
and the analysis and event reconstruction efficiency, $\veps_{\mathrm{ANA}}$. 
These effects are very difficult to estimate without a full Monte-Carlo simulation~\cite{esaf1,esaf2,TheaThesis,PesceThesis}, 
as they depend on the
contribution by all the photons at once.

Therefore naive expectations coming from single photons studies can be far too optimistic:
once more simple estimates may turn out to be optimistic.

%--------------------------------------------------------------------------------
\section{Design and optimisation of a space-borne apparatus}
\label{sec:Results}
%--------------------------------------------------------------------------------

%--------------------------------------------------------------------------------
\subsection{Orbit} \label{sec:Orbit}
%--------------------------------------------------------------------------------

The orbit parameters can be tuned to optimise the expected performances,
in particular the energy range and the instantaneous geometrical aperture.

The orbital height (semi-major axis of the ellipse) 
is one of the most important parameters.
In fact an higher altitude
implies a larger mass of observed target atmosphere (and therefore a 
better instantaneous geometrical aperture) 
but also an higher energy threshold
because of the smaller \EAS signal.

Varying the orbital height is useful in order to span a larger range of
energies: with an elliptic orbit the satellite spends more time at a
higher altitude, gaining in effective aperture, but it also spends some
time
at a lower altitude, decreasing the energy threshold, where a long
measuring time might be not necessary.

An orbit height spanning the range 
$ 400 \um{km} \approx H_\mathrm{MIN} \lesssim H \lesssim H_\mathrm{MAX} \approx 1000 \um{km} $,
for instance, would give the capability to extend the energy range, with respect to a fixed height, 
by taking data at different heights, that is shifting up and down the \UHECP energy range
with a scaling factor: 
$ ( H_\mathrm{MAX} / H_\mathrm{MIN} )^2 \approx 6 $. 
Moreover the instantaneous geometrical aperture at apogee would be the same factor larger than at
the perigee. On the positive side one should note that the angular granularity, 
at fixed $\DD{L}$, scales as $ H^{-1}$, and not as $H^{-2}$.

Another way to vary the height might be to use different almost circular orbits during the mission
lifetime (for example part of the time at a lower altitude and part of the time at an higher
altitude).  Natural orbit decay might be exploited as well.

One should remember that the orbit lifetime depends on the height and it strongly depends on the
epoch of the Solar cycle (that is on the epoch of the Mission), which
affects the atmospheric density.

\begin{figure}[htb]
	\centering
		\includegraphics[width=0.9\textwidth]{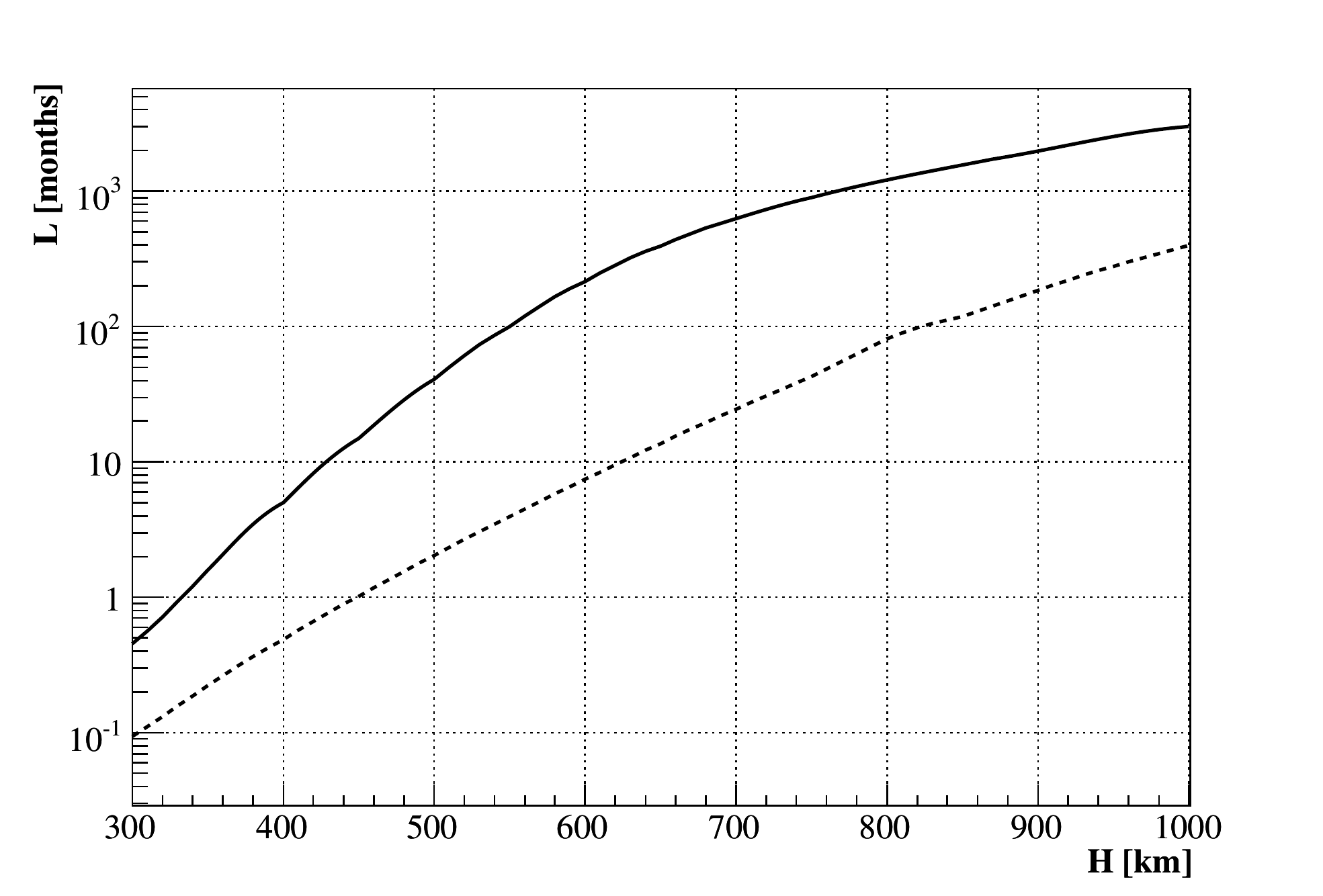}
	\caption{Lifetime $L$ (in months) of a satellite (mass $\sim$ 3000 kg; total
		area $\sim$ 150 m$^2$) versus the orbital height $H$. Solid
		line: solar cycle minimum; dotted line: solar cycle
		maximum.}
	\label{fig:lifetime}
\end{figure}

Another key point to keep in mind is that the ballistic coefficient of
the satellite is expected to be rather low,
due to the large expected area-to-mass ratio.
Therefore, very roughly, the orbit lifetime\footnote{Formula and data for calculation can be found in~\cite{wertz}.} is expected to be of the order of a few days for 
$ H \approx \un[300]{km} $, a few weeks for $ H \approx \un[400]{km} $
and some years at $ H \approx \un[700]{km} $ (see
figure~\ref{fig:lifetime}).

The orbit inclination should be chosen in order to span as much as possible the Earth surface
in order to obtain a sky coverage as uniform as possible.

An important parameter that depends on the orbit is the duty cycle, that should be as large as
possible.  For this purpose, the orbit design must be further optimised in order to avoid as much
as possible the light pollution from the man-made sources at ground and from the lightings (that are
concentrated mainly on lands) and the auroras.

Note also that the effects of a short 
duty cycle can be recovered increasing the mission lifetime.

The orbit design can be also optimised in order to access as much time as
possible over fixed targets at ground (ground \EAS experiments, weather
stations, ground-based calibration sources, etc.). 
In particular, the accesses to other \EAS experiments, namely the Pierre Auger Observatory, are very
useful for the apparatus calibration.

%--------------------------------------------------------------------------------
\subsection{Optical system design}
%--------------------------------------------------------------------------------

The two basic optical parameters, affecting both the performance and the engineering of the
experiment, are the Entrance Pupil (EP) area, $A_\mathrm{EP}$, and the Field of
View (\FoV) of the optics, defined as the half-angle of the \FoV circular cone, $\Gmax$.

The large desired aperture calls for an \Fnumb as small as possible in order to reduce the
size of the FS.
However the large desired aperture implies, in any case, a large FS size.

%--------------------------------------------------------------------------------
\subsubsection{Entrance pupil of the optics}
%--------------------------------------------------------------------------------

The apparatus sensitivity to low signals is affected by a basic and unique sizeable 
parameter: the optics aperture, that is the entrance pupil area, $A_\mathrm{EP}$.  
Increasing the aperture, to within external constraints, 
gives a guaranteed improvement of the performance: 
it is just a matter of technology and money to build such a large
apparatus.

As a very large aperture optics is needed a deployable optics
is most probably required.

The optics aperture, also fixes the size of the apparatus.
The lower limit for $A_\mathrm{EP}$
is basically set by the requirement on the energy threshold, while the 
upper limit is fixed by external constraints.

The total number of detected photons is roughly proportional to solid angle subtended by the optics
entrance pupil as it is seen by the \EAS:
$ \approx A_\mathrm{EP}/H^2$.
Therefore, for a required energy threshold, the choice of
$A_\mathrm{EP}$ and orbit height is driven by the number of photons that one wants to observe.

%--------------------------------------------------------------------------------
\subsubsection{\FoV of the optics}
%--------------------------------------------------------------------------------

The large desired aperture implies in any case a large FS size. 

In order to gain instantaneous geometrical aperture,
either a larger orbit height or a tilt with respect to nadir might 
be more effective than increasing the \FoV.
In fact these options might have two beneficial effects: to keep the FS size limited 
and to ensure a better optics performance, as the performace of the optics
tends to become worse, normally, when the \FoV increases.
In fact the large required \FoV and the necessity to keep the optics simple typically give a PSF
which is not good enough to detect the lateral \EAS dimensions.
Even if one manages to get a good enough PSF the required number of channels would increase unacceptably.
In our discussion we will consider a \FoV aperture (half-angle) $15\degr \le \Gmax \le 25\degr$.

%--------------------------------------------------------------------------------
\subsubsection{Reflective or refractive ?}
%--------------------------------------------------------------------------------

There are virtues and defects for using refractive versus reflective optics in this kind of experiment.
A refractive system is possibly difficult to deploy 
and a complex supporting structure is required due to the large dimensions.
A reflective system would suffer from the obscuration due to the FS 
and the very limited room for the FS Photo-Detector. 
The latter might be a real issue, depending on the kind of sensors one is going to use.

Ona main drawback of a reflective system is that the curvature of the FS
(that is center of curvature of the FS on the opposite side with respect to the
incoming light)
is such that the filling of the FS with the array of sensors is more
difficult and the filling factor tends to be worse than for a refractive
system, where the center of curvature of the FS is on the same side with respect to the
incoming light.

%--------------------------------------------------------------------------------
\subsection{Effect of the observation angle (location on the FS)}
%--------------------------------------------------------------------------------

At the edge of \FoV ($\gamma \simeq \Gmax$) the \EAS triggering and 
reconstruction is much more difficult than near the center of the \FoV.
In fact due to the larger \EAS distance
the irradiance (energy/photons per unit area per unit time) decreases as $\approx \cos^{2}{\gamma}$.
Moreover
the effective entrance pupil typically decreases as $\cos{\gamma}$:
this is the so-called obliquity
factor of the optics, due to the fact that the size of the entrance
pupil is seen reduced by a factor $ \cos{\gamma} $.
As a result the number of photons received on FS by an EAS of fixed
energy scales at least as $ \cos^{3}{\gamma} $.

In general the real optics triggering efficacy/efficiency,
$\veps_{\mathrm{O}}[\gamma]\veps_{\mathrm{B}}[\gamma]$,
also decreases while increasing the field-angle.
This also includes the effect of the absorption losses inside the refractive materials.

Moreover an \EAS detected at the edge of the \FoV would suffer 
from a larger atmospheric attenuation $T_\mathrm{A}[\gamma]$ (see
equation~\eqref{eq:atmtra}) due to the longer and more inclined path.

As a smaller number of photons is detected from any \EAS when it is observed at the edge of \FoV,
with respect to the same \EAS observed at the center of \FoV,
one can roughly say that the \FoV center (more photons from the \EAS, smaller geometrical acceptance) 
is better for low energy events and low photon flux while
the \FoV edge (less photons from the \EAS, larger geometrical acceptance) 
is better for high energy events and high photon flux rates.

%--------------------------------------------------------------------------------
\subsubsection{Signal and \RB roll-off with field-angle}
%--------------------------------------------------------------------------------

It is important to compare the dependence of both signal and \RB on the field-angle.

Let us assume an ideal optics, that is an optics such that all the photons hitting the entrance of
the optical system are focused inside the desired bucket. For such an optics:
$ \veps_\mathrm{O}^\prime[\gamma] \equiv \veps_\mathrm{O}[\gamma] \veps_\mathrm{B}[\gamma] = \cos{\gamma}$.
Beware that for a real optics $\veps_\mathrm{O}^\prime[\gamma] $ might typically drop faster with $\gamma$.

The field-angle dependence for a given source is given by 
\beq 
	S[\gamma] \sim
	T_\mathrm{A}[\gamma] \veps_\mathrm{O}^\prime[\gamma] \cos^2{\gamma}
	\longrightarrow \gamma (0 \rightarrow 25\degr) : (1.0 \rightarrow 0.72)
\punto
\eeq
On the other hand the field-angle dependence for the \RB is given by 
\beq
	B[\gamma] \sim  
	\veps_\mathrm{O}^\prime[\gamma] 
	\longrightarrow \gamma (0 \rightarrow 25\degr) : (1.0 \rightarrow 0.87)
\punto
\eeq

Both signal and \RB decrease at increasing field-angles, but in a different way. In
fact 
the field-angle dependence for the signal to \RB ratio is:
\beq
	\cfrac{ S }{ B } [\gamma]\sim   T_\mathrm{A}[\gamma]\cos^2{\gamma}
	\longrightarrow  \gamma (0 \rightarrow 25\degr) : (1.0 \rightarrow 0.79)
\punto
\eeq

The dependence on the field-angle of both the signal and the
signal-to-background ratio renders useless the increase of the optics \FoV above a
certain extent for a nadir pointing instrument. 
Similar considerations apply to tilted instruments.  

Many parameters (such as the trigger settings) require tuning as a function of the field-angle, 
that is as a function of the radial distance on the FS. 

One important result is this that, as 
the efficiency curve as a function of the energy roughly scales in energy 
as the inverse number of signal photons detected, 
a detection energy threshold at least $\approx 2$ times higher is expected 
for events detected at $\gamma = 25\degr$ with respect to events detected 
on-axis~\cite{TheaThesis,PesceThesis}.

As a conclusion the 
performances drop significantly at large angles from nadir 
due to the larger \EAS distance, the larger atmospheric absorption
and the worse optics photon collection efficiency 
(the latter actually depends on the angle from the optical axis, not from nadir).

%--------------------------------------------------------------------------------
\subsubsection{Other considerations}
%--------------------------------------------------------------------------------

The \FoV, together with the satellite altitude, determines the geometrical aperture
(discussed in the section \ref{subsec:aperture}) and therefore the
number of events to be detected. The $D_\mathrm{EP}$ defines the detector
collection area and its sensitivity, but also fixes the apparatus size and,
consequently, the mass and the volume of the payload.

The diameter $D_\mathrm{PD}$ of the FS is a function of the
\FoV (half-angle $\Gmax$) and of the focal length $F$:
\begin{equation}
D_\mathrm{PD}= 2 F \sin \Gmax
\end{equation}
If we introduce the optics $f$-number, $\Fnumb \equiv F/D_\mathrm{EP}$ then
\begin{equation}
D_\mathrm{PD}= 2 \Fnumb D_\mathrm{EP} \sin \Gmax
\end{equation}
Note that for $\Gmax=25\degr$ and $\Fnumb = 1$ the FS is
as large as the entrance pupil. A further discussion of these arguments can
be found in~\cite{mazzinghi}.

As the apparatus is basically photon-limited, the entrance pupil shall be as
large as possible, up to the limit allowed by external, technological and practical constraints.
Therefore the size of the optics will be approximately set to the maximum allowable size.
Therefore a reasonable assumption for any real apparatus is that the FS is
not larger than the optics: if the latter is approximated by the optics
entrance pupil, as it would be desirable for a high-efficiency optics, this
implies the rough estimate: $2 \Fnumb \sin \Gmax \lesssim 1$

%--------------------------------------------------------------------------------
\subsection{The photo-detector on the FS}
%--------------------------------------------------------------------------------

As a consequence of the limits on $f$-number and on the pixel size,
a very large area FS is required: the choice of the sensor has to account for this.
This basically means that compact and light sensors are required.

The  pixel size is driven by two competing requirements:
going far away from the Earth (or tilting) requires a smaller pixel size and 
increasing the optics aperture (and therefore the focal length) implies a larger pixel size.

The overall photo-detection efficiency of the photo-detector is one of the most
important parameters affecting the performance.
However, needless to say, it cannot increase larger than one.
Typical, realistic overall photo-detection efficiencies in the range 
$ \pton{ 0.1 \div 0.2 }$
can be quoted for these kind of apparatus. 
Even in the most optimistic approach it is not realistic to expect more than a
factor $ \approx 3 $ improvement, in the near future, provided 
huge efforts can be devoted to the development of suitable sensors with higher Quantum Efficiency.
In fact the main source of inefficiency, as of today, is typically the rather low
Quantum Efficiency, affecting the Total Photo-Detection efficiency of
the sensors.
However also the geometrical acceptance and filling factor of the arrya
are a crucial issue for this kind of application.

A lot of other factors affects the overall efficiency.
They are however all factors already very close to one (typically $ \approx 0.9$),
but the product of many of them can make the overall efficiency drop.
In order to make a significant change all of them should be substantially improved
because they are already very close to one: there is little hope to gain anything substantial.

%--------------------------------------------------------------------------------
\subsubsection{The parameters of the Photo-Detector}
%--------------------------------------------------------------------------------

 Approximate relations, useful to estimate the relevant photo-detector parameters,
 are summarised in this section.
 The relations in this section will be essentially geometrical, more refined
 estimates would require a precise design and full simulations.

 The characteristics of the optics have a strong impact on the photo-detector
 design.
 The Point Spread Function (PSF) of the optics has to match, approximately,
 the photo-detector pixel size. 
 In fact a finer granularity would allow a better reconstruction,
 provided enough photons are collected, possibly allowing to measure the
 PSF itself. However a tradeoff with cost and complexity (driven by the
 number of channels) is unavoidable. on the other had, obviously, a
 pixel size much larger than the PSF would not exploit all the optics
 performance, wasting the efforts for the optics.

The
 FS shape and dimensions are determined by the optics.
 At this stage, therefore, only rough estimates of the corresponding
 photo-detector parameters will be attempted.

 A more careful evaluation can only be carried on once the design of the optics can
 be better defined.

 The desired number of pixels of the photo-detector can be
 estimated, from the desired parameters, by the relation
 \begin{equation}\label{eq:nchans-e}
    N \approx \cfrac{\pi H^2 \tan^2{\gamma}}{\DD{L}^2 }
    \punto
 \end{equation}

 The required photo-detector pixel size, $\delta$, corresponding to observing
 a length $\DD{L}$ on the Earth surface,
 can be estimated by the relation
 \begin{equation}\label{eq:pixels-1}
    \delta \approx \cfrac{f \DD{L}}{H}
    \virgola
 \end{equation}
 where $f$ is the focal length of the optics.

 The photo-detector surface has to approximate the focal
 surface of the optics. The latter can be assumed, to a first
 approximation, 
to have a spherical shape with radius
 equal to the focal length of the optics, $f$, and
 maximum angular aperture $\beta$. Its area is then given by the relation
 \begin{equation}\label{eq:adet}
    A_\mathrm{det} = 2 \pi f^2 \pqua{1-\cos{\beta}}
    \punto
 \end{equation}
 The FS maximum diameter is given by
 \begin{equation}
    D_{f} = 2 f \sin{\beta}
    \punto
 \end{equation}
 Note that, at least to a first approximation, $\beta\simeq\Gmax$.

 The approximate maximum number of pixels which can be fitted on
 the FS of the photo-detector is given by the relation
 \begin{equation}\label{eq:nchans-fs}
    n \approx \cfrac{ A_\mathrm{det} }{\delta^2}
    \punto
 \end{equation}
 
 Alternatively the desired pixel dimension on the FS
 can be estimated, in terms of the FS parameters and the desired
 number of pixels one wants to see at the Earth, by the relation
 \begin{equation}\label{eq:pixels-2}
    d \approx \sqrt{\cfrac{ A_\mathrm{det}}{N}}
    \punto
 \end{equation}

 Note that, to the present level of approximation and with the present
 parameters, the two
 relations~\ref{eq:pixels-1} and ~\ref{eq:pixels-2} are roughly equivalent,
 given the relations~\ref{eq:nchans-e} and~\ref{eq:adet}.

 The optics also determines the distribution of incidence angles of the photons
 on the FS, which has some impact on the photo-detector design.
 The marginal ray angle is determined by the $f\#$ of the optics and
 is given by the relation
 \begin{equation}
    \tan{\theta_\mathrm{max}} \approx \cfrac{1}{2 f\#}
    \punto
 \end{equation}
 The angular granularity of the photo-detector, $\DD{\alpha}$,
 is given by the relations
 \begin{equation}
    \Delta\alpha \approx \cfrac{\DD{L}}{H} \simeq \cfrac{\delta}{f}
    \punto
 \end{equation}
 The solid angle coverage of every pixel, $\Delta\Omega$, is given by
 \begin{equation}
    \Delta\Omega \approx \cfrac{\DD{L}^2}{H^2} \simeq \cfrac{\delta^2}{f^2}
    \simeq \pqua{\Delta\alpha}^2
    \punto
 \end{equation}

 The approximate defocusing in the direction parallel to the
 FS, $\Delta w$, produced by a small displacement $\Delta
 z$ in the direction perpendicular to the FS,
 is given by the relation
 \begin{equation}\label{eq:defocusing}
    \Delta w \approx \Delta z \tan{\theta_\mathrm{max}} = \cfrac{\Delta z}{2 f\#}
 \ll \delta    
\punto
 \end{equation}

In order to reduce the effect of defocusing on the large FS a good fit
between the ideal optics FS and the real FS must be implemented. In
general it is not trivial to accomplish this because the optics FS has a
complex geometrical shape.

%--------------------------------------------------------------------------------
\subsection{Number of detected photons, efficacy and energy resolution}
\label{sec:NumPh}
%--------------------------------------------------------------------------------

The required energy resolution ($\Delta E/E \sim (20\% \div 30\%)$) calls for a relative error due to the Poisson statistics of
the detected photons not larger than $ \sigma_{N}/N \sim\pton{0.10 \div  0.15} $, 
assuming an equal contribution from statistical and
systematics errors, as it is appropriate for a well-designed apparatus. This
implies that at least one hundred photons 
must be detected from any \EAS at the lowest energies, with full triggering efficiency.
Obviously the number of detected photons impacts significantly on the other
observables as well.

Assuming an ideal optics and the reference \EAS
one can easily derive the requirements on the optical triggering
efficacy from the requirement on the energy resolution.
As the angular extension of the reference \EAS is short on the FS (see
section~\ref{sec:vislength}) 
the properties of the apparatus can be considered uniform.

\begin{figure}[htbp]
	\centering
		\includegraphics[width=0.9\textwidth]{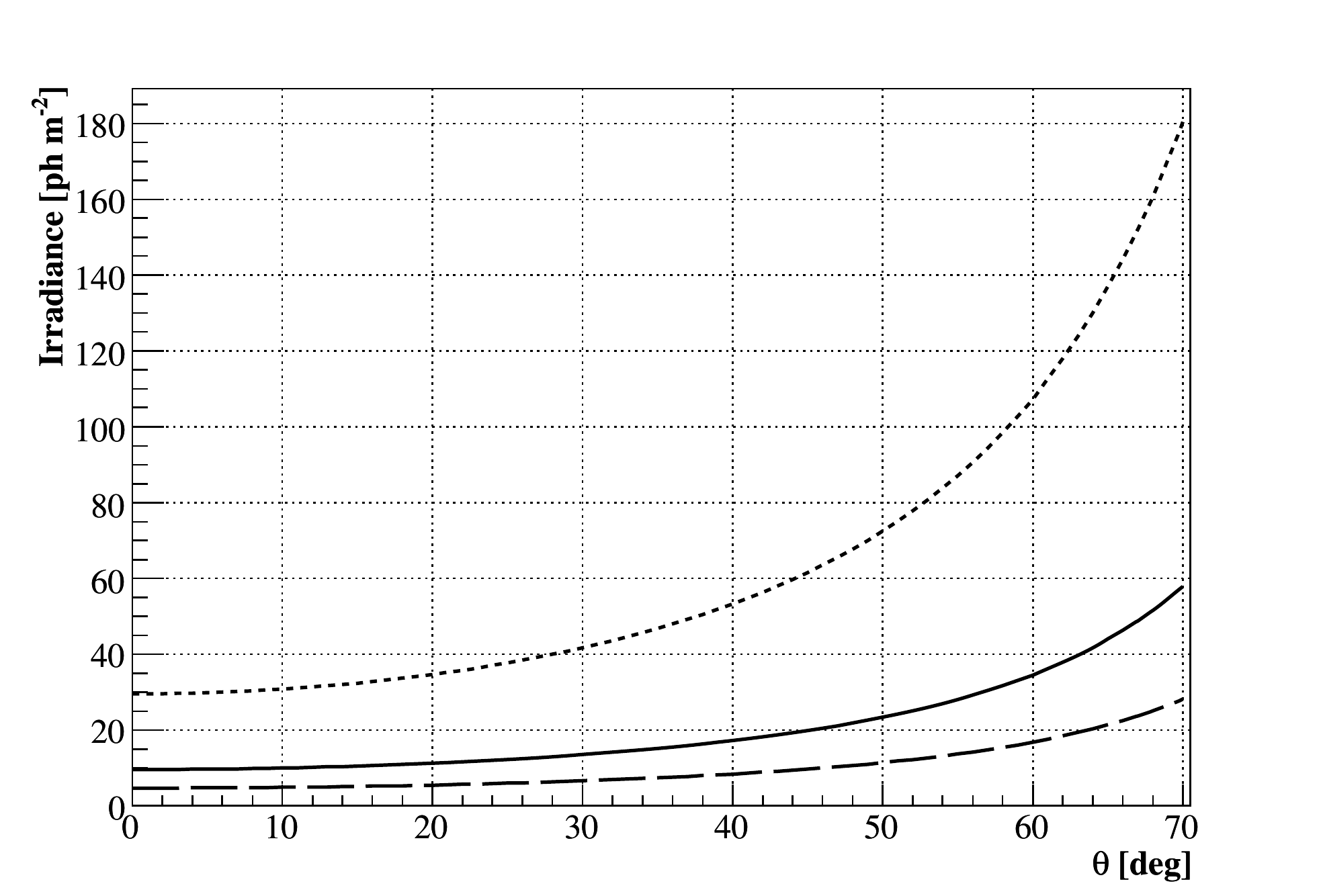}
		\caption{Signal time-integrated irradiance of an \EAS as a function of $\theta$ ($\psi_\mathrm{az}=90\degr$ and $\gamma=15\degr$). Solid line: $H=700\;\text{km}$; dotted line: $H=400\;\text{km}$; dashed line: $H=1000\;\text{km}$. }
		\label{fig:irrvstheta3h}
\end{figure}

\begin{figure}[htbp]
	\centering
		\includegraphics[width=0.9\textwidth]{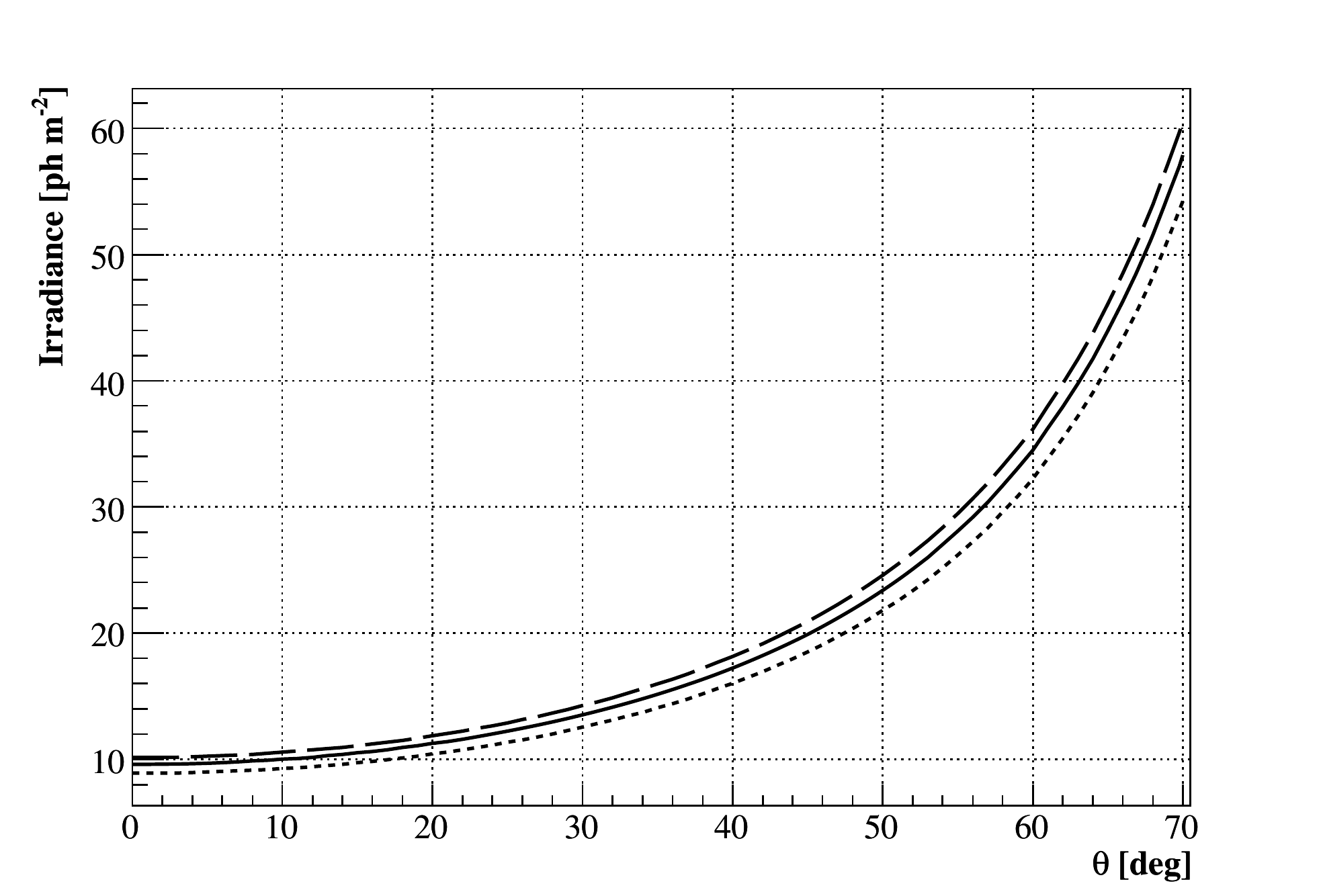}
		\caption{Signal time-integrated irradiance of an \EAS as a function of $\theta$ ($H=700\;\text{km}$ and $\psi_\mathrm{az}=90\degr$). Solid line: $\gamma=15\degr$; dotted line: $\gamma=20\degr$; dashed line: $\gamma=10\degr$. }
\label{fig:irrvstheta3gamma}
\end{figure}

The time-integrated irradiance of the signal reaching the instrument, as a function of $\theta$, is shown
in the figures~\ref{fig:irrvstheta3h} and~\ref{fig:irrvstheta3gamma}.
From the time-integrated irradiance we can infer the required overall triggering
efficacy and the minimum apparatus diameter (see section~\ref{sec:Estimates}).
This result does not depend on the implementation of the optical system and the
photo-sensor but some assumptions are required in order to proceed.
The required overall triggering
efficacy for observing the same \EAS obviously scales as $H^2$.


%--------------------------------------------------------------------------------
\subsection{The length and duration of the visible \EAS image}\label{sec:vislength}
%--------------------------------------------------------------------------------

The angle subtended by the visible \EAS image on the FS of the
apparatus as well as its time duration, which are fundamental parameters in the
experiment design,
are easily estimated by determining the first and
last detected points of the \EAS.

If one assumes to be able to subtract the background, the shape of
the detected photon hits is well described by the Gaisser-Hillas
function~\eqref{eq:gaisser-hillas}, which is a Gamma distribution in the variable
$ \widehat{X} \equiv X - X_0 $:
\begin{gather}\label{eq:gamma}
	g(\widehat{X}) = 
	\frac{1}{\Gamma(\alpha)}\beta^{-\alpha}{\widehat{X}}^{\alpha-1}\exp\pton{-\widehat{X}/\beta} \propto
        {\widehat{X}}^{\pton{ \Xris  / \lambda } }
	\exp\pton{  -  \widehat{X}  / \lambda } 
	\\
	\text{where} \qquad \alpha \equiv \frac{\Xris}{\beta} + 1,
	\spazio
	\beta \equiv \lambda,
	\spazio
	\Xris \equiv \Xmax - X_0
\punto
\end{gather}
The mean and the
standard deviation of this gamma distribution are therefore
\begin{gather}
	\left\langle X \right\rangle = \alpha \beta = \Xmax - X_0 + \lambda \nonumber \\
	\sigma_X = \beta \sqrt{\alpha} = \sqrt{\lambda(\Xmax - X_0 + \lambda)} \nonumber
\end{gather}

As we require to observe at least $ N \simeq  100 $ photons in order to
get a good enough energy resolution (see section~\ref{sec:NumPh}),
the minimum and maximum value and the range of $ \widehat{X}$ in the
distribution are:

\begin{gather}
	\min\pton{\widehat{X}}\approx \left\langle \widehat{X} \right\rangle-2\sigma_{\widehat{X}} \nonumber \\
	\max\pton{\widehat{X}}\approx \left\langle \widehat{X} \right\rangle+3\sigma_{\widehat{X}} \nonumber \\
	\text{range}\pton{\widehat{X}}\approx 5\sigma_{\widehat{X}} \nonumber
\virgola
\end{gather}
as it can be quickly determined via simulations of gamma distributions
using the physical parameters in section~\ref{sec:Reference}.

It should be noted that for $ N = 100 $ the range estimation is
$\approx 5\sigma$ for a gaussian distribution also but in that case the
distribution is obviously symmetrical with respect to the mean value.
However the gamma distribution arising from the current physical
parameters has a relatively small skewness\footnote{The skewness of a gamma distribution is given by 
$\gamma_1=2/\sqrt{\alpha}\approx0.55$ for a \sci{1}{19} eV proton EAS.} and kurtosis excess\footnote{The 
kurtosis excess of a gamma distribution is given by 
$\gamma_2=6/\alpha\approx0.46$ for a \sci{1}{19} eV proton EAS.} and therefore it is very similar to a gaussian.

Other values are summarized in table~\ref{tab:GammaStats}.
The results show that, as expected, the range of the observed photon
distribution (that is the observed image length) does not change by a
huge amount when changing the photon detection capability.
The dependence on the number of observed photons can be fitted by a
second degree polynomial.

Simulations with different parameters show that, as expected, the range of the observed photon
distribution (that is the observed image length) is not strongly
affected by little changes of the physical parameters of the \EAS.

\begin{table}[htb]
    \centering
        \begin{tabular}{cccc}
	\hline	
	$N$		& range		
	& $\left\langle \widehat{X} \right\rangle$-min 
	& max-$\left\langle \widehat{X} \right\rangle$ \\
	\hline	
	$N=10$		& $(3.05 \pm 0.01)\sigma$	& $1.4\sigma$ & $1.7\sigma$	\\
	$N=100$		& $(5.01 \pm 0.01)\sigma$	& $2.0\sigma$ & $3.0\sigma$	\\
	$N=1000$	& $(6.53 \pm 0.01)\sigma$	& $2.4\sigma$ & $4.1\sigma$ 	\\
	$N=10000$	& $(7.83 \pm 0.01)\sigma$	& $2.7\sigma$ & $5.1\sigma$ 	\\
	\hline
        \end{tabular}
        \caption{Range estimation for a gamma function with parameters
        corresponding to a typical \UHECP induced \EAS.}
\label{tab:GammaStats}
\end{table}

One must also take into account that a nearly vertical \EAS
is truncated by the ground.

\begin{figure}[htbp]
	\centering
		\includegraphics[width=0.9\textwidth]{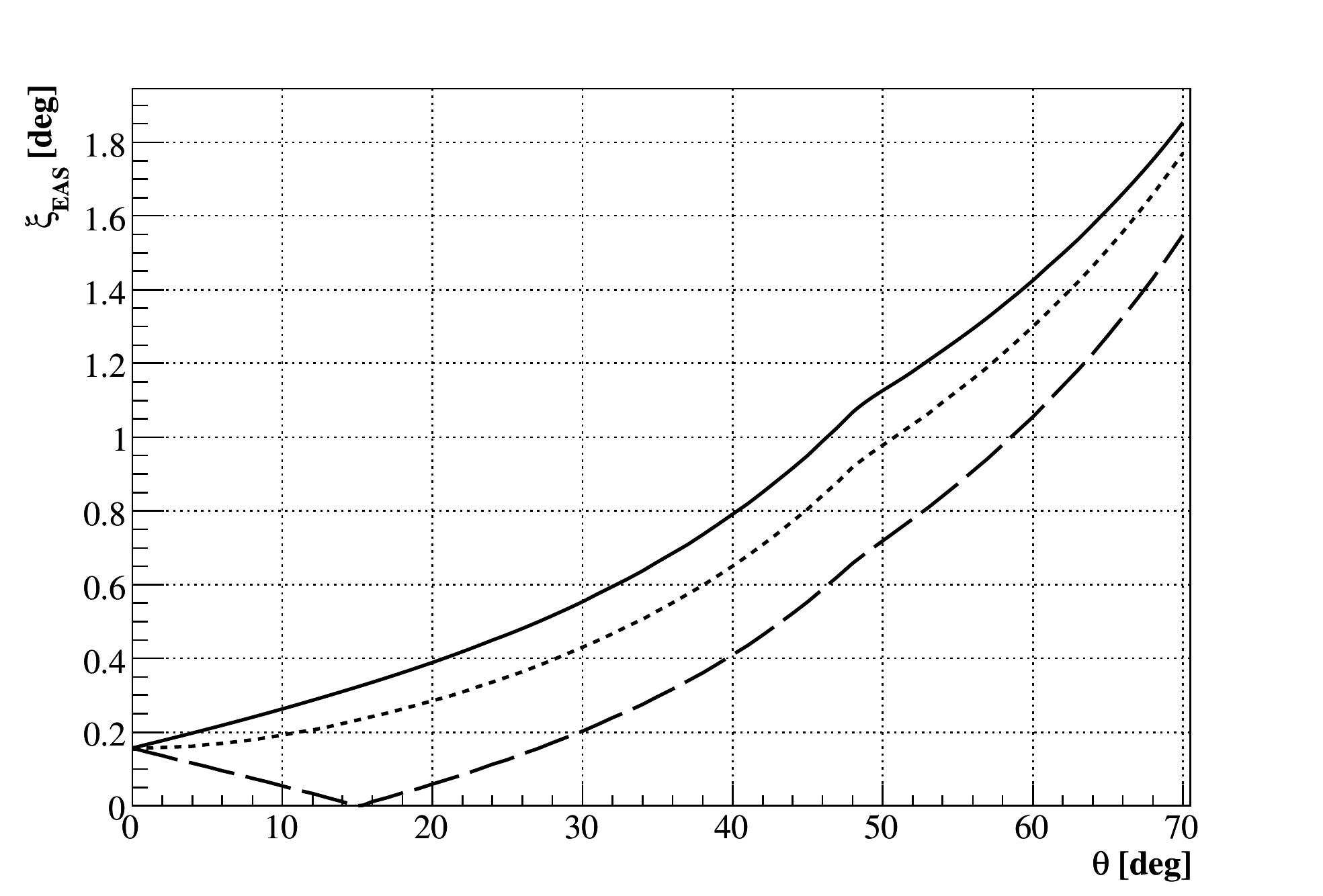}
		\caption{Angular extension of an EAS on the FS as a function of $\theta$ ($H=700\;\text{km}$ and $\gamma=15\degr$). Solid line: $\psi_\mathrm{az}=0\degr$; dotted line: $\psi_\mathrm{az}=90\degr$; dashed line: $\psi_\mathrm{az}=180\degr$. }\label{fig:angle3psi}
\end{figure}

\begin{figure}[htbp]
	\centering
		\includegraphics[width=0.9\textwidth]{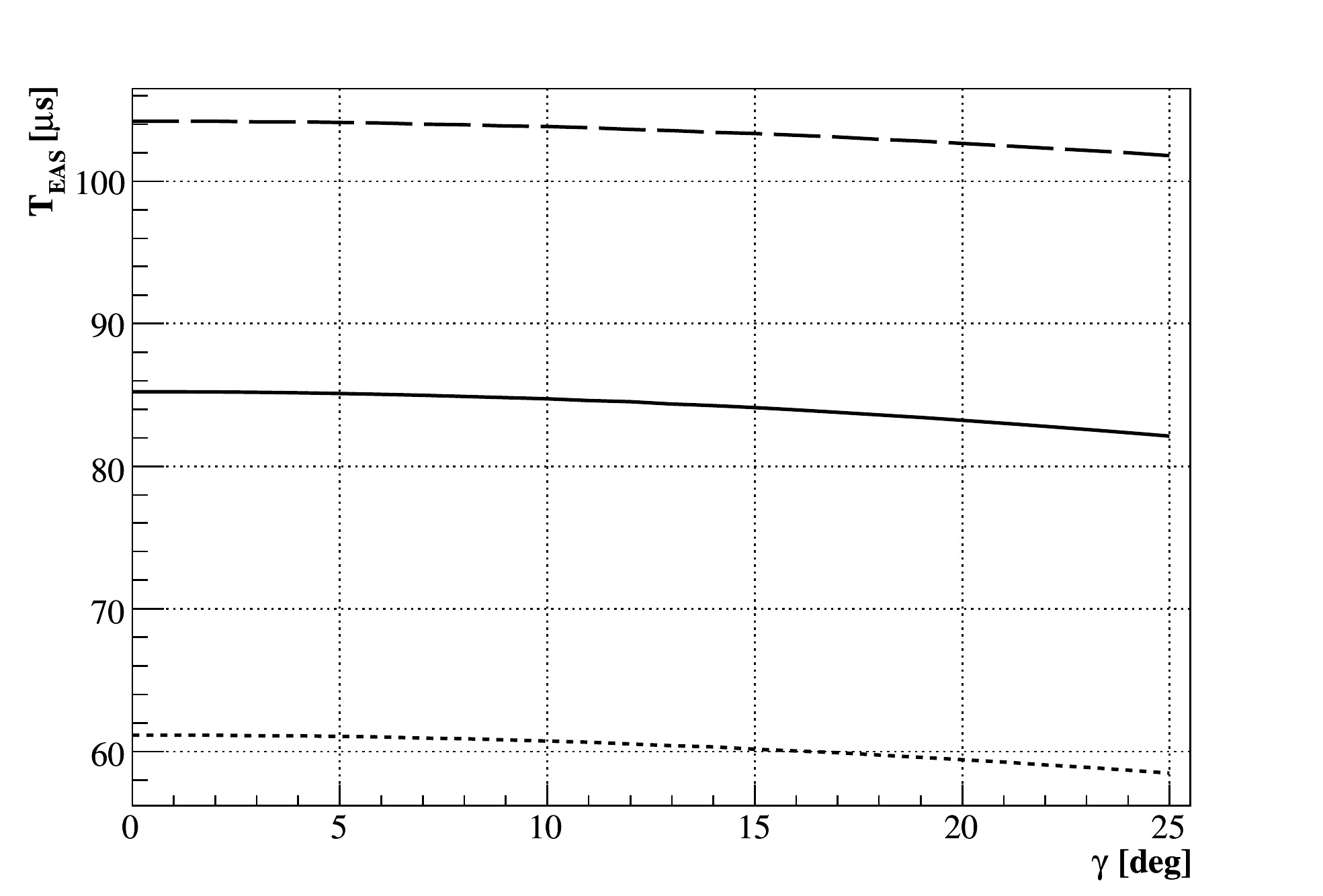}
	\caption{Time duration of an EAS on the FS as a function of $\gamma$ ($H=700\;\text{km}$ and $\psi_\mathrm{az}=90\degr$). Solid line: $\gamma=15\degr$; dotted line: $\gamma=20\degr$; dashed line: $\gamma=10\degr$. }
		\label{fig:durvsgamma}
	\end{figure}

It is now easy to estimate the angle $\EasAngle$ subtended by the \EAS
image on the FS
and the \EAS image duration $\EasTimeLength$ on the FS. 
The dependence on the \EAS zenith angle $\theta$ are shown respectively in
figures~\ref{fig:angle3psi}$\div$\ref{fig:durvsgamma}.

\begin{figure}[htbp]
	\centering
		\includegraphics[width=0.9\textwidth]{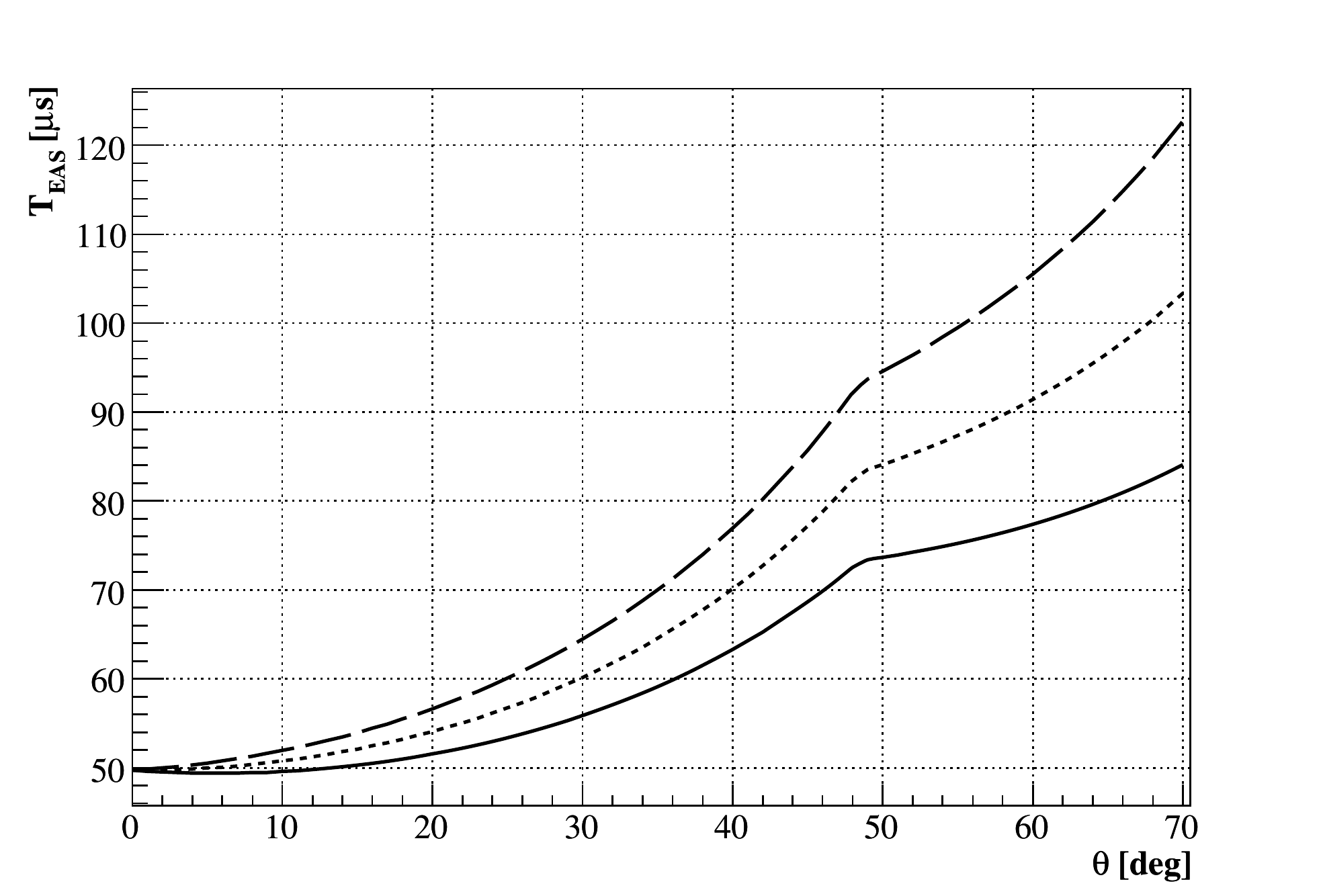}
		\caption{Time duration of an EAS on the FS as a function of $\theta$ ($H=700\;\text{km}$ and $\gamma=15\degr$). Solid line: $\psi_\mathrm{az}=0\degr$; dotted line: $\psi_\mathrm{az}=90\degr$; dashed line: $\psi_\mathrm{az}=180\degr$. }\label{fig:dur3psi}
\end{figure}
\begin{figure}[htbp]
	\centering
		\includegraphics[width=0.9\textwidth]{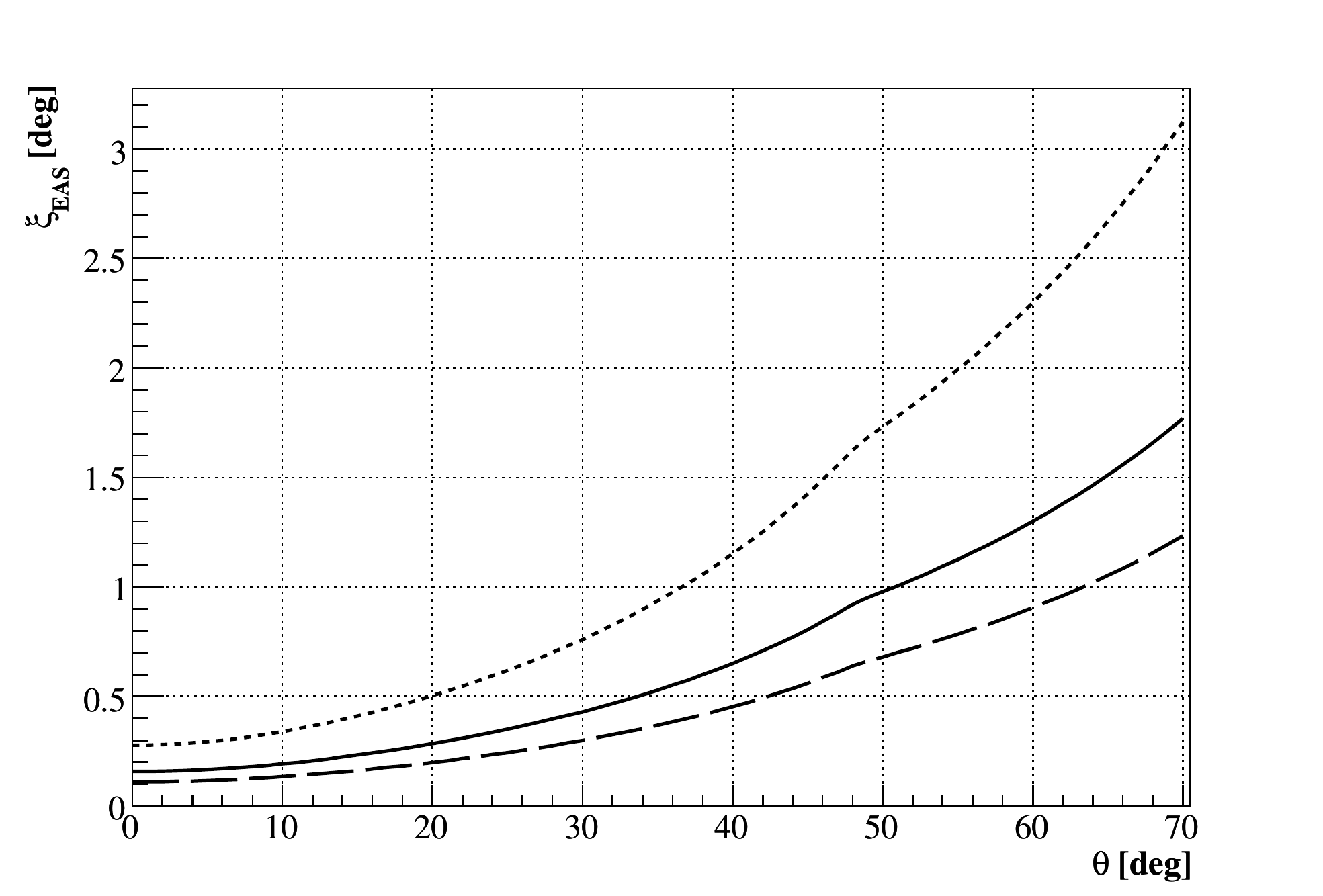}
		\caption{Angular extension of an EAS on the FS as a function of $\theta$ ($\psi_\mathrm{az}=90\degr$ and $\gamma=15\degr$). Solid line: $H=700\;\text{km}$; dotted line: $H=400\;\text{km}$; dashed line: $H=1000\;\text{km}$. }
		\label{fig:anglevstheta3h}
\end{figure}
\begin{figure}[htbp]
	\centering
		\includegraphics[width=0.9\textwidth]{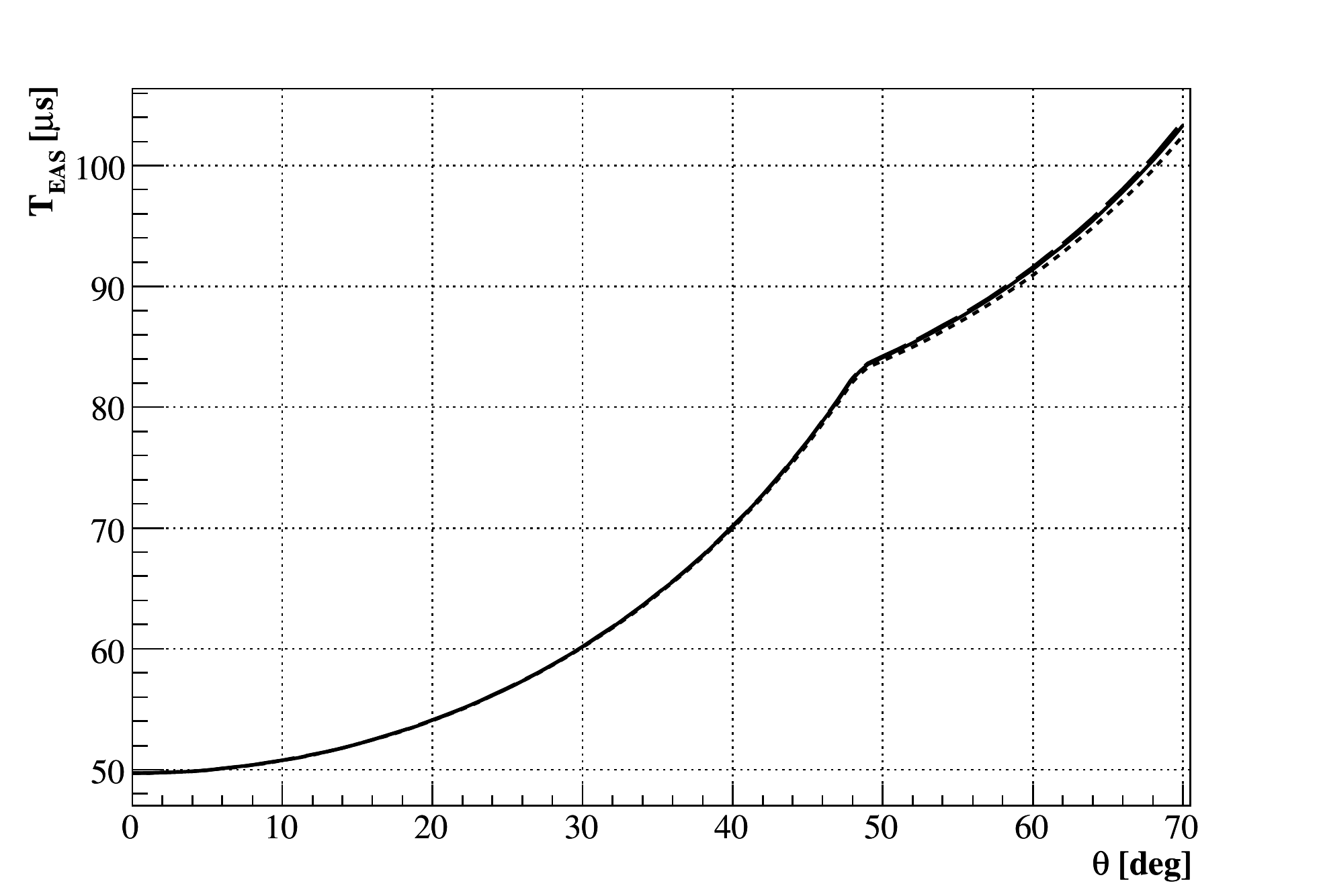}
	\caption{Time duration of an EAS on the FS as a function of $\theta$ ($\psi_\mathrm{az}=90\degr$ and $\gamma=15\degr$). Solid line: $H=700\;\text{km}$; dotted line: $H=400\;\text{km}$; dashed line: $H=1000\;\text{km}$. }
	\label{fig:durvstheta3h}
\end{figure}
\begin{figure}[htbp]
	\centering
		\includegraphics[width=0.9\textwidth]{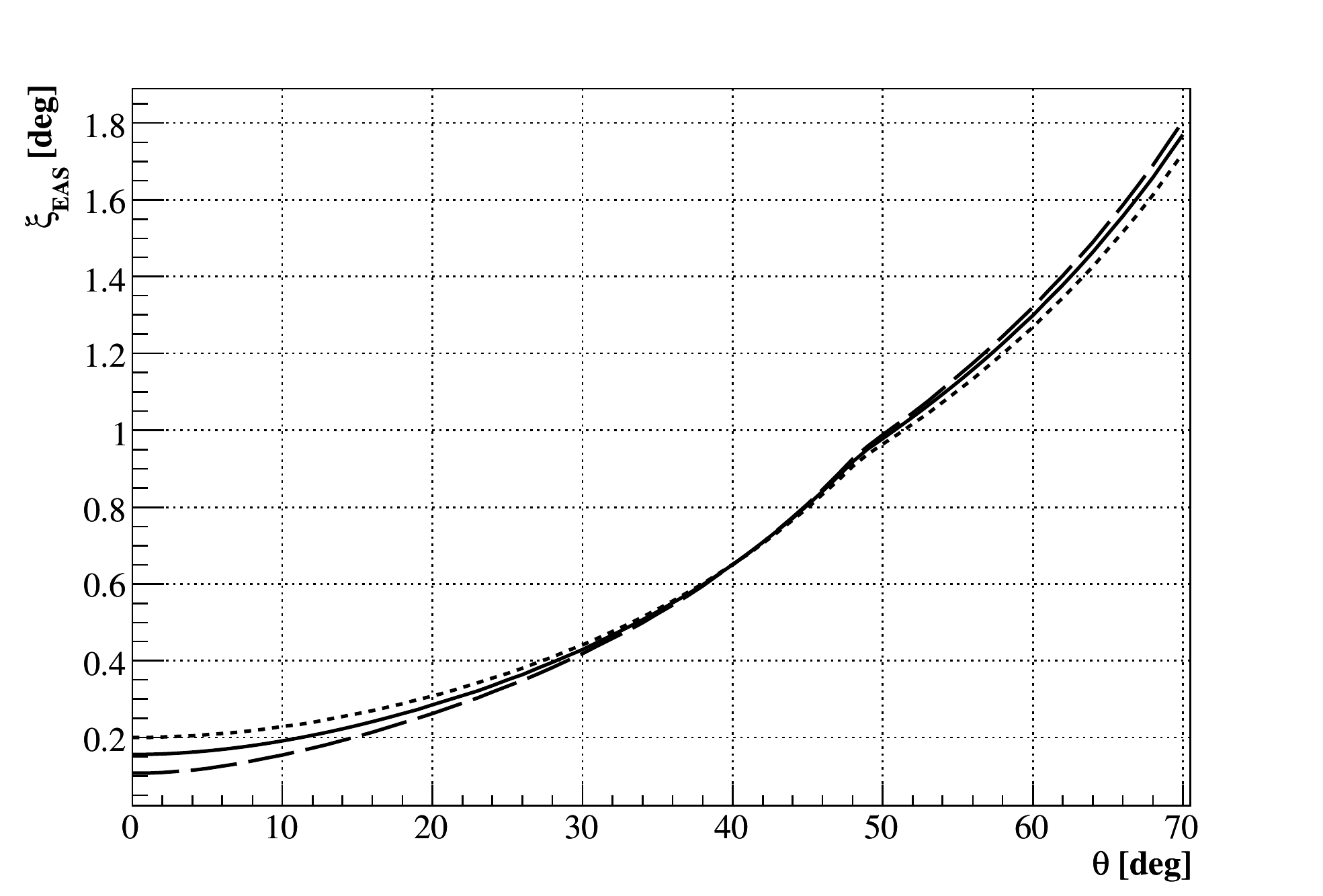}
		\caption{Angular extension of an EAS on the FS as a function of $\theta$ ($H=700\;\text{km}$ and $\psi_\mathrm{az}=90\degr$). Solid line: $\gamma=15\degr$; dotted line: $\gamma=20\degr$; dashed line: $\gamma=10\degr$. }
\label{fig:anglevstheta3gamma}
\end{figure} 
\begin{figure}[htbp]
	\centering
		\includegraphics[width=0.9\textwidth]{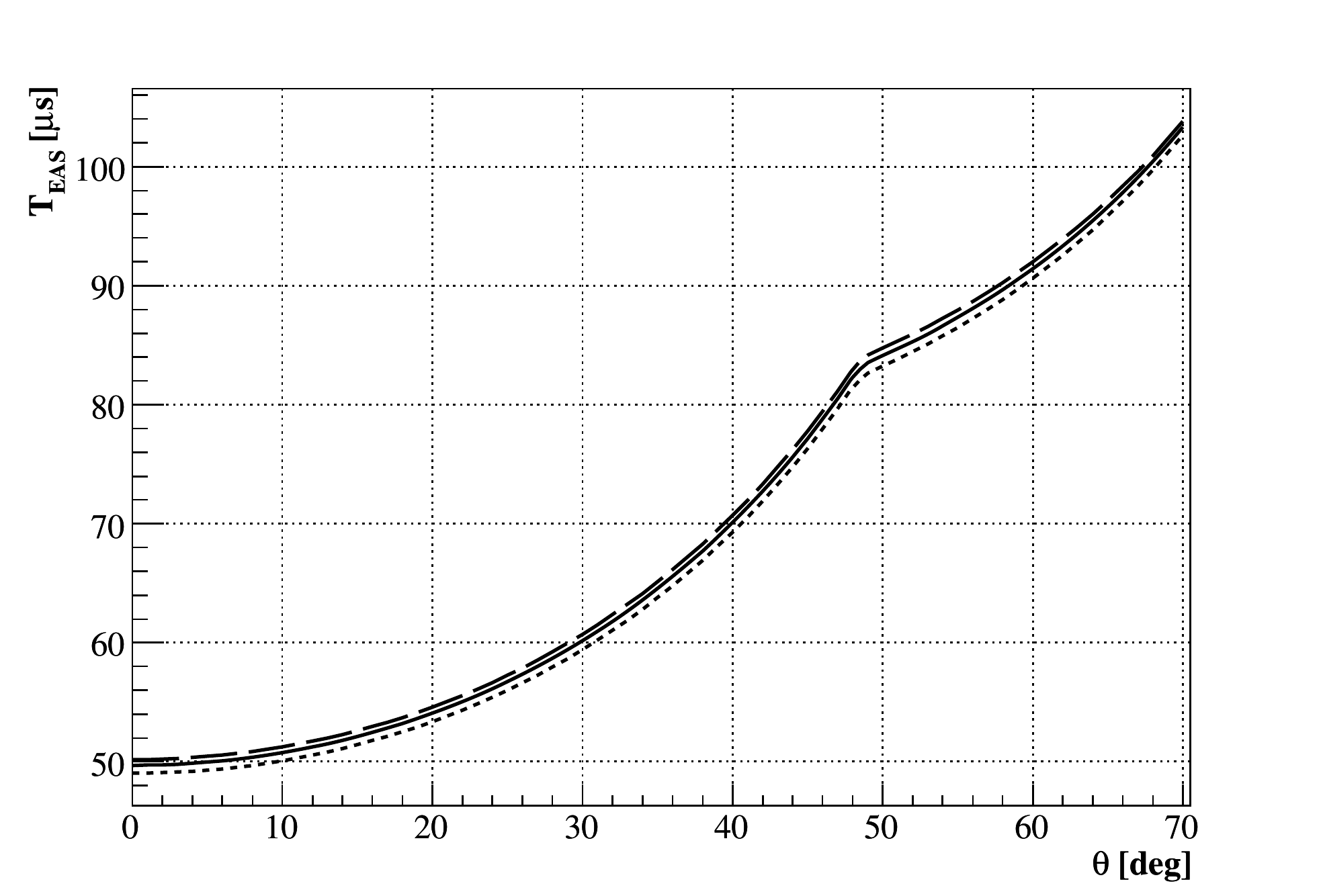}
	\caption{Time duration of an EAS on the FS as a function of $\theta$ ($H=700\;\text{km}$ and $\psi_\mathrm{az}=90\degr$). Solid line: $\gamma=15\degr$; dotted line: $\gamma=20\degr$; dashed line: $\gamma=10\degr$. }
	\label{fig:durvstheta3gamma}
\end{figure}
\begin{figure}[htbp]
	\centering
		\includegraphics[width=0.9\textwidth]{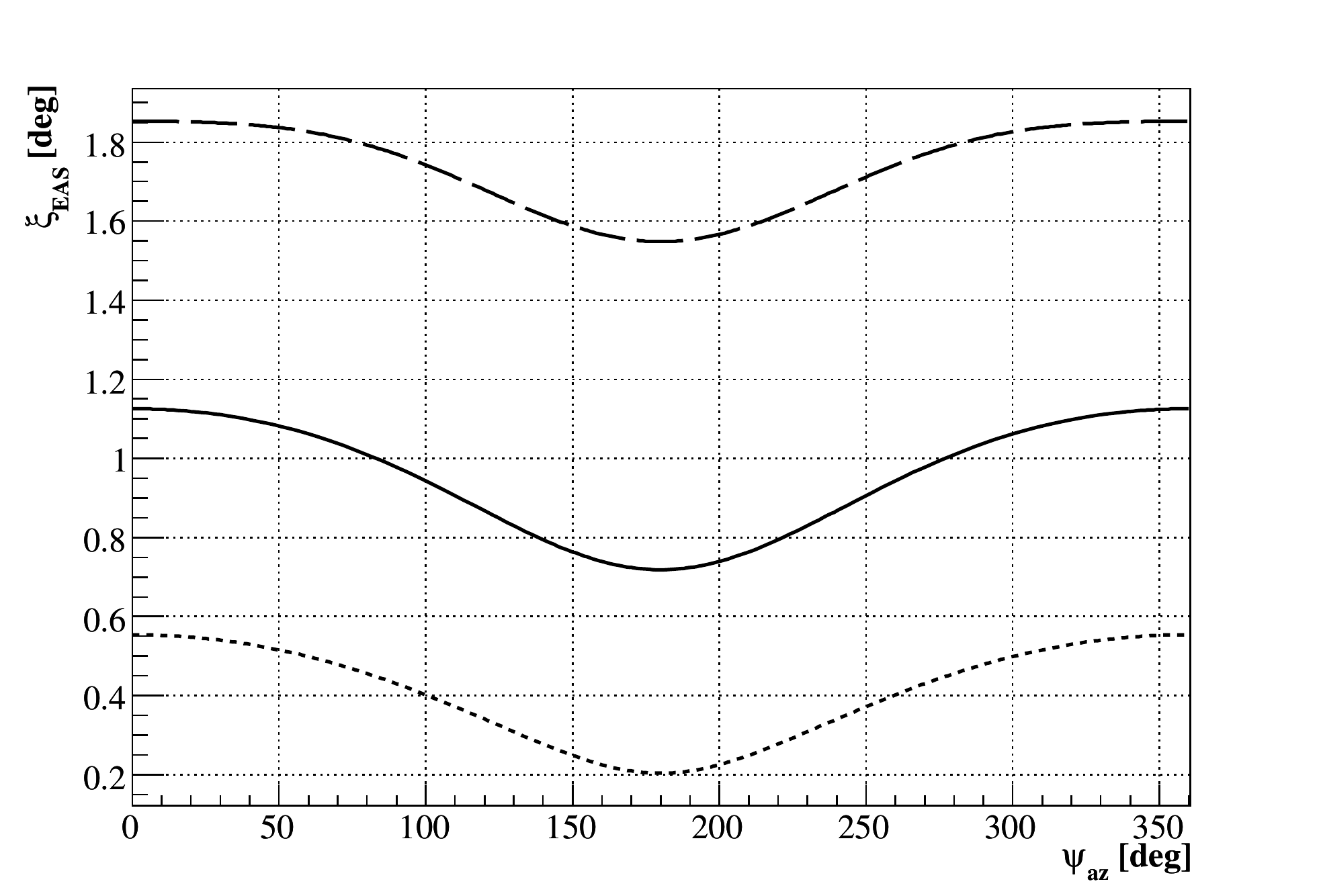}
		\caption{Angular extension of an EAS on the FS as a function of $\psi_\mathrm{az}$ ($H=700\;\text{km}$ and $\gamma=15\degr$). Solid line: $\theta=50\degr$; dotted line: $\theta=30\degr$; dashed line: $\theta=70\degr$. }
		\label{fig:angle3theta}
\end{figure}
\begin{figure}[htbp]
	\centering
		\includegraphics[width=0.9\textwidth]{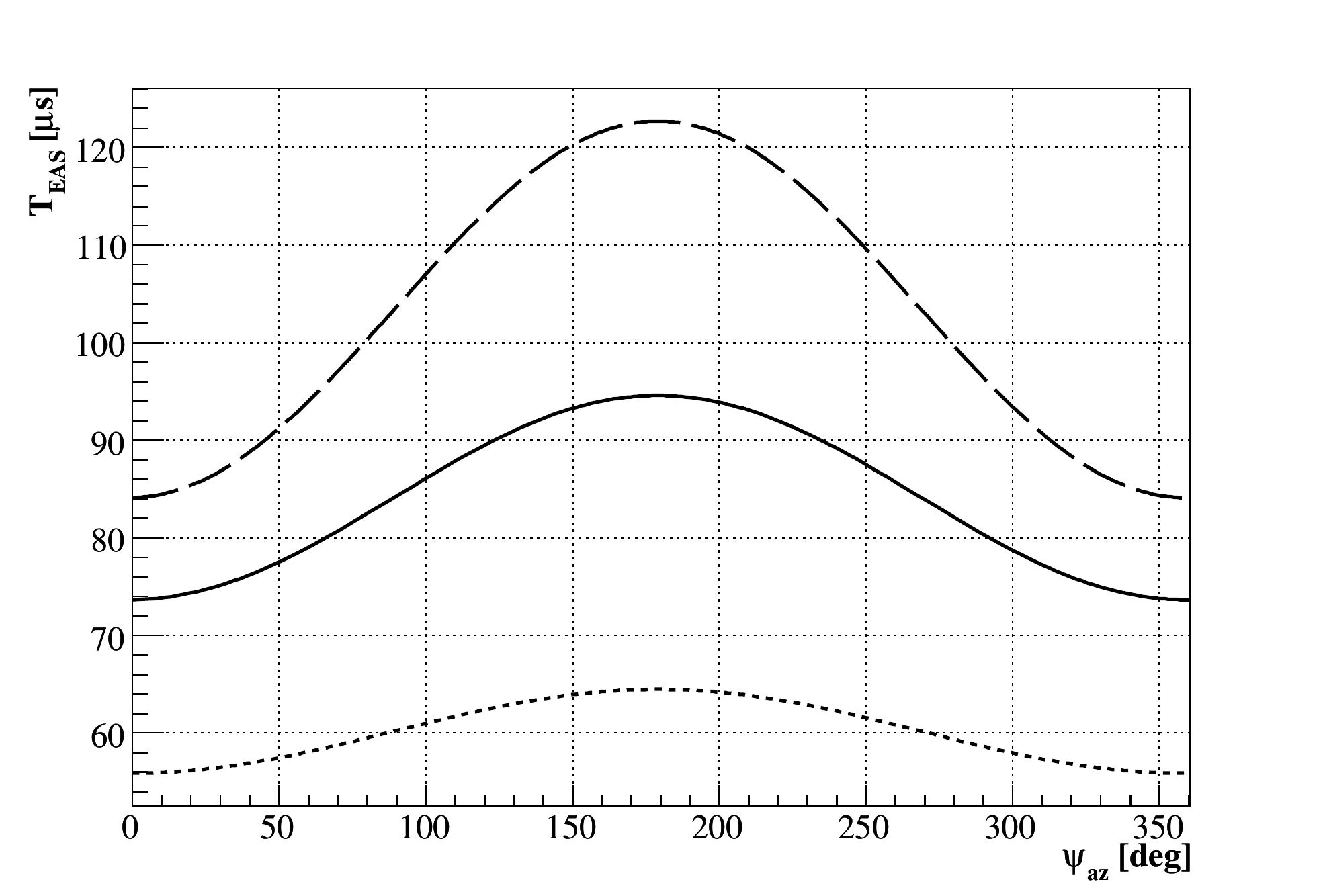}
		\caption{Time duration of an EAS on the FS as a function of $\psi_\mathrm{az}$ ($H=700\;\text{km}$ and $\gamma=15\degr$). Solid line: $\theta=50\degr$; dotted line: $\theta=30\degr$; dashed line: $\theta=70\degr$. }
	\label{fig:dur3theta}
\end{figure}
\begin{figure}[htbp]
	\centering
		\includegraphics[width=0.9\textwidth]{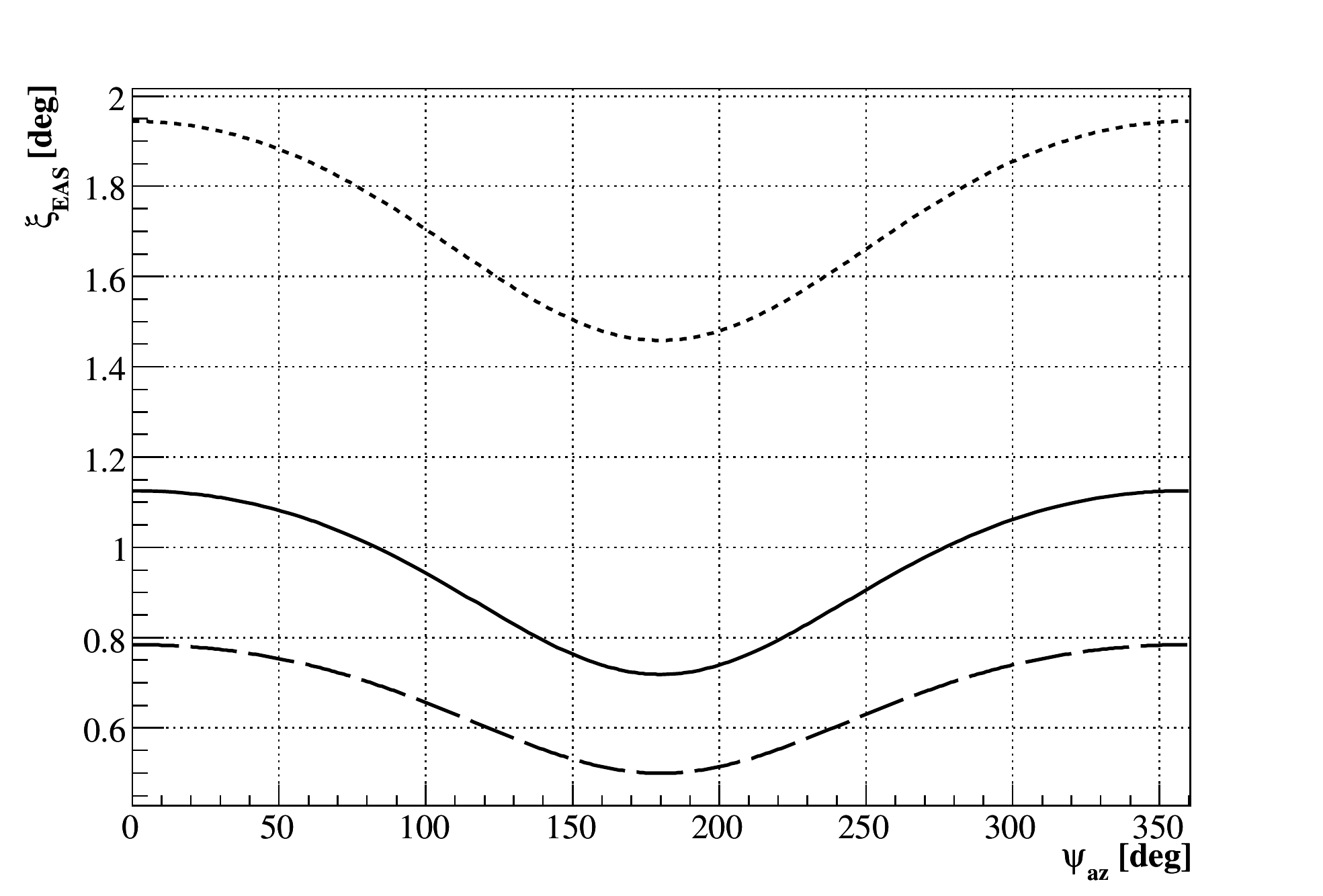}
		\caption{Angular extension of an EAS on the FS as a function of $\psi_\mathrm{az}$ ($\theta=50\degr$ and $\gamma=15\degr$). Solid line: $H=700\;\text{km}$; dotted line: $H=400\;\text{km}$; dashed line: $H=1000\;\text{km}$. }
		\label{fig:angle3h}
	\end{figure}
\begin{figure}[htbp]
	\centering
		\includegraphics[width=0.9\textwidth]{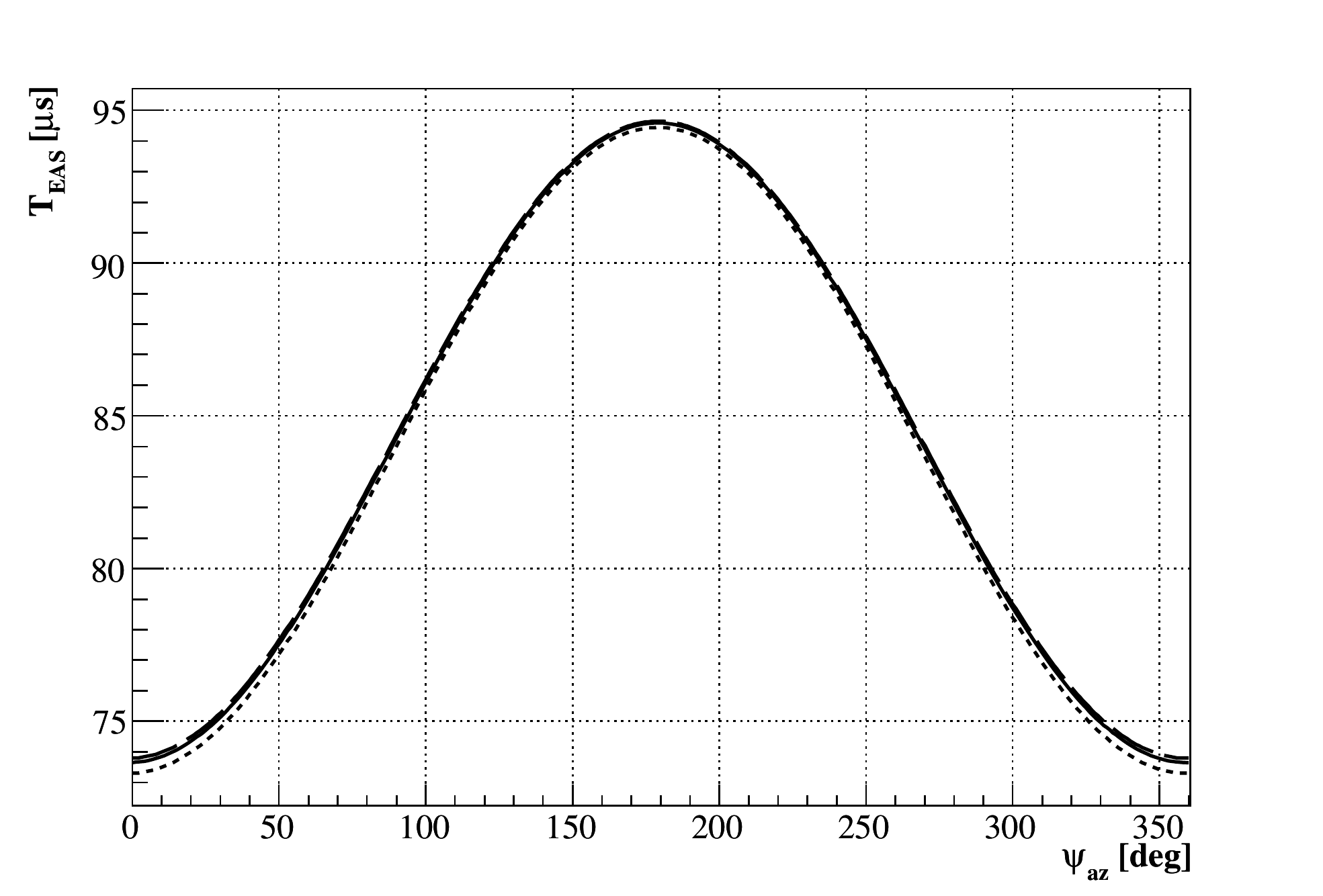}
		\caption{Time duration of an EAS on the FS as a function of $\psi_\mathrm{az}$ ($\theta=50\degr$ and $\gamma=15\degr$). Solid line: $H=700\;\text{km}$; dotted line: $H=400\;\text{km}$; dashed line: $H=1000\;\text{km}$. }
		\label{fig:dur3h}
	\end{figure}
	\begin{figure}[htbp]
	\centering
		\includegraphics[width=0.9\textwidth]{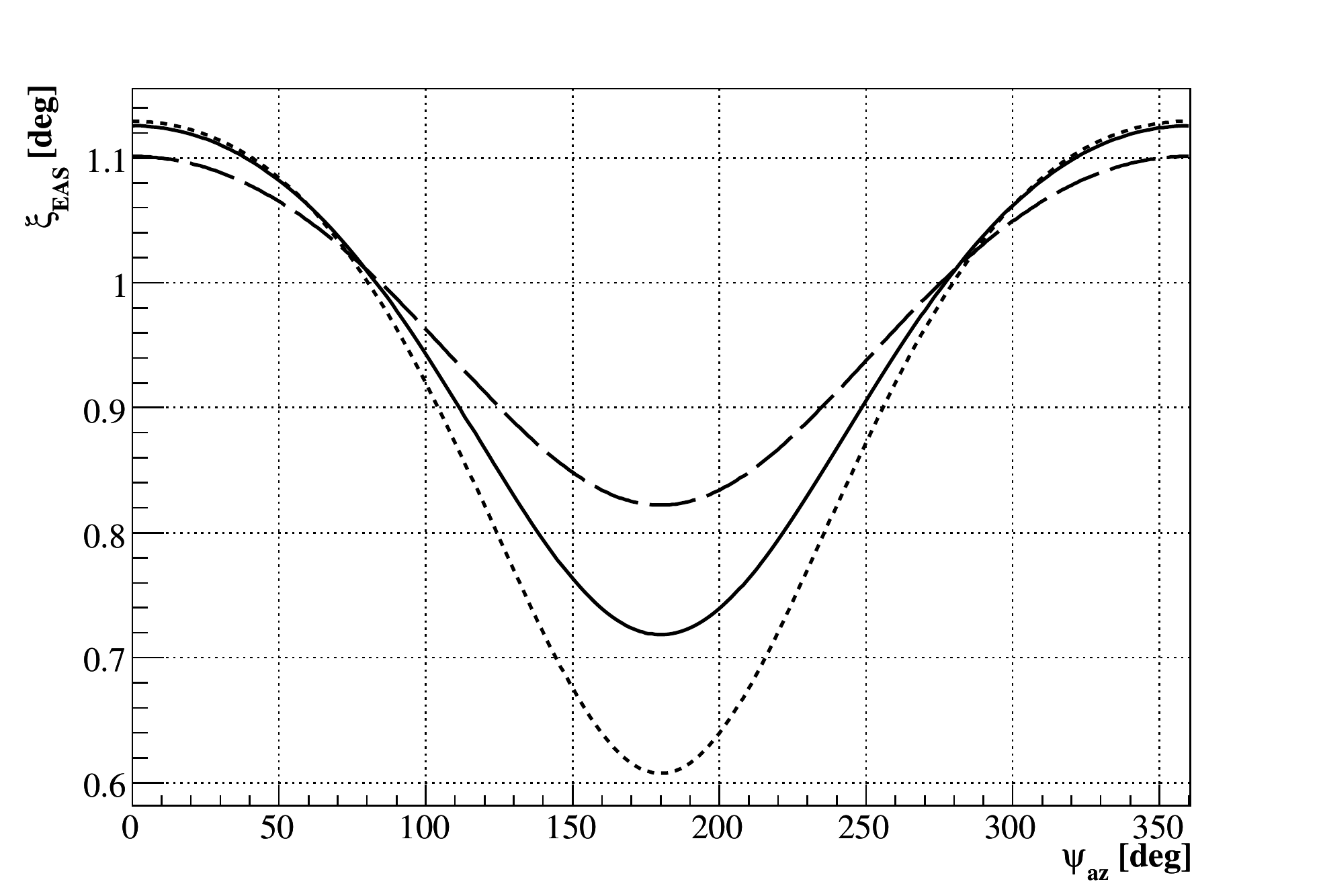}
		\caption{Angular extension of an EAS on the FS as a function of $\psi_\mathrm{az}$ ($H=700\;\text{km}$ and $\theta=50\degr$). Solid line: $\gamma=15\degr$; dotted line: $\gamma=20\degr$; dashed line: $\gamma=10\degr$. }
		\label{fig:angle3gamma}
	\end{figure}
		\begin{figure}[htbp]
	\centering
		\includegraphics[width=0.9\textwidth]{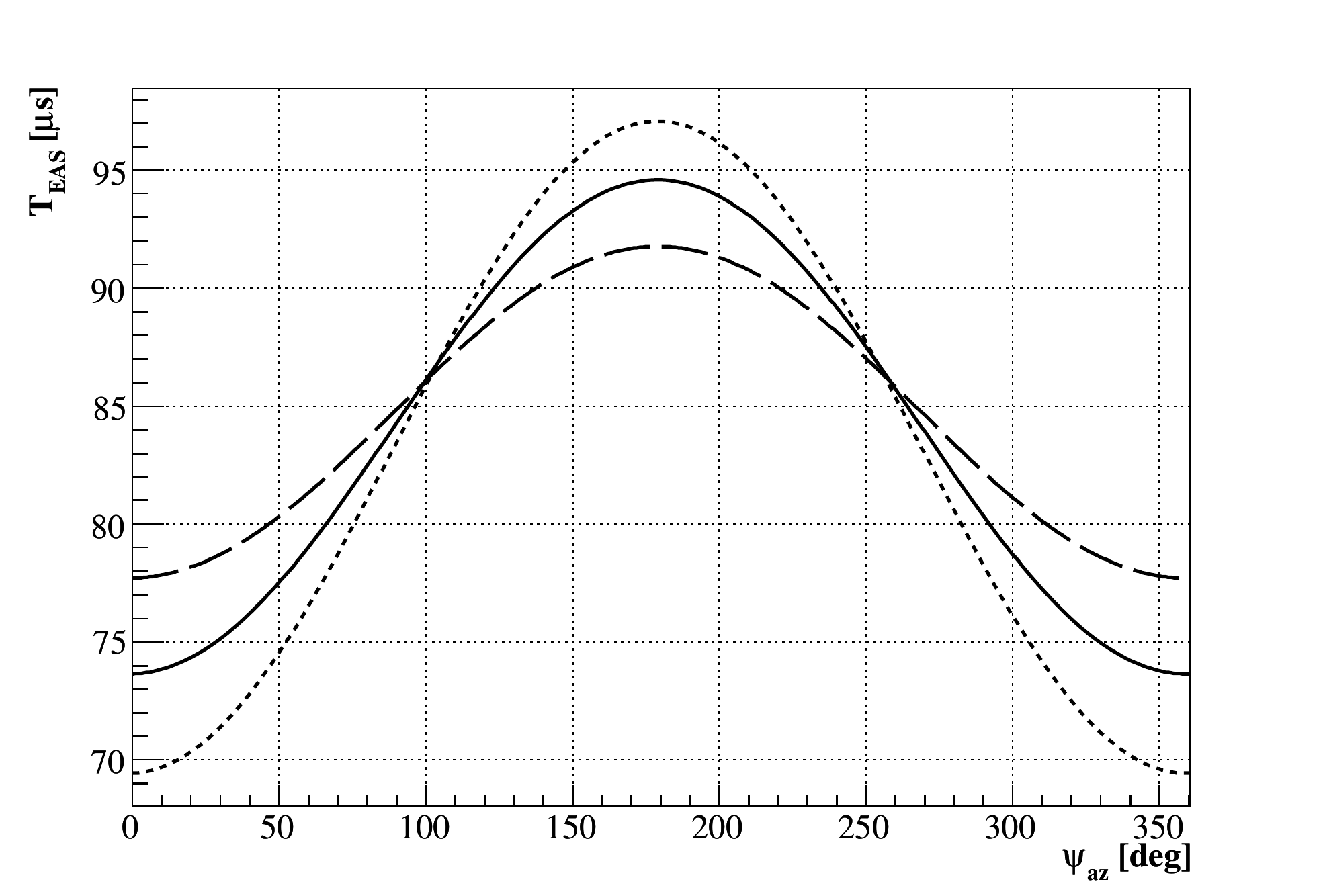}
		\caption{Time duration of an EAS on the FS as a function of $\psi_\mathrm{az}$ ($H=700\;\text{km}$ and $\theta=50\degr$). Solid line: $\gamma=15\degr$; dotted line: $\gamma=20\degr$; dashed line: $\gamma=10\degr$. }
		\label{fig:dur3gamma}
	\end{figure}
	\begin{figure}[htbp]
	\centering
		\includegraphics[width=0.9\textwidth]{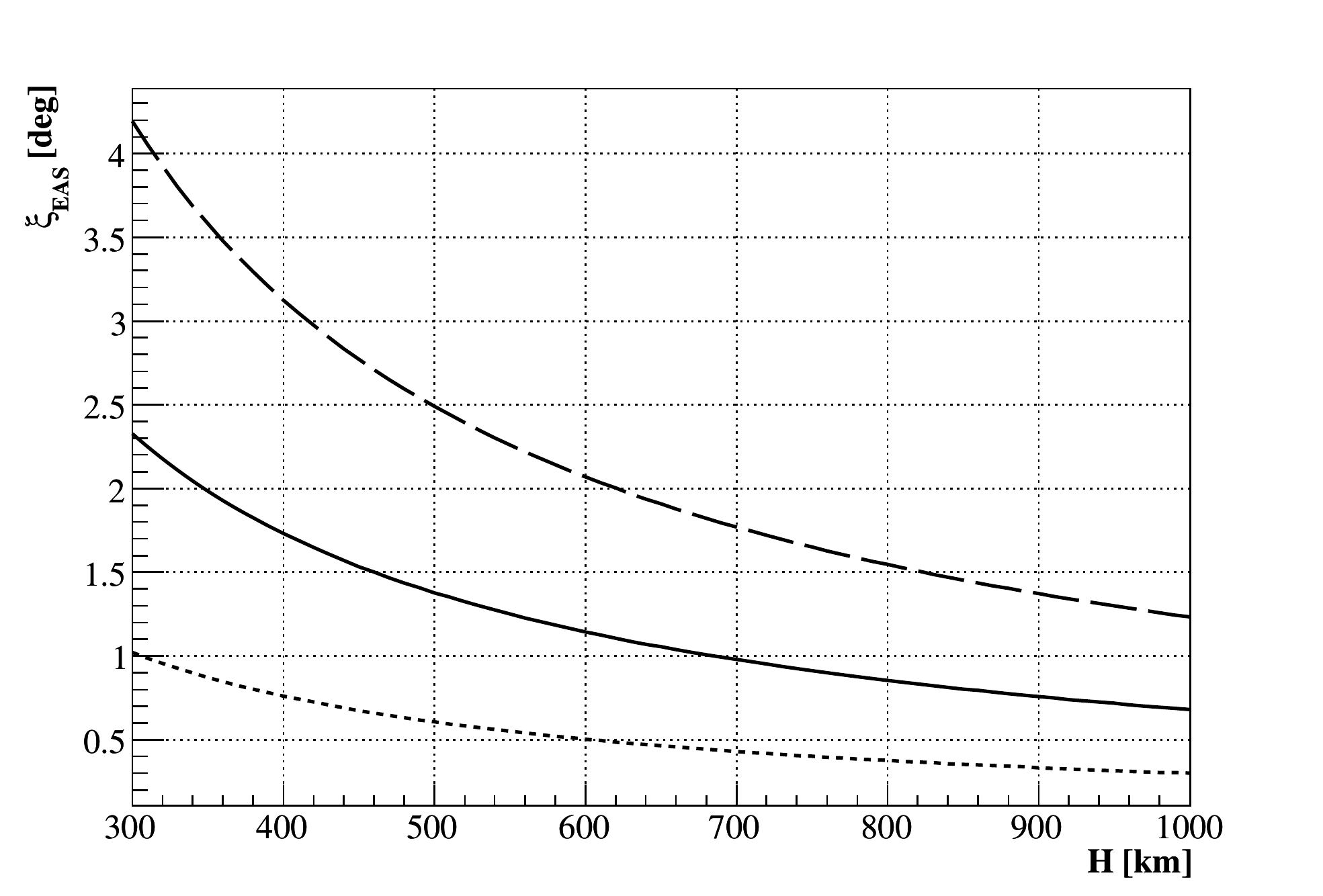}
		\caption{Angular extension of an EAS on the FS as a function of $H$ ($\psi_\mathrm{az}=90\degr$ and $\gamma=15\degr$). Solid line: $\theta=50\degr$; dotted line: $\theta=30\degr$; dashed line: $\theta=70\degr$. }
		\label{fig:anglevsh}
	\end{figure}
		\begin{figure}[htbp]
	\centering
		\includegraphics[width=0.9\textwidth]{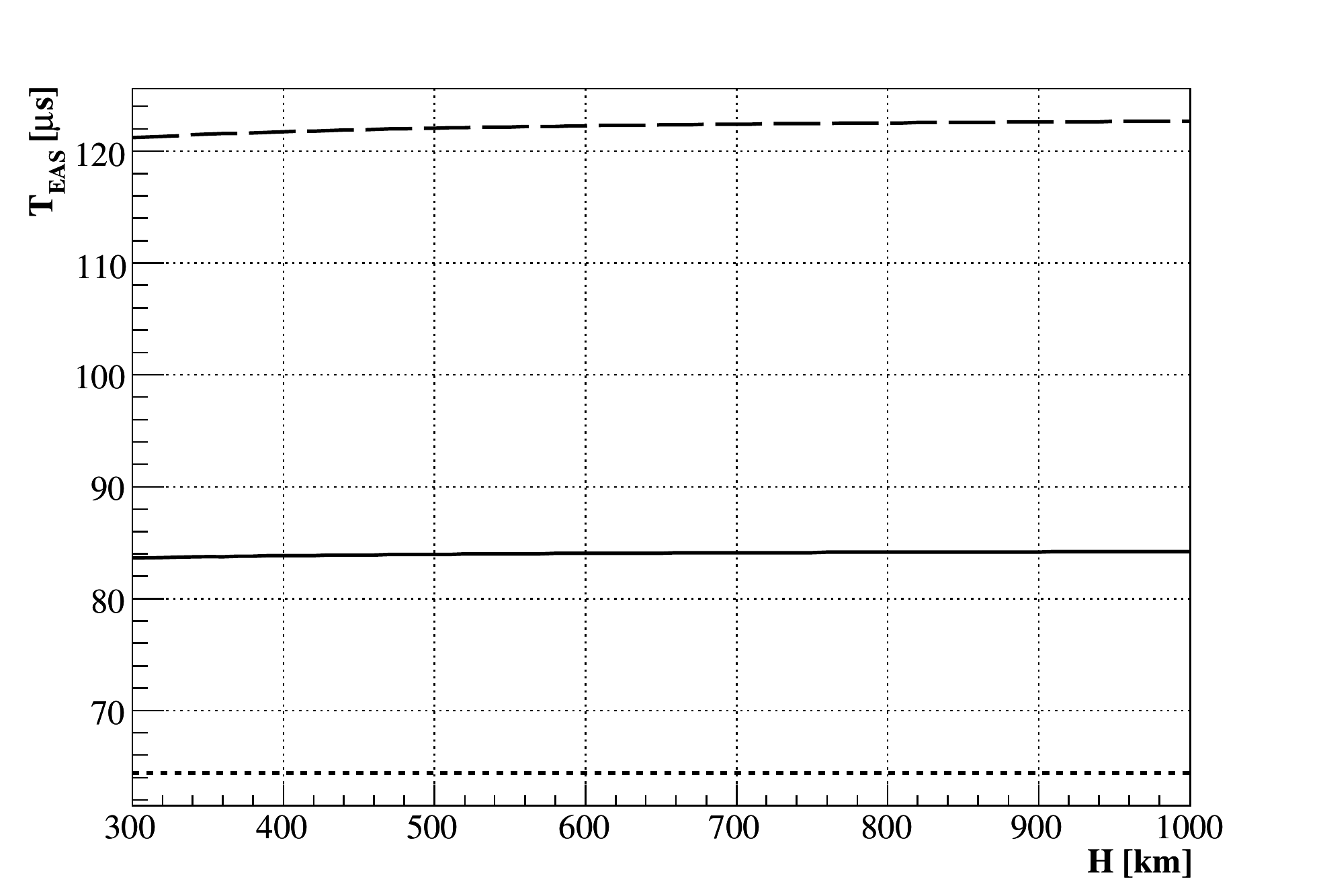}
		\caption{Time duration of an EAS on the FS as a function of $H$ ($\psi_\mathrm{az}=90\degr$ and $\gamma=15\degr$). Solid line: $\theta=50\degr$; dotted line: $\theta=30\degr$; dashed line: $\theta=70\degr$. }
		\label{fig:durvsh}
	\end{figure}
		\begin{figure}[htbp]
	\centering
		\includegraphics[width=0.9\textwidth]{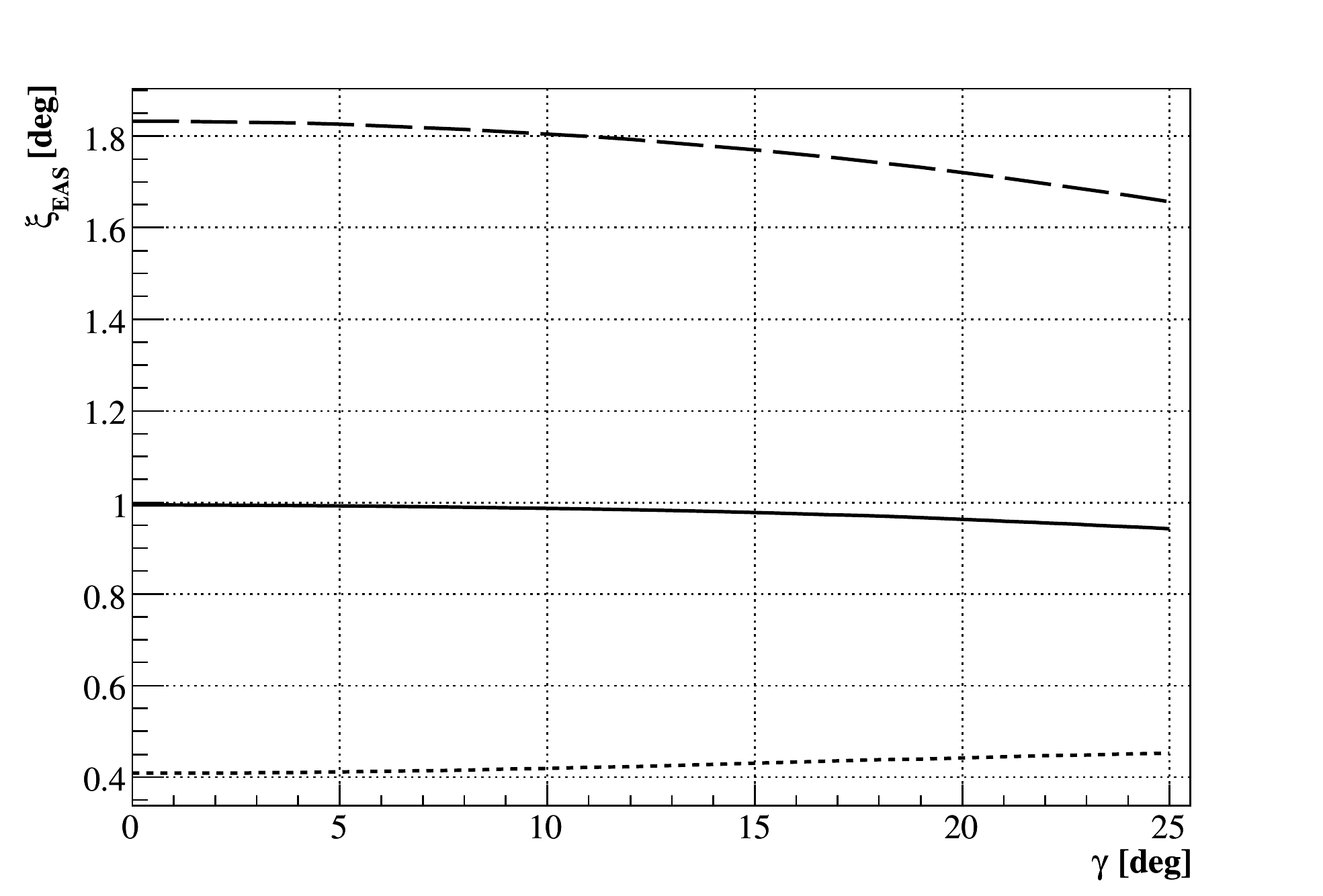}
		\caption{Angular extension of an EAS on the FS as a function of $\gamma$ ($H=700\;\text{km}$ and $\psi_\mathrm{az}=90\degr$). Solid line: $\gamma=15\degr$; dotted line: $\gamma=20\degr$; dashed line: $\gamma=10\degr$. }
		\label{fig:anglevsgamma}
	\end{figure}

As it is clear from these figures, $\EasAngle \lesssim 4\degr$,
justifying the assumptions of section~\ref{sec:GenAss}. Note also that,
for $\theta\lesssim50\degr$, the \EAS are truncated because the \EAS
hits the ground, as it is evident from the cusps in the
figures. Nevertheless the truncation does not affect the reference \EAS,
for which we requested $N=100$ photons.  Some numerical values for the
reference \EAS are given in section~\ref{sec:SignalEst}.

It is obvious that the time duration of the \EAS image is almost
independent on both $H$ and $\gamma$ for a fixed \EAS geometry
(figures~\ref{fig:durvstheta3h},~\ref{fig:durvstheta3gamma},~\ref{fig:durvsh}
and~\ref{fig:durvsgamma}) while the angular length, for the same \EAS,
will scale as $\sim\cos\gamma/H$ (figures~\ref{fig:anglevsh}
and~\ref{fig:anglevsgamma}).

%--------------------------------------------------------------------------------
\subsection{Aperture}
\label{subsec:aperture}
%--------------------------------------------------------------------------------

%--------------------------------------------------------------------------------
\subsubsection{Area observed at the Earth}
%--------------------------------------------------------------------------------

In the case of a nadir pointing apparatus
the geometrical area spanned by the \FoV at the surface of the Earth is 
the area of a spherical cap,
\begin{equation}
	A_0 = 2 \pi R_{\oplus}^2 \pton{ 1-\cos\beta_\mathrm{M} }
\virgola
\end{equation}
in terms of $H$, the orbital height,
$R_{\oplus}$, the Earth radius, $\Gmax$ the half-angle \FoV of the apparatus,
and $\beta_\mathrm{M}$, the angle at Earth center between nadir and the \FoV
border which is given by:
\begin{equation}
	\beta_\mathrm{M} = \arcsin \left( \frac{R_{\oplus}+H}{R_{\oplus}}\sin\Gmax\right) - \Gmax
\punto
\end{equation}

From $A_0 $ it is easy to estimate the total target mass, $M$, using the value
of the vertical
column density of air ($\sim 1033 \um{g/cm^2}$).
In table~\ref{tab:Area} some numbers are given for different values of
$H$ and \Gmax. An approximated value for the area, using the flat Earth
approximation is 
\begin{equation}\label{eq:AreaFlatEarth}
	A_0 = \pi H^2 \tan^2 \Gmax \punto
\end{equation}

\begin{table}[htb]
	\centering
		\begin{tabular}{cccc} \hline
			H [km]    &  \Gmax [\degr] &  $A_0$ [\un[\sci{1}{5}]{km^2}] & M [\un[\sci{1}{15}]{kg}] \\ \hline
			400       &  20  & 0.67  &   0.7 \\
			700       &  20  & 2.07  &   2.1 \\
			1000      &  20  & 4.25  &   4.4 \\
			700       &  15  & 1.11  &   1.2 \\
			700       &  25  & 3.43  &   3.5 \\ \hline
		\end{tabular}
	\caption{Area observed at the Earth and atmosphere mass target.}
	\label{tab:Area}
\end{table}

%--------------------------------------------------------------------------------
\subsubsection{Instantaneous geometrical aperture}
%--------------------------------------------------------------------------------

The instantaneous geometrical aperture is defined as:
\begin{equation}
    \GeoAperture \equiv \int_{A} \int_{\Omega} \hat{v}(\theta,\varphi)\cdot\hat{n}\diffl \Omega \diffl A
    \virgola 
\end{equation}
in terms of the normalized velocity vector, $\hat{v}(\theta,\varphi)$,
of the \EAS and the normal unit vector to
the surface, $\hat{n}$.

The effective aperture, for a flat Earth, of a nadir pointing apparatus,
only considering
\EAS which reach ground inside the \FoV, is approximately given by the relation:
\begin{equation}
	\EffAperture =
	\eta _\mathrm{o} \eta _\mathrm{c} \pton{ 1-\tau_{\mathrm{dead}}} \GeoAperture \approx
	\eta _\mathrm{o} \eta _\mathrm{c} \pton{ 1-\tau_{\mathrm{dead}} } \pi^2 H^2 \tan^2\Gmax
   \virgola
\end{equation}
in terms of the orbital height, $H$, the half-angle \FoV $\Gmax$, 
the observational duty cycle $\eta_\mathrm{o}$, the dead time $\tau_\mathrm{dead}$ 
and the cloud coverage efficiency ($\eta_\mathrm{c}\sim 0.5$), 
quantifying the effect of real cloud coverage on the \EAS detection efficiency.
The expression above also gives the effective high energy (asymptotic) aperture,
that is the effective aperture when the total detection efficiency equals one.

%--------------------------------------------------------------------------------
\subsubsection{Tilting of the apparatus}
%--------------------------------------------------------------------------------

The instantaneous geometrical aperture can increase if the apparatus is
tilted with respect to the local nadir by some angle $\TiltAngle$. In order to
evaluate it one needs to estimate the area observed at ground
with a tilted apparatus. As the geometrical solution of the problem to
find the intersection area of the circular \FoV cone with the spherical
Earth surface is not trivial at all, the easiest way is to use a simple
Monte-Carlo integration.

\begin{figure}[htbp]
	\centering
		\includegraphics[width=0.9\textwidth]{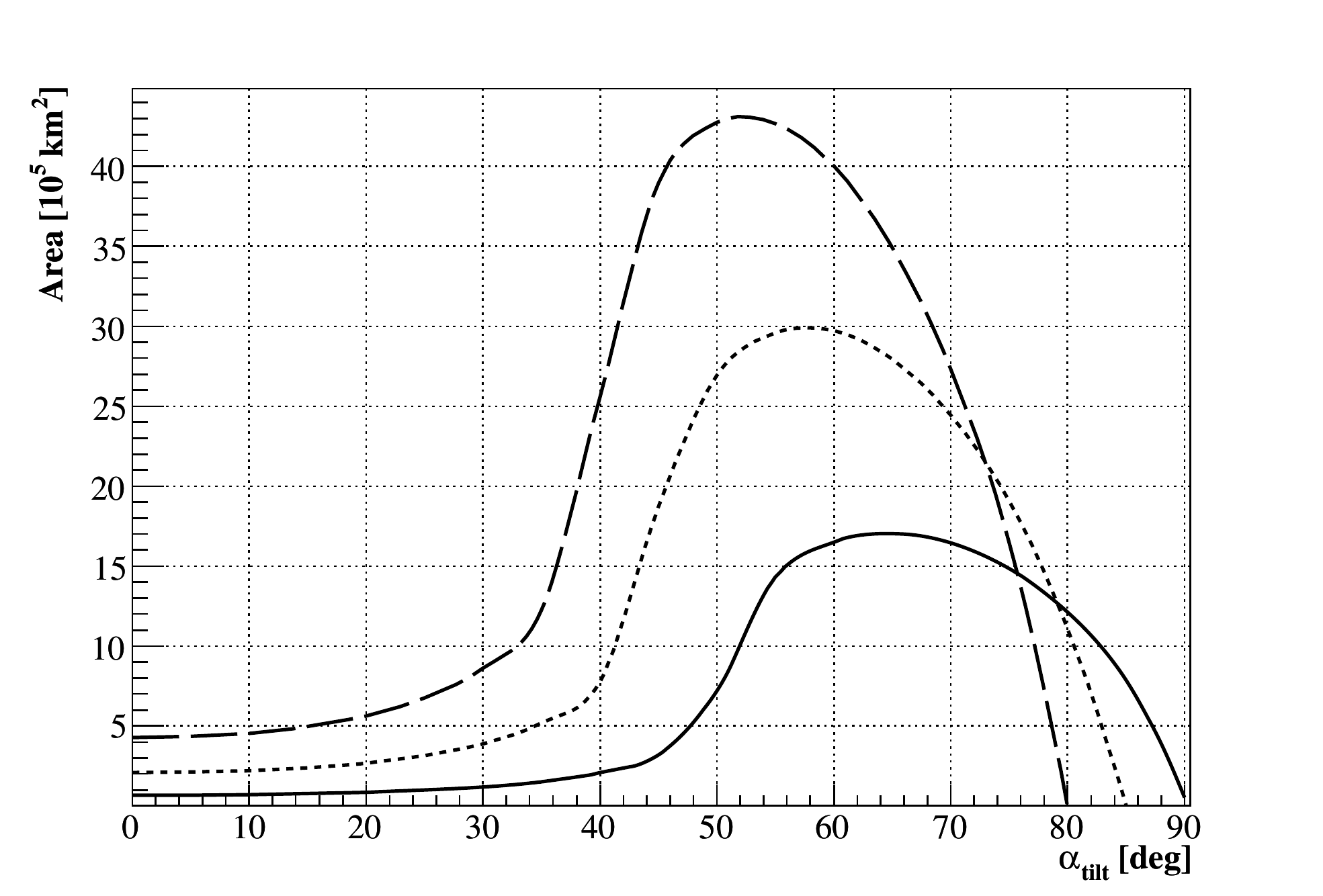}
	\caption{Area seen by the detector at the Earth surface versus
		$\TiltAngle$ for $\Gmax = 20\degr$. Solid line: $H =
		\un[400]{km}$ ($\beta_\mathrm{hor}\approx70\degr$),
		dotted line $H = \un[700]{km}$
		($\beta_\mathrm{hor}\approx64\degr$), dashed line: $H =
		\un[1000]{km}$ ($\beta_\mathrm{hor}\approx60\degr$).}
	\label{fig:tiltmulti}
\end{figure}

\begin{figure}[htbp]
	\centering
		\includegraphics[width=0.9\textwidth]{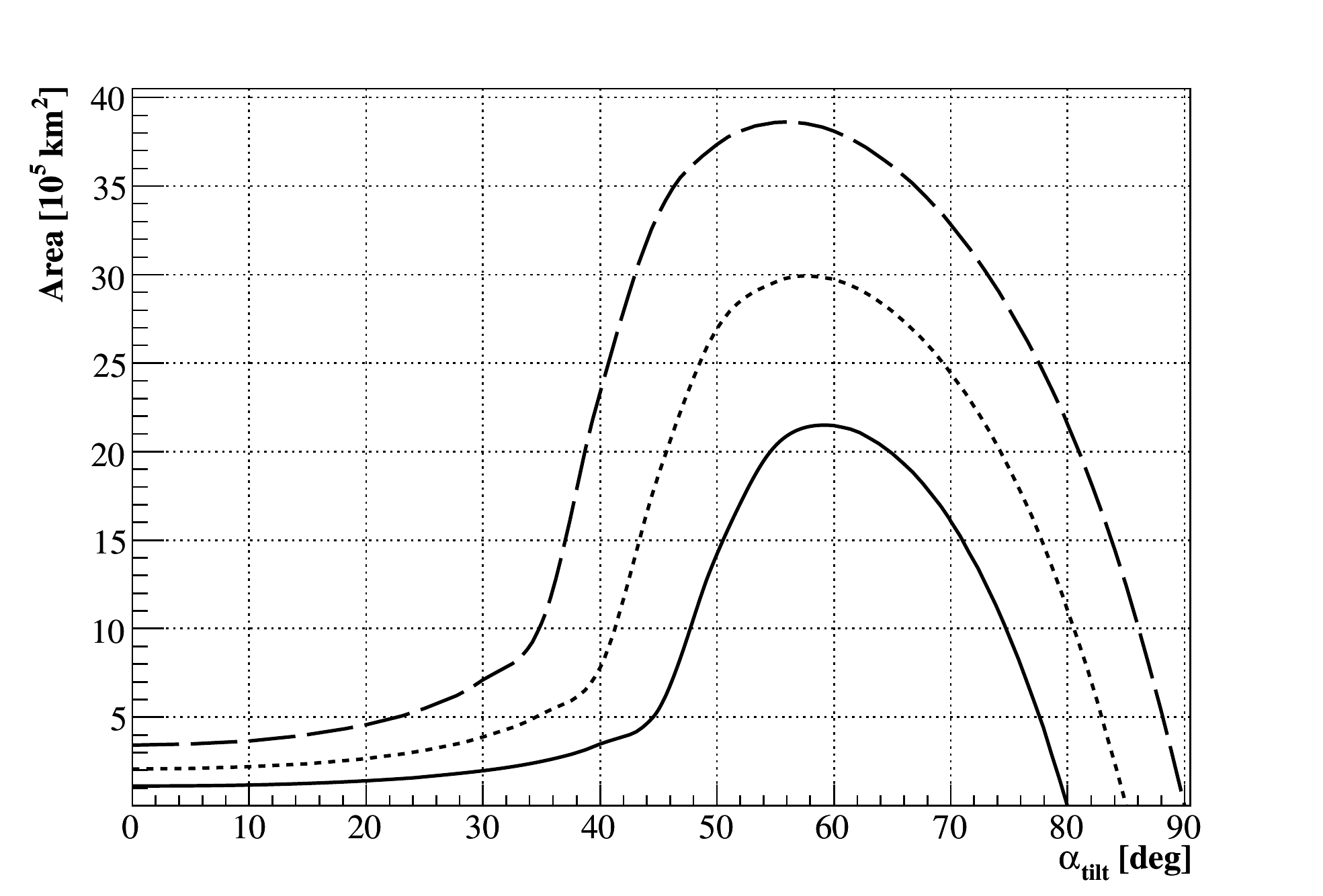}
	\caption{Area seen by the detector at the Earth surface versus
		$\TiltAngle$ for $H = \un[700]{km}$. Solid line: $\Gmax
		= 15\degr$, dotted line $\Gmax = 20\degr$, dashed line:
		$\Gmax = 25\degr$.}
	\label{fig:tiltmultigamma}
\end{figure}

The Monte-Carlo results for the intersection area, as a function of $\TiltAngle$, are shown
in the figures~\ref{fig:tiltmulti} and~\ref{fig:tiltmultigamma}. 

It should be
noted that the corresponding horizon angle is 
\[\beta_\mathrm{hor} = \arcsin\frac{R_{\oplus}}{R_{\oplus}+H}\approx 
\begin{cases}
	 70\degr & \text{at}\;H=400\um{km}, \\
	 64\degr & \text{at}\;H=700\um{km},\\
	 60\degr & \text{at}\;H=1000\um{km} 
\end{cases}
\virgola
\]
which must be taken into account when considering the possible titling
angles.

Tilted mode allows to tune the instantaneous geometrical aperture, up to
a factor $(3 \div 5)$.
While titling the apparatus is an effective way to increase the
instantaneous geometrical aperture this is not the only important issue
which comes into play.  In fact the main drawbacks of tilting are that
one is observing
\EAS at a larger and larger distance when looking at the far extreme of the \FoV, with a
drastically increasing atmospheric absorption.
This means that the effective energy threshold in the far part of the
\FoV increases.
Moreover the \FoV is highly non uniform and the angular resolution at the far
extreme becomes worse unless the pixel size is reduced.
The large \FoV of the optics, together with the large atmospheric
target observed, would require an excellent stray-light control for a
tilted apparatus due to
the large amount of light entering the \FoV.
Tilting, together with the large \FoV, would also strongly affect the
duty cycle, 
as the large area observed at the Earth would more often include
day-time areas.

On the other hand increasing the orbit height seems to be a more effective method
to increase the instantaneous geometrical aperture while minimizing the
drawbacks mentioned above.
In fact one would have no losses from the increased atmospheric transmission
and a more uniform \FoV and would benefit from smaller required \FoV in the optical design.

%--------------------------------------------------------------------------------
\subsubsection{Duty cycle}
%--------------------------------------------------------------------------------

It is not easy to estimate the duty-cycle of a space-based apparatus.
A dedicated measurement is possibly required.

The major limitation to the apparatus observational duty cycle\footnote{The
observational duty cycle is defined as the fraction of time with the apparatus
on, open and taking data.} comes from the Sun light and the Moon light. The
fraction of time in which the detector is unable to operate due to the Sun or
the Moon depends on the orbital elements and, during the moontime, on the
maximum background rate the detector for which data taking is still possible.
For an orbital height $H\sim \un[400]{km}$, an orbital inclination of $\sim
50\degr$ and requiring both the Sun and the Moon to be safely below the
horizon, the average duty cycle is $\eta _\mathrm{o}\sim 13\%$. If we accept an
additional background flux of
$\un[100]{ph}\um{m^{-2}}\um{ns^{-1}}\um{sr^{-1}}$ due to
the moonlight, we have $\eta _\mathrm{o}\sim 19\%$~\cite{Berat}.

The duty cycle may be influenced by other man-made or natural sources. They
will be discussed in section~\ref{subsec:NoiseBackground}

Observation duty cycle is driven by the acceptable background level.
Therefore it is, in principle, energy dependent but a very good
knowledge of the apparatus is required.

%--------------------------------------------------------------------------------
\subsection{Pixel size and angular resolution}
%--------------------------------------------------------------------------------

Any \EAS will be seen as a point moving inside the \FoV with a kinematics
(direction and an angular velocity) determined by the \EAS direction relative to
the line of sight from the instrument to the
\EAS instantaneous position.
The direction of the \EAS velocity vector, as seen by the detector,
can be decomposed into two components: one parallel to the line of sight 
and the other lying in the plane perpendicular to it.  
The \EAS develops approximately at the speed of light and its distance can be
considered as a known value, in the case of a space-epxeriment.
Therefore both of them can be reconstructed from 
the two-dimensional image on the FS plus the timing information.
The latter is inferred from the direction and angular velocity of the
\EAS inside the \FoV.

The pixel size, $\delta$, is driven by the Scientific Requirements and constrained by available
resources. 
In particular it affects:

\begin{itemize}

\item 
the \EAS reconstruction efficiency and the signal contamination: the number of signal
photons divided by the number of \RB photons on the
pixel roughly scales as $S/B \sim 1/\delta $ for a pixel much larger than the \EAS
track width on the FS (which basically depends on the optics PSF) 
while it saturates to a constant for a small enough pixel size;

\item 
the angular resolution;

\item 
the \Xmax resolution.

\end{itemize}

It should be kept in mind that due to the relatively small \EAS
transverse dimensions the \EAS image transverse dimensions on the FS will
be determined by the PSF only.

A pixel size much larger than the optics PSF would spoil the angular
resolution.
On the other hand a pixel size smaller than the optics PSF might be
useful, in case one has many photons per pixel to improve, via a suitable
fitting procedure, the angular
resolution.

A pixel size roughly of the same size of the optics PSF turns out to be,
usually, a good compromise.

The number of pixels has a strong impact on the Instrument budgets and
complexity. A trade-off on the pixel size is therefore very important.

An approximate and simplified analysis, leading to determine the required pixel
size, is presented below. Note that the following elementary analysis ignores
the effect of the background (making the angular resolution worse) and assumes
to have an unbiased estimator of the \EAS arrival direction. Therefore the
requirements which are derived must be considered as necessary requirements. On the other
hand the use of the diffusely reflected \Cherenkov flash might improve the
angular resolution.

%-------------------------------------------------------------------------------
\subsubsection{Angular resolution perpendicular to the line of sight}
%-------------------------------------------------------------------------------

The expected angular resolution $\Delta \beta_{\perp}$ on the \EAS direction
perpendicularly to the line of sight is readily estimated by assuming to
perform a linear fit. The error on the angle can be calculated from the standard
relations for a linear least squares fit as:
\begin{equation}
	\Delta \beta_{\perp} = \frac{\delta }{\sqrt {12}}\frac{1}{\sigma_\xi}\frac{1}{\sqrt N}
	\virgola
\end{equation}
where $N$ is the number of detected photons, $\sigma _{\xi}$ is the standard
deviation of the observed longitudinal photon distribution along the \EAS image 
(easily determined from  the methods presented in
section~\ref{sec:vislength}) and
$\delta$ is the uncertainty in the position on the FS which can be roughly
taken as the pixel size.

Using again the sample \EAS (table~\ref{tab:EASparameters}) and exploiting the
Gamma-like shape (Gaisser-Hillas function) of the \EAS longitudinal profile,
the relation between the observed (angular) \EAS length,
$\EasAngle$, (i.e. the range of the sampled values) and
$\sigma_{\xi}$ turns out to be $\EasAngle \approx 5
\sigma_{\xi}$ (for $N=100$), as discussed in section~\ref{sec:vislength}.
Therefore one finds that, in order to reach an angular resolution of the
order of $\Delta \beta_{\perp}\sim 1\degr$, a \FoV granularity of $\delta \sim 0.1\degr$ is required.

Note that the above result is consistent with the naive estimate:
\begin{equation}
	\Delta \beta_{\perp} \approx \frac{\delta}{\EasAngle}\frac{1}{\sqrt N},
	\punto
\end{equation}

\begin{figure}[htbp]
	\centering
		\includegraphics[width=0.90\textwidth]{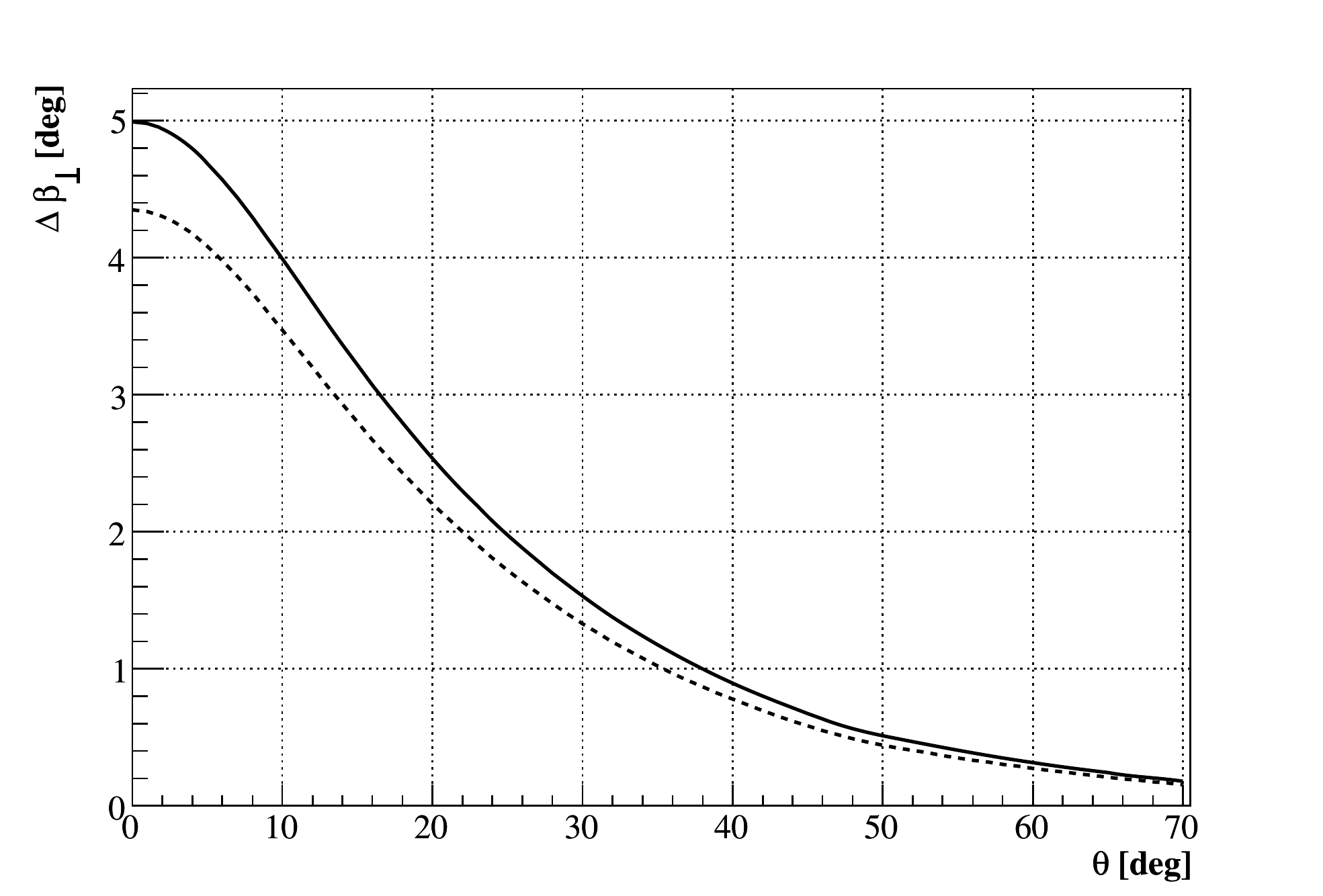}
		\caption{Angular resolution $\Delta \beta_{\perp}$ as a
		function of $\theta$ in the reference conditions. Solid line:
		$H=700\;\text{km}$; dotted line: $H=400\;\text{km}$ (the
		line for $H=1000\;\text{km}$ is very close to the first
		one). } 
	\label{fig:AngResPerp}
\end{figure}

The angular resolution $\Delta \beta_{\perp}$ as a function of $\theta$,
is shown in figure~\ref{fig:AngResPerp}. 

%--------------------------------------------------------------------------------
\subsubsection{Angular resolution parallel to the line of sight}
%--------------------------------------------------------------------------------

The relation between the observed angular velocity $\omega_{\EAS}$ and the
angle $\beta$ between the \EAS velocity vector and the line of sight
is the well known relation~\cite{Baltrusaitis:1985mx,Sommers:1995dm}:

\begin{equation}
    \EasAngularVelocity =
    \frac{c}{D} \pton{ {\frac{1-\cos \beta }{\sin \beta }} } = 
    \frac{c}{D}\tan \pton{ {\frac{\beta}{2}} }
\end{equation}
where $c$ is the speed of light and $D$ is the distance of the \EAS.

In the
present case, one assumes that the \EAS develops in the lower layers of the
atmosphere, within $\sim\un[15]{km}$ from the ground, so that $D$ is
approximatively known (the relative error is $\Delta D / D \lesssim 0.05 $). 
Due the non-linear relations
between $\omega_{\EAS}$ and $\beta$ the estimation of the best fit error is
more complex in this case. Therefore the error on $\beta$ is estimated by
assuming a simple measurement of the angular velocity of the \EAS. One obtains:

\begin{equation}
	\Delta \beta_{\parallel} =
	\frac{\Delta \EasAngularVelocity}{\EasAngularVelocity} =
	\left( {\frac{\delta }{\EasAngle} }+\frac{\delta_T}{\EasTimeLength } \right).
\end{equation}

For the typical \EAS of table~\ref{tab:EASparameters}, by assuming an \EAS
sampling time not larger than $\delta_T\sim 
\un[2.5]{\mu s}$, the second term is smaller than the first one. One finds $\Delta
\beta_{\parallel} \approx \un[0.15]{rad}\approx 10\degr$ by assuming again $\delta \sim
0.1\degr$. One might assume that, with a best fit, this results will scale as $N^{-1/2}$ obtaining
the desired $\Delta \beta_{\parallel} \approx  \un[0.015]{rad} \approx 1\degr$. Therefore one finds
that, in order to aim to get an angular resolution of the order of $\Delta \beta_{\parallel}\sim
1\degr$, one needs $\delta\sim 0.1\degr$  and $\delta_T \sim   \un[2.5]{\mu s}$.

Note that, as
long as $\delta_T$ is smaller than the pixel transit time the error is dominated by the pixel size and
not by timing. Moreover one might assume that, with a best fit, this results will scale as $N^{-1/2}$.

\begin{figure}[htbp]
	\centering
		\includegraphics[width=0.90\textwidth]{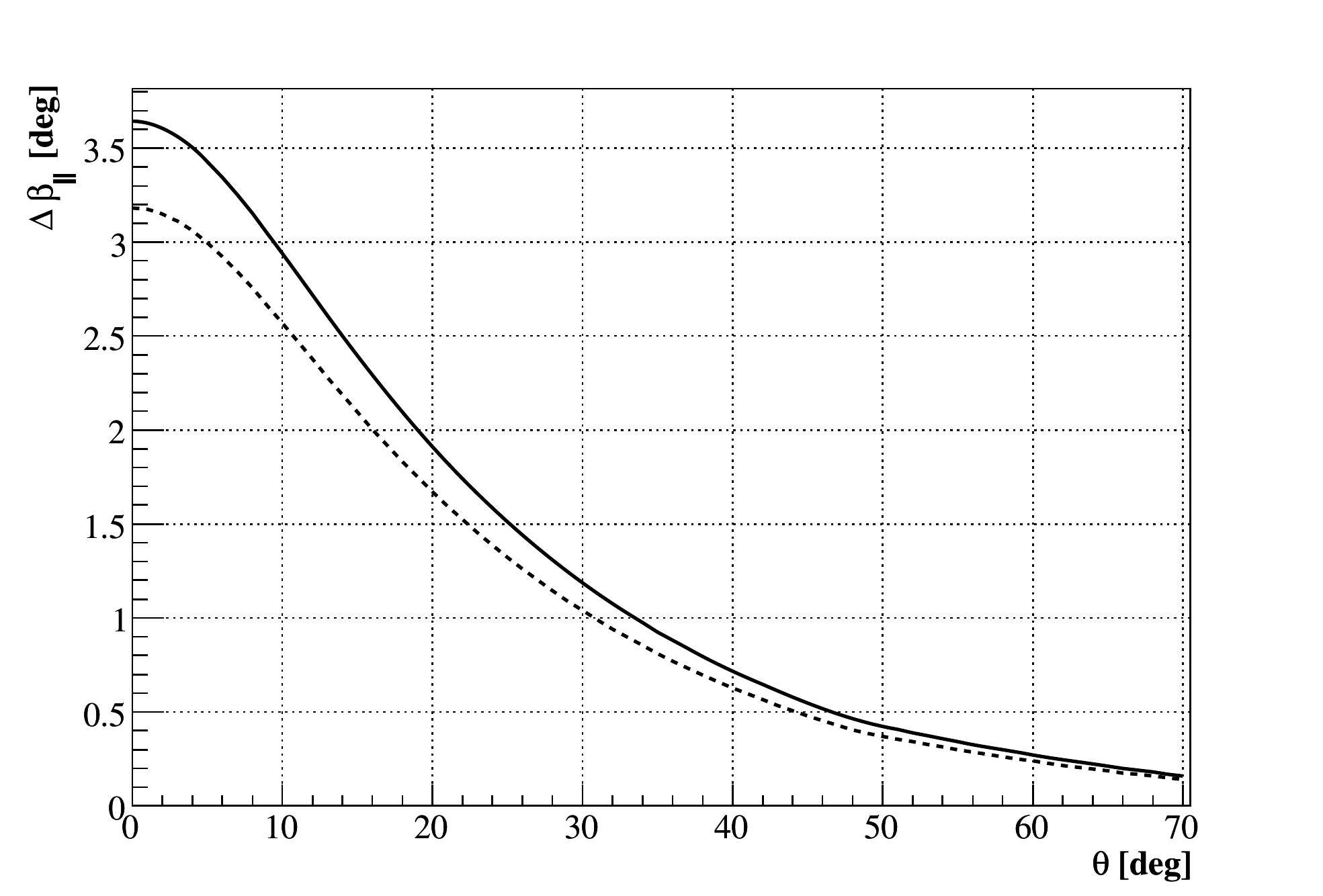}
		\caption{Angular resolution $\Delta \beta_{\parallel}$ as a function of $\theta$ in the reference conditions. Solid line: $H=700\;\text{km}$; dotted line: $H=400\;\text{km}$ (the line for $H=1000\;\text{km}$ is very close to the first one). 
		We assumed $\delta_T\sim\un[1]{\mu s}$.}
	\label{fig:AngResPar}
\end{figure}

The angular resolution $\Delta \beta_{\parallel}$ as a function of
$\theta$, is shown in figure~\ref{fig:AngResPar}.  
Some numerical estimates are given in the section~\ref{sec:AngResEst}.

%--------------------------------------------------------------------------------
\subsection{\Xmax resolution}

The requirement on $\DD{\Xmax} $ is satisfied following the previous
requirements. In fact the relation between slant depth $X$ and the
coordinate along the \EAS, $\ell$, is, 
in the approximation \eqref{atmodensity} and neglecting the Earth
curvature, is:
\begin{equation}
\left|\deriv{X}{\ell}\right|=\frac{|\cos \theta|}{h_0}X.
\end{equation}
Therefore for \Sref the $\Delta \ell$ corresponds $\Delta{\Xmax} \sim  35 \um{g/cm^2}$ ($ \Xmax \sim 824\um{g/cm^2}$) is
larger than $\un[0.5]{km}$. Therefore, neglecting the geometrical and kinematical details of the
\EAS development, it is larger than half the pixel size projected at the Earth\footnote{For an
  orbital height $H \sim \un[400]{km}$ and a pixel granularity of $\DD{\alpha} \sim 0.1\degr$, the pixel
  size projected on ground is $\sim \un[0.8]{km}$.}. The required \Xmax resolution is therefore
compatible with the pixel size. 
An improved resolution of $ \Xmax \sim 20\um{g/cm^2}$ might be reached by a suitable fitting procedure.

%--------------------------------------------------------------------------------
\subsection{Noise and background}
\label{subsec:NoiseBackground}
%--------------------------------------------------------------------------------

The \RB 
in the wavelength range $\mathrm{WR}$
is estimated from the results of the experiments 
BABY~\cite{baby},
NIGHTGLOW~\cite{nightglow}, 
Arizona-Airglow~\cite{arizona} 
and the Universitetsky-Tatiana microsatellite~\cite{tatiana}.

The noise level generated by all the parts of the experimental apparatus 
has to be kept well below the physical background level, due to the faintness of the air scintillation signal.

The total \RB rate intercepted by the apparatus 
(on the whole entrance pupil and full \FoV) is:
\beq
	\nu_\mathrm{tot}^\mathrm{b} = B A \pi \sin^2\pqua{ \Gmax }
\punto
\eeq

A typical value for the 
total rate of \RB hits detected by the apparatus is of the order of a few
hundreds of $\um{MHz}$ (see section~\ref{sec:NightGlowEst}).
Therefore the intrinsic noise of the whole apparatus (all sources) is required to be less
than a few GHz (over the whole Photo-Detector) in order to be negligible with
respect to the \RB. 
Note that this requirement should also include the
stray-light coming from lack of light-tightness of the apparatus.

In order to detect the faintest \EAS reaching the required energy threshold it
is mandatory to subtract online the \RB as its rate is
significant with respect to the air scintillation signal.
Therefore a continuous monitoring of the average background on a pixel-by-pixel basis
is, most likely, unavoidable to go low in energy in order to have a measure of
the \RB on the space-time scale comparable to the
space-time scale of the \EAS development.

As a preliminary step for a space-based mission it is therefore necessary to
carry on a detailed characterization of the \RB, 
on the space-time scale of the EAS development,
to improve our knowledge of it and improve background rejection~\cite{spacepart06}.

Many experiments have measured this quantity. However in order to devise a
method for online background subtraction it is mandatory to have a finer
characterization of the space-time behavior of the \RB
on the space-time scale of the EAS development: $ \DD{\alpha} \simeq 0.1\degr$
and $ \DD{t} \simeq \um{\mu s}$.
It is important to stress that the characterization should include measurements
along different directions from nadir to cope with off-field \RB.

%-------------------------------------------------------------------------------
\section{Some order of magnitude estimates}\label{sec:Estimates}
%-------------------------------------------------------------------------------

%-------------------------------------------------------------------------------
\subsection{Air scintillation signal}\label{sec:SignalEst}
%-------------------------------------------------------------------------------

Consider a typical hadron-induiced \EAS with 
$ E = 10^{19}\um{eV} $, 
at $\gamma= 15\degr$, with an ideal optics ($ \veps_\mathrm{O}^\prime[\gamma]=\cos\pqua{\gamma} $).

The signal time-integrated irradiance, $\diffl{\mathcal{N}}/\diffl{A}$, reaching the apparatus is given in table~\ref{tab:irradiance} for different values of $\theta$ and $H$ (the other parameter values are the one listed in table~\ref{tab:ReferenceConditions}).

Note the the irradiance scales as $\sim\left(\cos\gamma/H\right)^2$.

\begin{table}[htb]
\begin{center}
\begin{tabular}{c|ccc}
$\diffl{\mathcal{N}}/\diffl{A}\approx\:[\un{ph}\un{m^{-2}}]$ & $\theta=30\degr$ & $\theta=50\degr$ & $\theta=70\degr$ \\ \hline
$H=\un[400]{km}$  & 40 & 70 & 180 \\
$H=\un[700]{km}$  & 15 & 25 & 60 \\
$H=\un[1000]{km}$  & 5 & 10 & 30 \\ 
\end{tabular}
\caption{Time-integrated irradiance $\deriv{ \mathcal{N} }{A}$ of an \EAS.}\label{tab:irradiance}
\end{center}
\end{table}

The detection of such an \EAS obviously requires a large entrance pupil of many squared meters (see next section).

The typical angular length and the apparent time duration of the signal are given respectively in table~\ref{tab:anglength} and~\ref{tab:duration}.

\begin{table}[htb]
\begin{center}
\begin{tabular}{c|ccc}
$\EasAngle\approx\:[\degr]$ & $\theta=30\degr$ & $\theta=50\degr$ & $\theta=70\degr$ \\ \hline
$H=\un[400]{km}$  & 0.8 & 1.7 & 3.1 \\
$H=\un[700]{km}$  & 0.4 & 1.0 & 1.8 \\
$H=\un[1000]{km}$  & 0.3 & 0.7 & 1.2 \\ 
\end{tabular}
\caption{Angular length \EasAngle of an \EAS.}\label{tab:anglength}
\end{center}
\end{table}

\begin{table}[htb]
\begin{center}
\begin{tabular}{c|ccc}
$\EasTimeLength\approx\:[\un{\mu s}]$ & $\theta=30\degr$ & $\theta=50\degr$ & $\theta=70\degr$ \\ \hline
$H=\un[400]{km}$  & 60 & 84 & 102 \\
$H=\un[700]{km}$  & 60 & 84 & 103 \\
$H=\un[1000]{km}$  & 60 & 84 & 104 \\ 
\end{tabular}
\caption{Apparent time duration \EasTimeLength of an \EAS.}\label{tab:duration}
\end{center}
\end{table}

%-------------------------------------------------------------------------------
\subsection{Requirements on the optical triggering efficacy}\label{sec:TheoOptRequ}
%-------------------------------------------------------------------------------

Following the discussion in section~\ref{sec:NumPh} we can calculate a 
required overall photo-detection efficacy in the reference conditions
\[
 \PDEfficacy(\gamma=15\degr) \approx N_\mathrm{ph}\left(\deriv{ \mathcal{N} }{A}\right)^{-1}  \virgola
 \] 
where we remember that $N_\mathrm{ph}\simeq 100$. Using the values of the time-integrated irradiance given in the
previous section we obtain $\PDEfficacy(\gamma=15\degr)\approx \un[4]{m^2}$
when observing at $H=\un[700]{km}$, $\PDEfficacy(\gamma=15\degr)\approx \un[1.5]{m^2}$
at $H=\un[400]{km}$ and $\PDEfficacy(\gamma=15\degr)\approx \un[10]{m^2}$ 
at $H=\un[1000]{km}$.

If one conservatively assumes a safe overall (including filters, electronics, photo-sensor and filling factor) efficiency $\veps_\mathrm{PD}\sim0.1$, we obtain an optical triggering efficacy, from the~\eqref{eq:PDEfficacyTerms},
\begin{equation}\label{eq:OptEfficacy:Req}
    \veps_\mathrm{O}'(\gamma=15\degr)\approx \frac{\PDEfficacy}{\veps_\mathrm{PD}} \approx
    \begin{cases}
     \un[15]{m^2}  &\text{at}\:H=\un[400]{km} \\
     \un[40]{m^2}  &\text{at}\:H=\un[700]{km} \\
     \un[100]{m^2} &\text{at}\:H=\un[1000]{km} \\
    \end{cases}   
    \punto
\end{equation}

The triggering optical efficacy at $\gamma=0\degr$, $A\veps_\mathrm{O}'(0\degr)$, i.e. the entrance pupil area,
is not only the effective collection area but it is also an estimate of the
physical area of the optics, actually an optimistic one, and 
consequently a requirement on the minimum size of the telescope.  

Translated into
minimum area and diameter for the optics we obtain, for an ideal optics in which $\mathcal{E}(\gamma)=S\cos\gamma$,
\begin{equation}\label{eq:ADest}
    \begin{split}
        A_\mathrm{min}&=\OptEfficacy^\mathrm{tri}(0\degr)\approx\frac{\OptEfficacy^\mathrm{tri}(15\degr)}{\cos 15\degr}\approx 				
        \begin{cases}
        \un[16]{m^2}  &\text{at}\:H=\un[400]{km} \\
     \un[42]{m^2}  &\text{at}\:H=\un[700]{km} \\
     \un[104]{m^2} &\text{at}\:H=\un[1000]{km} \\
        \end{cases} \\
        D_\mathrm{min}&=2\sqrt{\frac{A\veps_\mathrm{O}'(0\degr)}{\pi}}\approx
        \begin{cases}
        \un[4.5]{m}  &\text{at}\:H=\un[400]{km} \\
     \un[7.5]{m}  &\text{at}\:H=\un[700]{km} \\
     \un[12]{m} &\text{at}\:H=\un[1000]{km} \\
        \end{cases}        
    \end{split}
    \punto
\end{equation}

This result cannot be applied without accounting for the various sources of
inefficiencies in a real optical system.
Based on existing designs these inefficiencies can be assumed to reduce the
amount of photons in the bucket depending on many factors.
Therefore, the previous lower limit on $D_\mathrm{min}$ might rise by a factor of $\gg 1$,
as confirmed by Monte-Carlo simulations~\cite{esaf1,esaf2,TheaThesis,PesceThesis}.

%-------------------------------------------------------------------------------
\subsection{Granularity and angular resolution}\label{sec:AngResEst}
%-------------------------------------------------------------------------------

If we require a spatial granularity on Earth
surface $\DD{\ell} \lesssim \un[1]{km}$, we
can fix the corresponding pixel granularity at different orbital heights:

\begin{equation}\label{eq:DeltaAlphaEst}
    \Delta\alpha \approx \cfrac{\DD{\ell}}{H} \approx
    \begin{cases}
      0.10\degr &\text{at}\:H=\un[400]{km} \\
      0.06\degr &\text{at}\:H=\un[700]{km} \\
      0.04\degr &\text{at}\:H=\un[1000]{km} \\
    \end{cases}   
    \punto
\end{equation}

Using the values of table~\ref{tab:anglength}, one can easily see that
at $\theta=50\degr$, the \EAS is about 17 pixels long; 
so there are about 6 photons per pixel. 

The pixel size on the FS is, from~\eqref{eq:pixels-1}, $\delta\approx
D\Fnumb \DD{\ell}/H\approx\un[5]{mm}$, using the minimum diameter given
in~\eqref{eq:ADest} and a $\Fnumb\sim 0.5$. The value of $\delta$
doesn't depend on $H$ since  
$D_\mathrm{min}\propto H$.

With these numbers one can see that, in the reference conditions, the angular resolution perpendicular and parallel to the line of sight are both $\sim 0.5\degr$. The total angular resolution is then  $\DD{\beta}_\mathrm{tot}=\sqrt{\DD{\beta_\perp}^2 + \DD{\beta_\parallel}^2}\lesssim 1\degr$.

%-------------------------------------------------------------------------------
\subsection{The random background (RB)}
\label{sec:NightGlowEst}
%-------------------------------------------------------------------------------

The total \RB rate intercepted on the whole entrance pupil by the apparatus in the reference conditions, 
with $\Gmax=20\degr$ (corresponding to \un[0.38]{sr} full \FoV) is given in table~\ref{tab:RB} together with the
corresponding number of pixels and the total \RB rate detected per pixel (assuming a total photon efficiency, optics plus
photo-detector, $ \veps_\mathrm{T} \approx 0.1 $).

\begin{table}[htb]
\begin{center}
\begin{tabular}{cccc}
H [km] &  $N_\mathrm{pix}$   &  $\RB_\mathrm{tot}$ [THz] & $\RB_\mathrm{pix}$ [MHz] \\ \hline
400    &  \sci{1.2}{5}       &  2.8                      & 2.3 \\
700    &  \sci{3.5}{5}       &  7.6                      & 2.2 \\
1000   &  \sci{7.8}{5}       &  19.0                     & 2.4 \\
\end{tabular}
\end{center}
\caption{Number of pixels, total \RB rate ($=B A \pi \sin^2\pqua{ \Gmax }$) and \RB rate per pixel 
($=B A \pi \sin^2\pqua{ \Gmax }  \veps_\mathrm{T} / N$). We used the values of $A$ and $\DD{\alpha}$ from~\eqref{eq:ADest} 
and~\eqref{eq:DeltaAlphaEst} respectively.}
\label{tab:RB}
\end{table}

This gives:
 one order of magnitude more \RB than signal photons 
superimposed on the typical \EAS (all space-time length) and
roughly the same number of signal and \RB photons near the \EAS maximum.
The acceptable background level also depends on the energy of the \EAS.
However to allow for background dependent observations implies a very precise knowledge 
of the Apparatus sensitivity as a function of the background level.

%--------------------------------------------------------------------------------
\section{Conclusions}
\label{sec:Conclusions}
%--------------------------------------------------------------------------------

The design of a space-based apparatus for the detection of the \EAS
produced by \UHECP is a challenging task, requiring a careful design and
a preparation based on both preliminary studies and measurements and
various technological demonstrators.

Some sort of path-finder and/or technology demonstrator is certainly needed.

%--------------------------------------------------------------------------------
\section*{Acknowledgments}
\addcontentsline{toc}{section}{Acknowledgments}
%--------------------------------------------------------------------------------

The authors wish to thank the members of the \EUSO Collaboration and in particular
Osvaldo Catalano (INAF/IASF-Palermo),
Lloyd Hillman (University of Alabama Huntsville) deceased,
Didier Lebrun (LPSC, Grenoble),
Piero Mazzinghi (INOA, Firenze) and 
Sergio Bottai (INFN, Firenze)
for many useful discussions and suggestions.

The pioneering work~\cite{arisaka} of Katsushi Arisaka (UCLA, Los Angeles) is acknowledged.

\nocite{*}

\bibliographystyle{unsrt}
\bibliography{space_biblio}

\end{document}